\documentclass{aa}  

\usepackage{graphicx}
\usepackage{txfonts}
\usepackage{xcolor}
\usepackage{hyperref}
\usepackage{placeins}
\begin{document} 

   \title{White dwarf-neutron star binaries: a plausible pathway for long-duration gamma-ray bursts from compact object mergers?}


   \author{A.~A. Chrimes
          \inst{1}\fnmsep\inst{2}\thanks{ESA Research Fellow},\
          N. Gaspari\inst{2},\
          A.~J. Levan\inst{2}\fnmsep\inst{3},\
          M.~M. Briel\inst{4,5},\
          J.~J. Eldridge\inst{6},\
          B.~P. Gompertz\inst{7},\
          G. Nelemans\inst{2,8,9},\
          A.~E. Nugent\inst{10}\fnmsep\inst{11},\
          J.~C. Rastinejad\inst{10}\fnmsep\inst{11}
          \and W.~G.~J. van Zeist\inst{2,6,12}         
          }

   \institute{European Space Agency (ESA), European Space Research and Technology Centre (ESTEC), Keplerlaan 1, 2201 AZ Noordwijk, the Netherlands \\
              \email{ashley.chrimes@esa.int}
         \and
            Department of Astrophysics/IMAPP, Radboud University, PO Box 9010, 6500 GL Nijmegen, The Netherlands
         \and
            Department of Physics, University of Warwick, Gibbet Hill Road, CV4 7AL Coventry, United Kingdom 
         \and
            Département d’Astronomie, Université de Genève, Chemin Pegasi 51, CH-1290 Versoix, Switzerland
         \and
            Gravitational Wave Science Center (GWSC), Université de Genève, CH1211 Geneva, Switzerland
         \and
            Department of Physics, Private Bag 92019, University of Auckland, Auckland 1010, New Zealand
         \and
            Institute for Gravitational Wave Astronomy and School of Physics and Astronomy, University of Birmingham, B15 2TT, UK
         \and
            SRON, Netherlands Institute for Space Research, Niels Bohrweg 4, 2333 CA, Leiden, The Netherlands
         \and
            Institute of Astronomy, KU Leuven, Celestijnenlaan 200D, B-3001, Leuven, Belgium
         \and
            Center for Interdisciplinary Exploration and Research in Astrophysics, Northwestern University, 1800 Sherman Ave., Evanston, 60208, IL, USA
         \and
            Department of Physics and Astronomy, Northwestern University, 2145 Sheridan Road, Evanston, 60208-3112, IL, USA
         \and
            Leiden Observatory, Leiden University, Einsteinweg 55, 2333 CC, Leiden, The Netherlands
             }

   \date{Received April 11, 2025; accepted August 14, 2025}


 
  \abstract
   {Two long-duration gamma-ray bursts (GRBs) were recently discovered with kilonovae, the signature of $r$-process element production in a compact binary merger, rather than supernovae. This has forced a re-evaluation of the long-established dichotomy between short bursts ($<2$s, arising from compact binary mergers) and long bursts ($>2$s, a class of massive star core-collapse event).}
   {We aim to determine whether white dwarf-neutron star (WDNS) and white dwarf-black hole (WDBH) mergers are plausible explanations for long-duration compact merger GRBs, in terms of their galactocentric merger offsets and cosmological rates.}
   {We model the host galaxies of GRBs\,211211A and 230307A, and employ binary population synthesis, to predict the merger offset distributions of compact binaries. We compare with the observed (projected) offsets of GRBs\,211211A and 230307A. We also investigate the evolutionary pathways to WDNS and WDBH mergers, predict their cosmological rates, and compare with inferred volumetric GRB rates.}
   {We find that WDNS mergers occur at lower host offsets than binary neutron star mergers, but that in the specific cases of GRBs\,211211A and 230307A, the observed offsets are consistent with either scenario. We predict that WDNS mergers occur at a similar rate to binary neutron star mergers and long GRBs, and that WDBH mergers are a factor of ten rarer, with the caveat that these rates currently carry uncertainties at the order of magnitude level.}
   {We have demonstrated, solely in terms of galactocentric offsets and event rates, that WDNS mergers are a plausible explanation for GRBs\,211211A and 230307A, and long duration gamma-ray bursts from compact object mergers more generally. WDNS binaries have lower systemic velocities than binary neutron stars, but longer delay times, and ultimately merge with an offset distribution that is not measurably different without large sample sizes. Therefore, offsets and rates alone cannot currently distinguish between compact binary progenitor models for supernova-less long duration GRBs.}

   \titlerunning{WDNS binaries for long GRBs from compact mergers}
   \authorrunning{A. A. Chrimes et al.}
   \maketitle
%

\section{Introduction}
Gamma-ray bursts (GRBs) are extragalactic transients which can be broadly described, observationally, as prompt flashes of gamma-rays followed by multi-wavelength synchrotron afterglows \citep[e.g.][]{2004RvMP...76.1143P}. Their durations are typically measured in terms of T$_{90}$, the time taken for 90\% of the prompt $\gamma$-ray flux to arrive. The GRB duration distribution shows bimodality which can be characterised by two log-normals \citep{1993ApJ...413L.101K}. While these distributions overlap, and the positions of the peaks and their overlap depends on the photon energies and instrument sensitivities \citep[e.g.][]{2020ApJ...893...46V}, the standard nomenclature is to describe bursts with T$_{90} < 2$s as short bursts, and those with T$_{90} > 2$s as long bursts. The dichotomy becomes even clearer when the spectral hardness of the emission is considered: short GRBs tend to be spectrally harder than long GRBs \citep[e.g.][]{2016ApJS..223...28N}. 

Over the past 25 years, a picture emerged in which short and long bursts had two distinct classes of progenitor. The association of long bursts with type Ic-broad line supernovae \citep[][]{1998Natur.395..670G,2003Natur.423..847H}, along with their preferential occurrence in the bright regions of typically low-mass, low-metallicity and star-forming galaxies \citep{2006Natur.441..463F,2020MNRAS.491.3479C,2024arXiv241010378F}, and the presence of wind-like circumstellar medium profiles in many cases \citep{2006MNRAS.367..186E,2006A&A...460..105V}, firmly established their nature. These are stripped-envelope, likely rapidly rotating, massive star core-collapse (`collapsar') events. In this scenario, the collapsing star launches relativistic jets from the newly formed compact object. If a jet is oriented along our line of sight, we observe a GRB.

Short GRBs, meanwhile, were suspected to be the result of compact binary mergers, likely binary neutron star mergers, in which jets are launched from the combined remnant produced in the merger. The varied host galaxy population - including some ancient, elliptical galaxies \citep[e.g.][]{2022ApJ...940...57N} - and the prevalence of short GRBs occurring at large projected offsets from their hosts \citep{2002AJ....123.1111B,2013ApJ...776...18F,2014MNRAS.437.1495T,2022ApJ...940...56F,2022MNRAS.515.4890O}, was evidence in favour of the merger interpretation. In this scenario, a natal kick imparted in the formation of the (neutron star) remnants 
\citep{2005MNRAS.360..974H,2017A&A...608A..57V,2020MNRAS.494.3663I,2023MNRAS.519.5893K,2024A&A...689A.348D}
imparts the binary with linear momentum and hence a systemic velocity, which can reach hundreds of km\,s$^{-1}$ 
\citep[e.g.][]{2017ApJ...846..170T,2025arXiv250301429D}.
Combined with long gravitational wave in-spiral times, such binaries can eventually merge well outside the stellar light of their host galaxies \citep[e.g.][]{2010ApJ...722.1946B,2013ApJ...776...18F,2014MNRAS.437.1495T,2022ApJ...940...56F,2022MNRAS.515.4890O,2024MNRAS.527.1101G,2024A&A...692A..21G}. Further evidence for the binary neutron star (BNS) merger scenario came from the discovery of kilonovae in short GRB optical light-curves \citep[e.g.][]{2013Natur.500..547T,2018ApJ...860...62G,2021ApJ...916...89R,2025ApJ...979..190R}, the tell-tale signature of heavy $r$-process elements recently produced in a neutron-rich environment. The BNS model was finally confirmed beyond doubt by the coincident detection of the gravitational wave signal GW170817 \citep{2017PhRvL.119p1101A} and the short-duration GRB170817A \citep[e.g.][]{2017ApJ...848L..13A,2018NatAs...2..751L}, along with an associated kilonova, AT2017gfo \citep[e.g.][]{2017ApJ...848L..17C,2017ApJ...848L..27T,2017ApJ...848L..12A}.

As of the early 2020s, a clear picture had emerged: long duration, spectrally soft bursts are from collapsars, and short duration, spectrally hard bursts are from binary neutron star mergers. However, two events have since challenged this interpretation of the GRB population. The first, GRB\,211211A, had an unambiguously long-duration of 51s. In the bimodal model for GRB durations described by two Gaussians, this is far down the tail of the `short' GRB distribution \citep{2023ApJ...954L...5V}. It was also spectrally soft, and therefore - based on the prompt emission alone - indistinguishable from a regular long-duration burst. However, it lacked a supernova to deep limits \citep[strong constraints were possible thanks to a redshift of $z=0.076$, in the lowest $\sim$1\% of GRBs][]{2016ApJ...817....7P}, and instead displayed a kilonova in the post-burst light-curve remarkably similar to GW170817's kilonova AT2017gfo \citep{2022Natur.612..223R,2022Natur.612..232Y,2022Natur.612..228T}. Combined with a projected offset of $\sim$8\,kpc, placing it well outside a low-mass, star-forming galaxy - GRB211211A clearly demonstrated that compact object mergers can in fact produce canonically long-duration GRBs. The second event, GRB230307A, was similarly long duration/spectrally soft (T$_{90}$=35s), low-redshift ($z=0.065$), lacking a supernova and again showing a kilonova instead, this time spectroscopically confirmed with JWST \citep{2024Natur.626..737L,2024Natur.626..742Y}. GRB\,230307A occurred at a large offset of $\sim$40\,kpc from its host galaxy. The association with this galaxy is based on probability of chance alignment arguments \citep{2002AJ....123.1111B}. If GRB\,230307A is placed at the redshift of this galaxy, its kilonova is comparable in terms of luminosity with AT2017gfo, supporting the association.

GRBs\,211211A and 230307A have demonstrated that long bursts can arise from compact binary mergers. This is prompting a re-evaluation of previous long-duration events which did not follow the long-collapsar/short-merger dichotomy by lacking supernovae to deep limits. Such events were assigned alternative explanations \citep[e.g. GRB191019A,][suggested as the merger of dynamically formed binary in a dense environment, given its offset of less than 100pc from the host nucleus]{2023NatAs...7..976L,2024ApJ...963..156W,2024ApJ...965L..20E,2024arXiv241204059S}, or were otherwise considered to be outliers \citep[e.g. GRB111005A,][also nuclear]{2018A&A...616A.169M}. Although challenges to the dichotomy have existed for nearly 20 years \citep[e.g. GRB060614,][lying at an offset of 0.7kpc]{2006Natur.444.1047F,2006Natur.444.1044G,2007MNRAS.374L..34K}, the recent events GRBs\,211211A and 230307A stand apart from these examples by occurring at large projected (and host-normalised) offsets from their host galaxies.

It is theoretically difficult to produce sustained accretion in a binary neutron star merger, since there is only $\sim$3--4\,M$_{\odot}$ of (extremely tightly bound) material present in the system \citep{2025arXiv250100239Z}. Replacing one of the components with a white dwarf is a possible solution to this problem, provided the binary doesn't stabilise as an ultra-compact X-ray binary \citep{2012A&A...537A.104V,2017MNRAS.467.3556B}, as stabilised systems can persist for gigayear timescales. White dwarf-neutron star (WDNS) and white dwarf-black hole (WDBH) mergers have been discussed in the context of (long-duration) GRBs without supernovae for well over 20 years \citep{1999ApJ...520..650F,1999ApJ...526..152F,2007MNRAS.374L..34K,2009A&A...498..501C,2018MNRAS.475L.101D}, and such models have seen a revival in interest following GRBs\,211211A and 230307A \citep[e.g.][]{2022Natur.612..232Y,2023ApJ...947L..21Z,2024MNRAS.535.2800L,2024ApJ...973L..33C}. Such mergers are only viable explanations for SN-less long GRBs if they fail to produce bright optical transients. A number of works have performed detailed hydrodynamic and radiative transfer simulations of WDNS and WDBH mergers. These typically predict faint and/or fast optical transients \citep[e.g.][]{2009PhRvD..80b4006P,2012MNRAS.419..827M,2016MNRAS.461.1154M,2019ApJ...885L..23M,2019MNRAS.486.1805Z,2020MNRAS.493.3956Z,2020MNRAS.497..246G,2022MNRAS.510.3758B,2023ApJ...956...71K,2025ApJ...984...77G}.

Nevertheless, the diversity of behaviour seen amongst kilonovae \citep[][]{2018ApJ...860...62G,2020MNRAS.493.3379R,2025ApJ...979..190R} may suggest that different populations of compact mergers contribute. BHNS mergers are a possible alternative \citep[e.g.][]{2020ApJ...895...58G}. The dynamical timescale suggests they should produce GRBs of similar duration to BNS mergers, but longer durations are possible if there is substantial fallback accretion \citep{2007MNRAS.376L..48R,2019MNRAS.485.4404D} or an otherwise particularly massive accretion disc \citep{2023ApJ...958L..33G}. Another suggestion is that compact binary mergers (typically BNS or WDNS) can produce magnetar-powered GRBs and kilonovae \citep{2008MNRAS.385.1455M,2012MNRAS.419.1537B,2013MNRAS.431.1745G,2014MNRAS.438..240G,2022MNRAS.516.2614A}, which may explain some of the properties of GRBs\,211211A and GRB230307A, including their extended duration \citep{2022Natur.612..232Y,2023ApJ...947L..21Z,2024ApJ...962L..27D,2024ApJ...964L...9W,2024A&A...681A..41M,2024ApJ...967..156P,2025arXiv250100239Z,2025ApJ...978...52A}. However, recent semi-analytic and numerical results suggest that WDNS mergers are unlikely to produce kilonovae \citep[e.g.][and references therein]{2023ApJ...956...71K,2025ApJ...984...77G}.

One approach to investigating feasible transient progenitor channels is through (binary) population synthesis. There is an extensive body of literature performing such studies in the context of binary compact object mergers \citep[see e.g.][for reviews]{2022MNRAS.516.5737B,2022LRR....25....1M}. Many codes employ stellar evolution prescriptions based on analytic stellar evolution approximations or fits to the results of detailed modelling. Examples of WDNS and WDBH birth and merger rate predictions with such codes include \citet[][SeBa]{2001AandA...375..890N,2012A&A...546A..70T,2018AandA...619A..53T}, \citet[][StarTrack]{2008ApJS..174..223B}, \citet[][COSMIC]{2020ApJ...898...71B,2024MNRAS.535.2800L}, \citet[][PARSEC]{2015MNRAS.451.4086S} and \citet[][SEVN]{2023MNRAS.524..426I}. Other codes use detailed binary evolution models. These have fewer simplifying approximations but are computationally far more expensive, limiting the feasibility of running many variations of the code with different modelling assumptions. Examples of work using these codes for compact binaries include POSYDON \citep[][]{2023ApJS..264...45F,2024arXiv241102376A}, which is built from a grid of detailed {\sc mesa} \citep{2011ApJS..192....3P} models, and the code employed in this paper, {\sc bpass} \citep{2019MNRAS.482..870E,2023MNRAS.524.2836V,2024arXiv241102563T,2025A&A...699A.172V}.

By folding merger rates as a function of metallicity and their delay times (since star-formation) through (metallicity-dependent) star-formation histories, we can predict event rates and compare with the inferred volumetric rates of different classes of transients. The predicted rates depend on the specific choices made when modelling the stellar evolution \citep[e.g. the common envelope efficiency,][]{2013A&ARv..21...59I}, the initial conditions \citep[e.g. binary parameter distributions and the initial mass function,][]{2020MNRAS.495.4605S,2023MNRAS.522.4430S,2024arXiv241001451D}, and the cosmic star formation history \citep[e.g.][]{2019MNRAS.482.5012C}. There are also uncertainties in the determination of the volumetric transient rates to which predictions are compared \citep[e.g.][]{2016ApJ...817....7P,2022ApJ...932...10G,2023A&A...680A..45S}.

An additional comparison can be made between the observed transient-host galaxy offset, and expectations for compact merger offsets, taking into consideration the star formation history and gravitational potential of the host galaxy \citep{1999MNRAS.305..763B,1999ApJ...526..152F,2006ApJ...648.1110B,2011MNRAS.413.2004C,2017ApJ...848L..22B,2022MNRAS.514.2716M,2024arXiv240904543W,2024A&A...692A..21G}. \citet{2018AandA...619A..53T} modelled the merger offsets of WDNS binaries in this way, using SeBa models, for a dwarf elliptical and spiral galaxy. In this paper, we investigate the viability of WDNS and WDBH mergers as a channel for compact mergers producing long-duration GRBs with kilonovae, in terms of their merger offsets and volumetric rates. For the first time, we combine binary population synthesis predictions with modelling of the host galaxies of two long-duration GRBs with kilonovae. We compare our predictions with (i) the observed galactocentric merger offsets of GRBs\,211211A and 230307A and (ii) the estimated cosmological rate of long and short GRBs. Recent estimates for the volumetric rate of SN-less long GRBs range from far less than 1\% of the total long GRB rate \citep{2022Natur.612..228T}, to as much as several tens of percent \citep{2024MNRAS.535.2800L,2024ApJ...963L..12P}. Given the challenges in detecting or ruling out the presence of kilonovae in most SN-less long GRBs (largely due to the typical distances and depth of observations), it is currently unclear what fraction of SN-less long GRBs also produce kilonovae (but see Levan et al. in prep). We therefore frame our rate predictions in terms of the total long GRB rate, to be as agnostic as possible about the relative contributions of core-collapse and merger events to this population.

We adopt a $\Lambda$CDM flat cosmology with H$_{\rm 0}=70$\,km\,s$^{-1}$\,Mpc$^{-1}$, $\Omega_{\rm m}=0.3$ and $\Omega_{\Lambda}=0.7$ \citep[between supernova type Ia and cosmic microwave background derived values,][]{2018ApJ...861..126R,2020A&A...641A...6P}, and all magnitudes are reported in the AB system \citep{1983ApJ...266..713O}.



\begin{table*}
	\centering
	\caption{Merger rate results for compact binaries containing a white dwarf and a neutron star/black hole, for the three natal kick prescriptions (Verbunt, Hobbs and Bray, left to right), and both metallicities considered in this work (0.5\,Z$_{\odot}$) above the horizontal line, 0.2\,Z$_{\odot}$ below it). $f_{\rm merge}$ is the fraction of binaries from each formation pathway that merge within a Hubble time. $f_{\rm bin}$ is the fraction of mergers within a Hubble time, for each class of binary (e.g. WDNS), arising from each pathway within that class (e.g. NS1-WD2). We also list the number of mergers from each pathway in a Hubble time, per Solar mass of stars formed. Although they dominate mergers within a Hubble time, at ages $\lesssim$10,Gyr, the direct (NS1 and BH1) channels are negligible and the reverse channel dominates. The WD1 pathways are therefore over-represented in the GRB host stellar populations studied in this work, which are $<$10\,Gyr old (see Figure \ref{fig:sfh}).}
	\label{tab:evolution}
	\begin{tabular}{l@{\hspace{2em}}ccc@{\hspace{2em}}ccc@{\hspace{2em}}ccc} 
		\hline
		\hline
            &
            \multicolumn{3}{c}{\citet{2017A&A...608A..57V}}\hspace{2em} & \multicolumn{3}{c}{\citet{2005MNRAS.360..974H}}\hspace{2em} &  \multicolumn{3}{c}{\citet{2016MNRAS.461.3747B}}\hspace{2em} \\
		  Binary  & $f_{\rm merge}$ & $f_{\rm bin}$ & $\log_{10}$($N$/M$_{\odot}$) & $f_{\rm merge}$ & $f_{\rm bin}$ & $\log_{10}$($N$/M$_{\odot}$)  & $f_{\rm merge}$ & $f_{\rm bin}$ & $\log_{10}$($N$/M$_{\odot}$) \\
        \hline
        \multicolumn{10}{c}{$Z$ = 0.5\,Z$_{\odot}$} \\
		\hline
            WD1-NS2 & 0.03 		& 0.22 	& -6.52 & 0.10 		& 0.28 	& -6.56	& 0.01 		& 0.14	& -6.69 \\
            NS1-WD2 & 10$^{-2.5}$	& 0.78 	& -5.98 & 0.01 		& 0.72 	& -6.14	& 10$^{-2.6}$	& 0.86 	& -5.92 \\
            WD1-BH2 & -- 	  	& -- 	& -- 	& -- 		& -- 	& -- 	& -- 		& -- 	& -- 	\\
            BH1-WD2 & 10$^{-3.3}$ 	& 1.00 	& -7.18 & 10$^{-2.9}$	& 1.00 	& -7.11 & 10$^{-3.3}$ 	& 1.00 	& -6.97 \\
        \hline
        \multicolumn{10}{c}{$Z$ = 0.2\,Z$_{\odot}$} \\
		\hline
            WD1-NS2 & 0.03 		& 0.12 	& -6.38 & 0.14 		& 0.23 	& -6.17	& 0.01 		& 0.05 	& -6.63 \\
            NS1-WD2 & 0.01 		& 0.88 	& -5.52 & 0.02 		& 0.77 	& -5.64 & 0.01 		& 0.95 	& -5.37 \\
            WD1-BH2 & -- 		& -- 	& --  	& -- 		& -- 	& -- 	& -- 		& -- 	& --  	\\
            BH1-WD2 & 10$^{-3.4}$  	& 1.00 	& -7.13	& 10$^{-3.4}$	& 1.00 	& -7.35	& 10$^{-3.2}$ 	& 1.00 	& -6.71 \\
		\hline
	\end{tabular}
\end{table*}

\section{Binary Population Synthesis}\label{sec:popsynth}
\subsection{Binary stellar evolution models}
We make use of the pre-calculated, publicly available\footnote{\url{bpass.auckland.ac.nz}} Binary Population and Spectral Synthesis v2.2.1 stellar models, which are described in full by \citet{2017PASA...34...58E} and \citet{2018MNRAS.479...75S}. These models have been used extensively for electromagnetic and gravitational wave transient progenitor models and rate estimates \citep{2019MNRAS.482..870E,2022MNRAS.514.1315B,2022MNRAS.511.1201G}, including studies dedicated specifically to binary black holes \citep{2023MNRAS.520.5724B}, binary neutron stars/short gamma-ray bursts \citep{2023NatAs...7..444S,2024A&A...692A..21G,2025arXiv250404825G}, core-collapse gamma-ray bursts \citep{2020MNRAS.491.3479C}, superluminous supernovae \citep{2021MNRAS.504L..51S}, pair instability supernovae \citep{2024arXiv240813076B} and supernova light-curves \citep{2018PASA...35...49E,2019PASA...36...41E}. There are also several publications comparing {\sc bpass} population synthesis predictions with the observed population of detached Galactic binaries \citep[including systems containing at least one white dwarf,][]{2023MNRAS.524.2836V,2024A&A...691A.316V,2025A&A...699A.172V,2024MNRAS.534.1707T,2024arXiv241102563T}, and forthcoming papers on {\sc bpass} predictions for X-ray binaries (Bray, Stanway \& Eldridge, submitted; Eastep et al. in prep).

The models can trace their lineage back to the STARS code \citep{1971MNRAS.151..351E}. In binaries, the primary (the initially more massive star) is evolved in detail, while the secondary is evolved with the rapid evolutionary algorithms of \citet{2002MNRAS.329..897H}. Models where the primary is evolved in detail are `primary models'. Once the primary becomes a remnant - a white dwarf (WD), neutron star (NS) or black hole (BH) - the subsequent model has the original secondary now evolving in detail, with the remnant in orbit. This is a `secondary' model. The models are provided at 12 metallicities $Z$\footnote{Metallicities $Z$ by mass fraction: 10$^{-4}$, 0.001, 0.002, 0.003, 0.004, 0.006, 0.008, 0.010, 0.014, 0.020, 0.030 and 0.040}, and are initialised with the binary parameter distributions of \citet{2017ApJS..230...15M}. We adopt the model set with a \citet{2001MNRAS.322..231K} initial mass function (IMF) where the slope $\alpha = -1.30$ below 0.5\,M$_{\odot}$, and $\alpha = -2.35$ between 0.5 and 300\,M$_{\odot}$. All pre-supernova binary orbits are assumed to be circular in {\sc bpass}, for full details of the set-up we refer the reader to \citet{2017PASA...34...58E}. Mass transfer rates are from \citet{2002MNRAS.329..897H}, and are limited by the thermal timescale of the accretor for non-degenerate stars. Accretion onto NSs is Eddington limited, while BH accretion is allowed to be super-Eddington \citep[see][]{2023MNRAS.523.1711G}. Common envelope evolution (CEE) treatment is similar to the $\gamma$ formalism \citep{2000A&A...360.1011N}; for more detail on the way CEE is implemented in {\sc bpass}, see \citet{2023NatAs...7..444S,2024MNRAS.534.1707T,2024A&A...691A.316V}.

At each metallicity, every model has a pre-calculated weighting based on the binary parameter distribution and IMF. This weighting corresponds to the number of each system expected in a 10$^{6}$\,M$_{\odot}$ stellar population. All metallicities are considered in the rate calculations (Section \ref{sec:rates}), while the $Z=0.004$ (0.2\,Z$_{\odot}$) and $Z=0.010$ (0.5\,Z$_{\odot}$) model sets are chosen for the two host galaxies (see Section \ref{sec:hostgalaxies}). We have a adopted metal mass fraction of 0.020 as Solar metallicity \citep[the BPASS default,][]{2017PASA...34...58E}, but note that this specific choice has no impact on the analysis.

The detailed models are deemed to end either with the beginning of neon burning, or for lower mass stars, with thermal pulses on the asymptotic giant branch \citep{2008MNRAS.384.1109E}. Core-collapse is deemed to occur if, at the end of the model, the star's total mass exceeds 1.5\,M$_{\odot}$, the CO core mass exceeds 1.38\,M$_{\odot}$ and the ONe core mass is non-zero. If these conditions are met, an energy injection of 10$^{51}$\,erg is used to determine the mass of the star which unbinds (the ejecta) and which remains bound (the remnant), as described by \citet{2004MNRAS.353...87E}. Core-collapse remnants less than 3\,M$_{\odot}$ are classified as neutron stars (although we note that the majority are well below this, close to 1.4\,M$_{\odot}$), those heavier than 3\,M$_{\odot}$ are black holes. If the conditions for core-collapse are not met, then the remnant is a white dwarf with a mass equal to the He core mass, or, if the CO core mass is non-zero, the mean of the He and CO core masses.

\subsection{Supernovae in binaries and post-SN evolution}
If the star undergoes core-collapse as decided above, we determine the outcome of the binary following \citet{1998A&A...330.1047T}. The effect of \citep[effectively instantaneous,][]{1961BAN....15..265B,1961BAN....15..291B} mass-loss and the natal kick to the remnant both play a role in determining whether the binary remains bound, and the subsequent orbit. For natal kicks, we use three prescriptions: the distribution of \citet[][the fiducial distribution adopted in this work]{2017A&A...608A..57V}, the distribution of \citet{2005MNRAS.360..974H} and the model of \citet[][see also \citealt{2023MNRAS.522.3972R}]{2016MNRAS.461.3747B,2018MNRAS.480.5657B}
which ties the kick velocity to the ejecta and remnant masses. For the first two, we randomly draw kicks from the distributions, and in each case the kick direction is random (isotropically distributed). For black holes, the fiducial approach is to reduce the natal kick by a factor of M$_{\rm BH}$/1.4\,M$_{\odot}$, but we recognise that the true distribution of black hole natal kicks is uncertain \citep{2016MNRAS.456..578M,2017MNRAS.467..298R,2019MNRAS.489.3116A,2025PASP..137c4203N}.
Natal kicks to white dwarfs are assumed to be negligible. In the formation of a compact binary, if the first remnant forms through core-collapse, we assume that the orbit afterwards is circular (due to the action of mass transfer and/or tides). We circularise the orbit by replacing $a_{0}$, the semi-major axis output by the model of \citet{1998A&A...330.1047T}, with the semi-latus rectum $a_{0}(1-e^{2})$, where $e$ is the eccentricity. After WD formation there is no eccentricity, since there is no (significant) natal or Blaauw (mass-loss) kick. Therefore, we only retain the eccentricity output from the model of \citet{1998A&A...330.1047T} if the second remnant forms through core-collapse.

For binaries which remain bound up to the point of having two remnants, we model the remaining orbital evolution as being dominated by gravitational wave emission following \citet{2021RNAAS...5..223M}. Here, $a_{0}$ is again the semi-major axis immediately post core-collapse (or WD formation), and the merger timescale $\tau_{\rm GW}$ is shortened by a factor which depends on the eccentricity (unless the second remnant to form is a WD, in which case we have $e=0$). We simply take the $\tau_{\rm GW}$ timescale as the merger time, neglecting tidal effects or mass transfer in the late stages of the in-spiral, and assume that all compact binaries will merge after $\tau_{\rm GW}$. This approximation is discussed in Section \ref{sec:criteria}. No assumption is made a priori about which mergers could produce a GRB and/or kilonova.

\begin{figure*}
\centering
\includegraphics[width=0.49\textwidth]{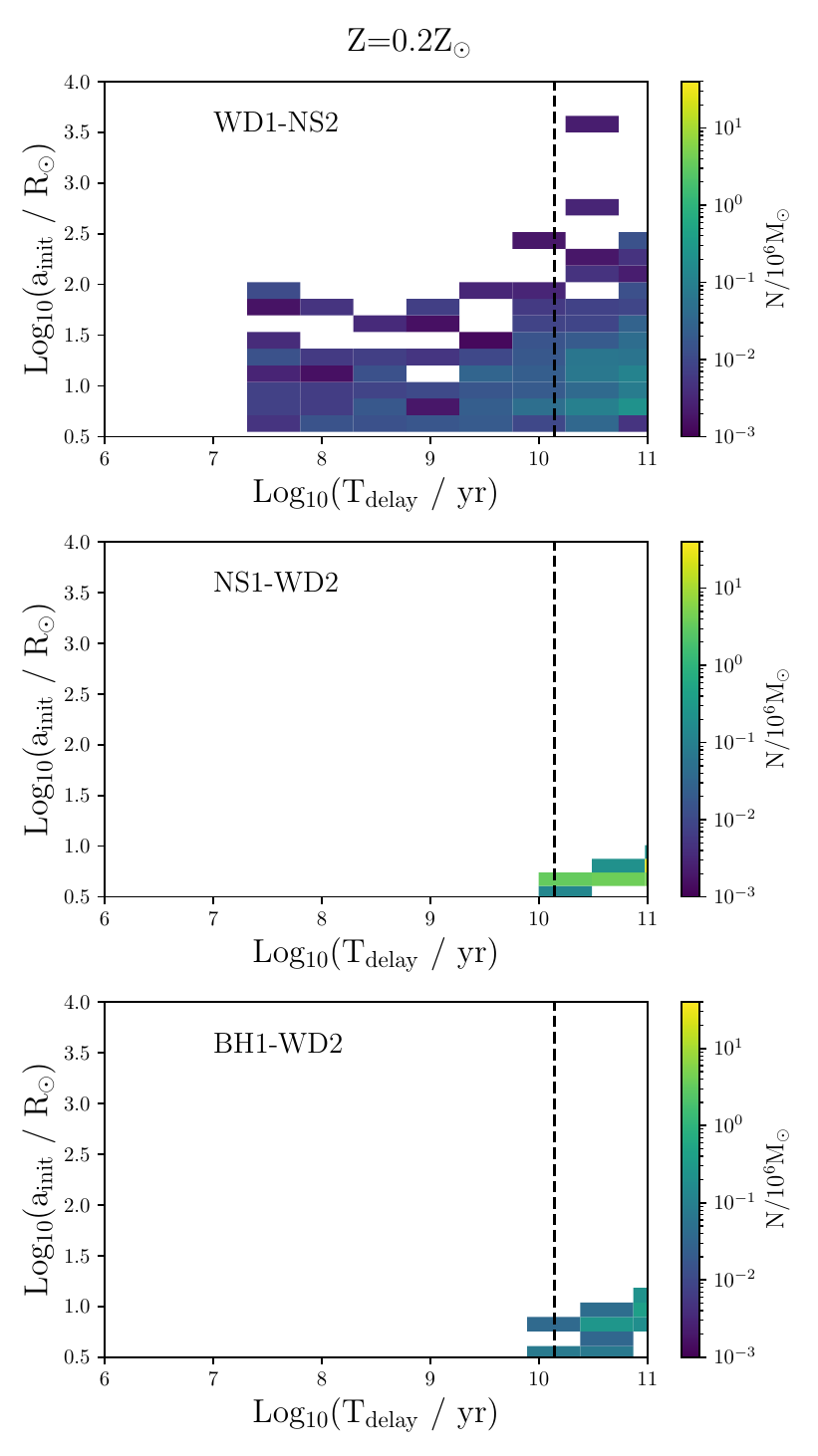}
\includegraphics[width=0.49\textwidth]{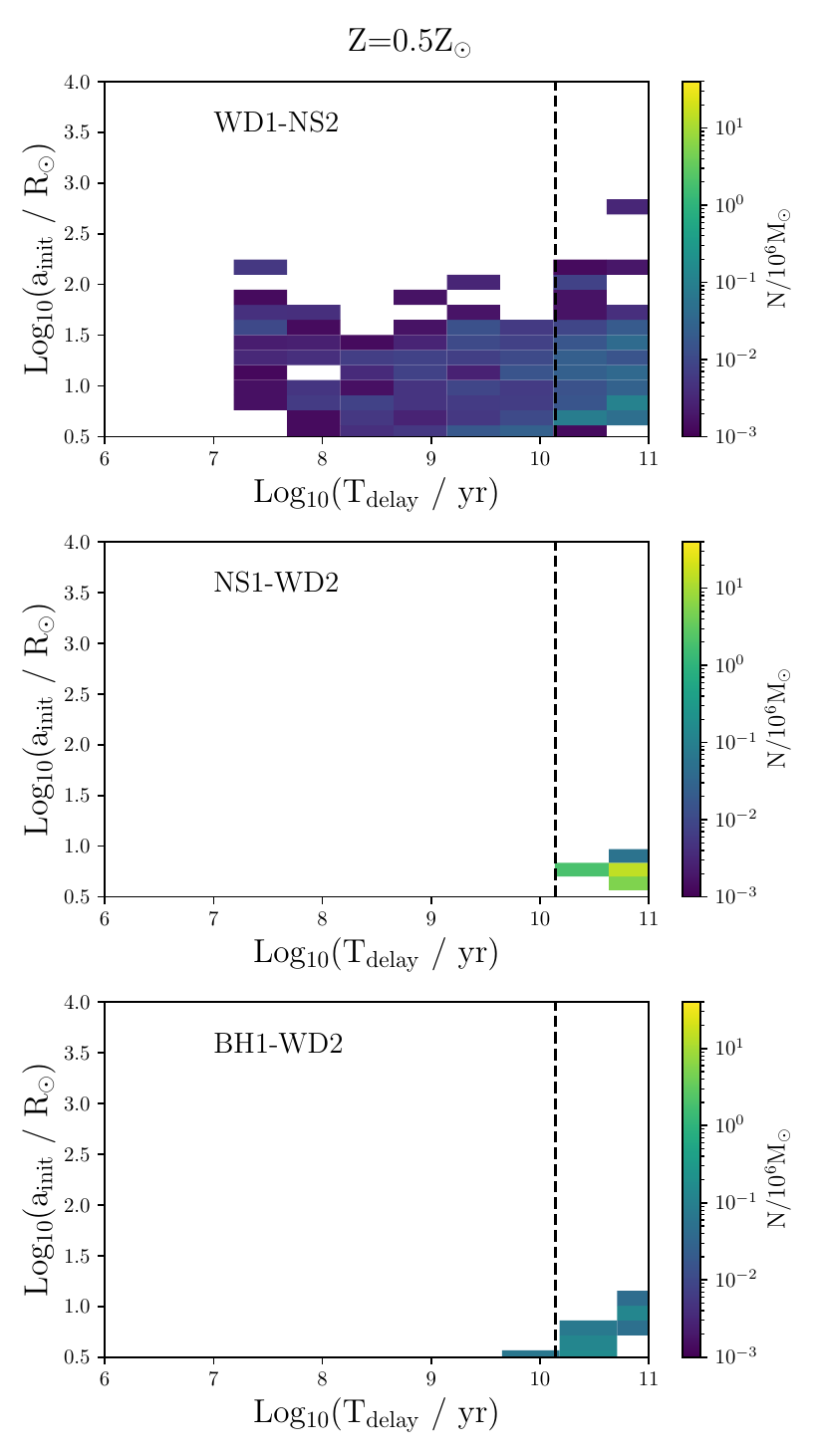}
\caption{The number of compact binary models containing a white dwarf and a core-collapse remnant for each initial semi-major axis $a_{0}$ (immediately post second remnant formation) and delay time $T_{\rm delay}$. The shading corresponds to the number of compact binaries produced in each 2D bin per 10$^{6}$\,M$_{\odot}$. The left-hand column shows results at one-fifth Solar metallicity, on the right at half-Solar. Top row: WDNS binaries in which the WD forms first. Eccentricity is retained from the supernova, which occurs second, producing a greater range of semi major axes with a tail extending to higher values at fixed delay time. Middle row: WDNS binaries in which the NS forms first. Bottom row: WDBH binaries in which the BH forms first. We find no WD1-BH2 models. In each case, a vertical dashed line indicates a Hubble time. The vast majority of such binaries merge on much longer timescales (see Table \ref{tab:evolution}). \citet{2017A&A...608A..57V} natal kicks are adopted, equivalent plots with \citet{2005MNRAS.360..974H} and \citet{2016MNRAS.461.3747B} kicks are given in Appendix \ref{app:A}.}
\label{fig:delay_sep}
\end{figure*}

\subsection{Formation pathways}\label{sec:pathways}
We now discuss the possible formation channels of WD-NS/BH binaries and the rates with which they lead to mergers in {\sc bpass}. Hereafter, the order in which the remnants form is denoted by a number (e.g. NS1-WD2 for NS first, WD second). We make reference to the four WDNS pathways identified in {\sc SeBa} by \citet{2018AandA...619A..53T} \citep[see also][]{2024MNRAS.530..844K}. Analogous WDBH systems are possible in each case \citep[e.g.][]{2001AandA...375..890N,2021ApJ...920...81S,2024MNRAS.535.2800L}. In each case, the total delay time - the time from star formation to merger - includes the stellar evolution and gravitational wave (GW) in-spiral timescales.

Pathway 1 is the direct WD1-NS2 pathway, where the initially more massive star becomes a white dwarf before the secondary becomes a neutron star, due to strong mass transfer onto the secondary. This pathway has delay times as short as a few tens of Myrs, because it requires tight orbits to initiate strong mass transfer and hence reverse the masses. The secondary only fills its Roche lobe once the primary has become a WD. The initial compact binary orbits from this pathway are tight, but also eccentric. This shortens the GW-driven in-spiral, and hence the total delay time, even further. The peak in the delay time distribution at a few tens of Myrs (see Figure \ref{fig:vsystdelay}) corresponds to the stellar lifetimes. The tail to longer delay times is due to the GW in-spiral time distribution. 

Pathway 2 (NS1-WD2) typically occurs in wider binaries with weak (or without) interactions, such that the primary maintains its status as the most massive star and becomes a neutron star first, with the secondary later evolving into a white dwarf. Delay times can be long for this pathway (up to and beyond a Hubble time) since the upper limit is defined by the lifetime of low-mass stars. Furthermore, the orbits are typically wider and (in our models) are assumed to be circular.

Pathway 3 (WD1-NS2) is a variation of pathway 1, but where the initial mass ratio is close to one, such that the evolutionary timescales for each star are similar and the secondary expands and fills its Roche lobe before the primary has evolved into a white dwarf. Ultimately, the primary still becomes a white dwarf before the secondary becomes a neutron star. Using the {\sc hoki} \citep{2020JOSS....5.1987S} Evolution Explorer ({\sc eve}), we find that among the WD1-NS2 channels (pathways 1 and 3), post WD formation at 0.2\,Z$_{\odot}$, $\sim$75\% of the models undergo mass transfer (of any kind) and $\sim$60\% of all models undergo common envelope evolution. At 0.5\,Z$_{\odot}$, 70\% of the systems interact post-WD formation and this is exclusively in the form of common envelope evolution.

Finally, pathway 4 (NS1-WD2) resembles pathway 2 in terms of outcome but is more like pathways 1 and 3 in terms of evolution. In this case the initially less massive star gains enough mass to have its evolution accelerated to the point that it undergoes core-collapse first, before the primary evolves into a white dwarf. This pathway is hard to identify in {\sc bpass} since the end states of primary models are defined by the remnant the primary produces: in other words, the secondary star cannot end its evolution first. Such systems will contribute to the WD1-NS2/BH2 models instead. We note that the time between WD and NS/BH formation in these systems is very short, so the order of formation is likely not critical to the evolution, and in either case the binaries are tight, such that GW in-spiral times will be short (compared with stellar evolution timescales) regardless of whether the eccentricity is retained or not. Again using {\sc eve}, we find that among NS1-WD2 models at 0.2\,Z$_{\odot}$, $\sim$70\% of the models interact post-NS formation, of which $\sim$half is through a common envelope and half through Roche lobe overflow (RLOF). At 0.5\,Z$_{\odot}$, $\sim$60\% interact, of which 2/3 is through a common envelope phase. 

We summarise the evolutionary properties of the {\sc bpass} WDNS and WDBH populations at two metallicities, and adopting the three natal kick prescriptions, in Table \ref{tab:evolution}. Pathways 1 and 3 are grouped together and are represented by WD1-NS2/BH2 models. Pathway 2 is represented by NS1/BH1-WD2 systems. As discussed above, pathway 4 is hard to identify in the current version of {\sc bpass} and will be contributing to WD1-NS2/BH2 models, along with pathways 1 and 3. This will be addressed in a future version of {\sc bpass}.

We find that NS1-WD2 channels represent $\sim$70-90\% of the WDNS rate (for mergers within a Hubble time and depending on the natal kicks adopted), even though the probability of any given NS1-WD2 system merging within a Hubble time is lower than for a WD1-NS2 system. This is because the WD1-NS2 channel is rarer, but such systems have tighter (and eccentric) orbits as explained above. The rarity of the WD1-NS2 channel, with respect to NS1-WD2, is in contrast with \citet{2018AandA...619A..53T} who find that WD1-NS2 mergers dominate the WDNS merger rate. This may be due, in part, to the assumed initial mass ratio distribution, which in the {\sc SeBa} study of \citet{2018AandA...619A..53T} is flat, whereas in {\sc bpass}, asymmetric binaries are given lower weightings \citep{2017ApJS..230...15M}. As high mass ratios (as well as tight orbits) are required for pathway 1, this pathway is consequently rarer in our population. We note that {\sc bpass} predicts relatively many WDNSs compared to other codes such as SeBa \citep{2023MNRAS.524.2836V,2024A&A...691A.316V,2024arXiv241102563T}, likely because common envelope ejection is more efficient and results in less orbital shrinkage.

An important caveat is that the above statements apply to all WDNS mergers within a Hubble time, however, the NS1-WD2 mergers are biased towards the upper end of this timescale (i.e. the lifetime of non-interacting low mass stars). Hence, for galaxies with stellar populations younger than $\sim$10\,Gyr, the WD1-NS2 channels will dominate the merger rate in practice.

Finally, we find that WDBH mergers are extremely rare \citep[in common with previous population synthesis results, e.g.][]{2001AandA...375..890N}. In terms of WDBH binaries being born, the rate is around half of the WDNS birth rate at 0.5\,Z$_{\odot}$, and close to parity at 0.2\,Z$_{\odot}$ \citep[see also e.g.][]{2023MNRAS.524.2836V}. This is due to metallicity-dependent mass-loss \citep[e.g.][]{2005A&A...442..587V,2008NewAR..52..419V} and hence higher progenitor masses at lower metallicity \citep{2010ApJ...715L.138B}. However, WDBH binaries are much less likely to merge within a Hubble time than WDNS binaries, and the WD1-BH2 channel does not occur at all in {\sc bpass} (at the metallicities investigated). Plotting the initial compact binary separations for WDNS and WDBH populations in Figure \ref{fig:delay_sep}, we can see that WDBH binaries start wider and hence far fewer merge within a Hubble time. This is because the progenitors of BHs are more massive, and typically larger. The complete lack of WD1-BH2 systems may be due to the extreme amount of mass transfer that would be required to reverse the evolutionary order in this case, exacerbated by the low weighting of models with high initial mass ratios. Furthermore, the initial model grids, in steps of 0.1 in mass ratio, undersample this region of parameter space.

\subsection{Delay times and systemic velocities}\label{sec:delays}
For our binary merger offset modelling, the delay time (since star-formation) and systemic velocity of the compact binaries are the key parameters of interest. In Figure \ref{fig:results} we show the systemic velocities and delay times for compact binary mergers, at 0.2 and 0.5\,Z${_\odot}$. We show WDNS (split into WD1-NS2 and NS1-WD2), WDBH and BNS mergers, calculated as described above. 

\begin{figure*}
\centering
\includegraphics[width=\textwidth]{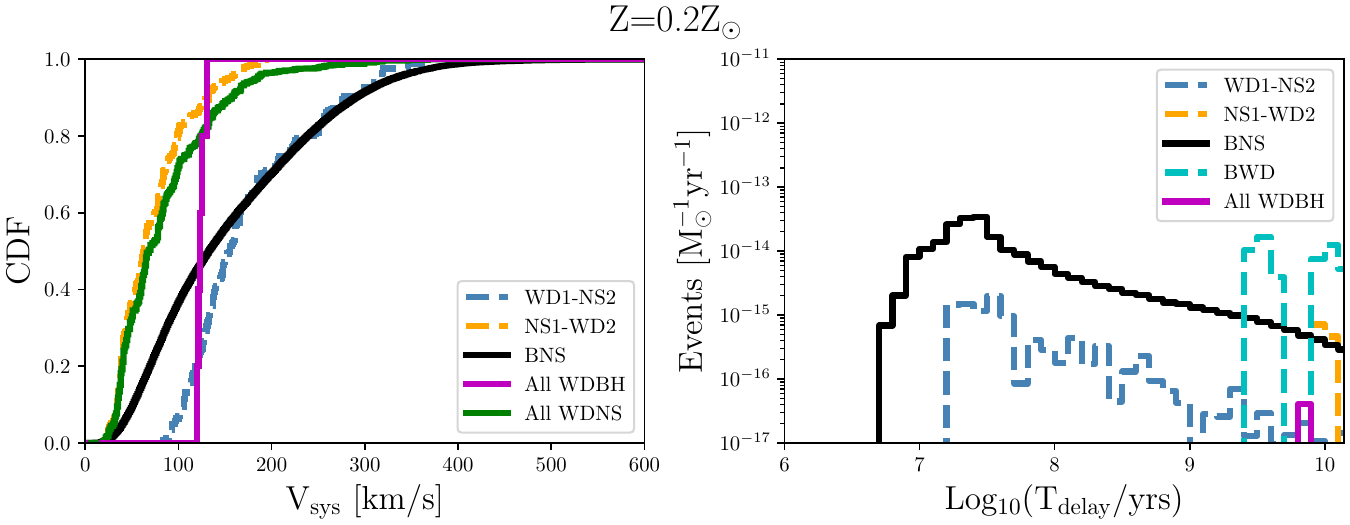}
\includegraphics[width=\textwidth]{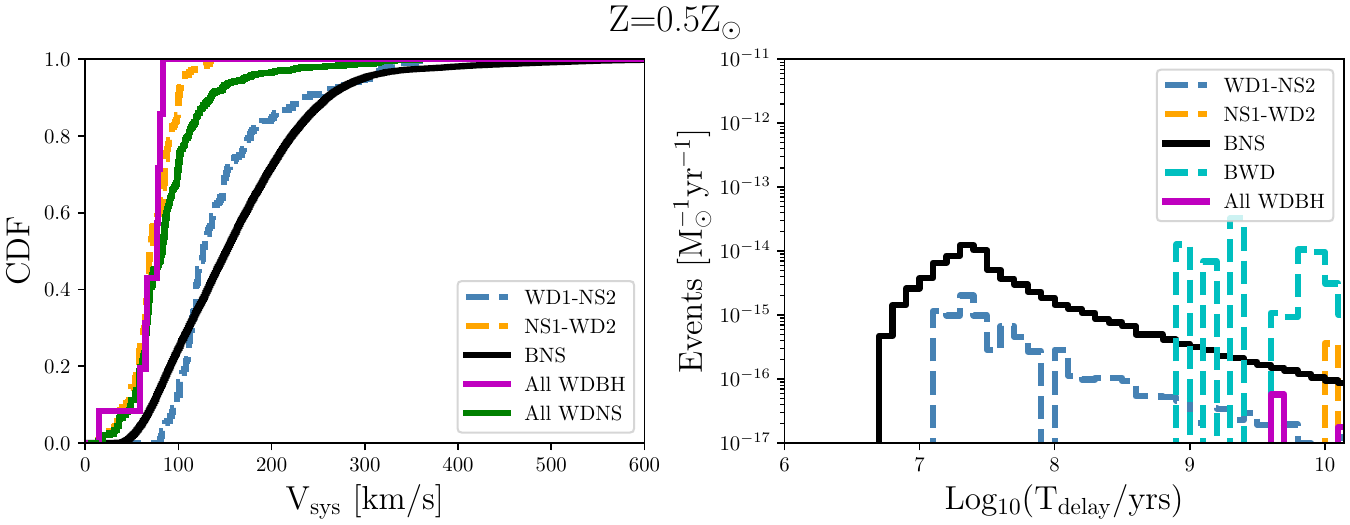}
\caption{Systemic velocities (at birth) for compact binaries (left), and their total delay time distributions (from ZAMS to merger, right), at 20\% Solar metallicity (top, similar to the host galaxy of GRB211211A) and 50\% Solar metallicity (bottom, matching the host of GRB230307A). These results adopt the \citet{2017A&A...608A..57V} distribution for neutron star natal kicks (results for alternative natal kicks are given in appendix figures \ref{fig:apx_hobbs} and \ref{fig:apx_bray}). Only systems which merge within a Hubble time are shown. We split the WDNS mergers into NS1-WD2 and WD1-NS2 channels; WDBH mergers are however much rarer and exclusively have the BH forming first. These velocities do not include any contribution from the initial galactic orbits of their progenitor systems (these are introduced in Section \ref{sec:offsets}) for performing offset simulations.}
\label{fig:vsystdelay}
\end{figure*}

Binaries in which the WD forms first are much faster, due to the lower mass in these systems at the time of supernovae, and have systemic velocities comparable to those of BNSs (see Figure \ref{fig:vsystdelay}). Our results are similar to those of other population synthesis studies in terms of the distribution shape and maximum velocities reached \citep[e.g.][]{2018AandA...619A..53T,2024MNRAS.530..844K}. The delay time distribution for WD1-NS2 systems follows 1/t beyond the progenitor lifetimes, as expected, where the delay is dominated by the GW-driven in-spiral time. For NS1-WD2 binaries - at least those which merger within a Hubble time - stellar evolution dominates the delay time. 

The delay time distributions presented here are shifted to longer delays compared with previous {\sc bpass} results \citep[e.g.][]{2019MNRAS.482..870E}. Consequently, there are fewer mergers within a Hubble time, and the volumetric event rates are lower. This is primarily due to improvements in the gravitational wave in-spiral time calculation\footnote{Specifically, the eccentric semi-major axis is now used in the gravitational wave in-spiral time calculation when the second remnant to form is a product of core-collapse, rather than the smaller circularised value.}, but also updates to the rejuvenation of secondary stars during mass transfer \citep{2022MNRAS.511.1201G,2023MNRAS.520.5724B}.

\section{Merger offsets}\label{sec:offsets}
In this Section, we model the host galaxies of GRBs\,211211A and 230307A, seed our binary population synthesis models in these galaxies in space and time, and predict the merger offset distribution for WDNS and BNS populations around these galaxies. This is then compared with the observed galactocentric offsets of GRBs\,211211A and 230307A. 

Our methodology follows that of \citet{2011MNRAS.413.2004C}, \citet{2024A&A...692A..21G} and \citet{2025arXiv250404825G}. The procedure involves the modelling of the star-formation history, for seeding the {\sc bpass} binaries in time; modelling the light profile, for seeing the binaries in space; and modelling the potential, including contributions from the stellar component and a dark matter halo. The trajectories of compact binaries can then be followed until the point of merger. The process is briefly described below, but for a full description we refer the reader to \citet{2025arXiv250404825G}. Details of the two host galaxies are provided in Table \ref{tab:host_galaxy}.

\subsection{Host galaxy models}\label{sec:hostgalaxies}

\begin{table}
	\centering
	\caption{Summary of host galaxy and GRB offset parameters. References are listed below the table. If the value is not referenced, it was calculated for this work. The host of GRB\,230307A is modelled with two S{\'e}rsic components (see text). For star-formation histories we use the non-parametric fits of \citet[][see Figure \ref{fig:sfh}]{2025ApJ...982..144N}.}
	\label{tab:host_galaxy}
	\begin{tabular}{ccc} 
		\hline
		\hline
		   & GRB211211A & GRB230307A \\
		\hline
            $z$ & 0.076 [1] & 0.065 [2] \\
            $M_B$ & -17.38 [3] & -18.77 \\
            $\log_{10}(M_{\star}$/M$_{\odot}$) & 8.84$^{+0.10}_{-0.05}$ [1] & 9.28$\pm$0.01 [2]\\
            t$_{\rm m}$/Gyr & 2.53$^{+1.24}_{-0.50}$ [4] & 1.00$^{+0.02}_{-0.01}$ [4] \\
            $r_{50}$/kpc & 1.6 [5] & 0.03 / 2.04 [2] \\
            S{\'e}rsic $n$ & 1.0 [1] & 0.5 / 1.1 [2] \\
            $Z$/Z$_{\odot}$ & 0.20$^{0.05}_{-0.08}$ [1] & 0.57$^{+0.03}_{-0.01}$ [2] \\
            $r_{\rm grb}$/kpc & 7.92$\pm$0.03 [1] & 38.9 [2] \\
		\hline
	\end{tabular}
 \newline
 [1] - \citet{2022Natur.612..223R}, [2] - \citet{2024Natur.626..737L}, [3] - \citet{2025arXiv250404825G}, [4] - \citet{2025ApJ...982..144N}, [5] - \citet{2022Natur.612..228T}
\end{table}

\subsubsection{Star-formation histories}
\begin{figure*}
\centering
\includegraphics[width=\textwidth]{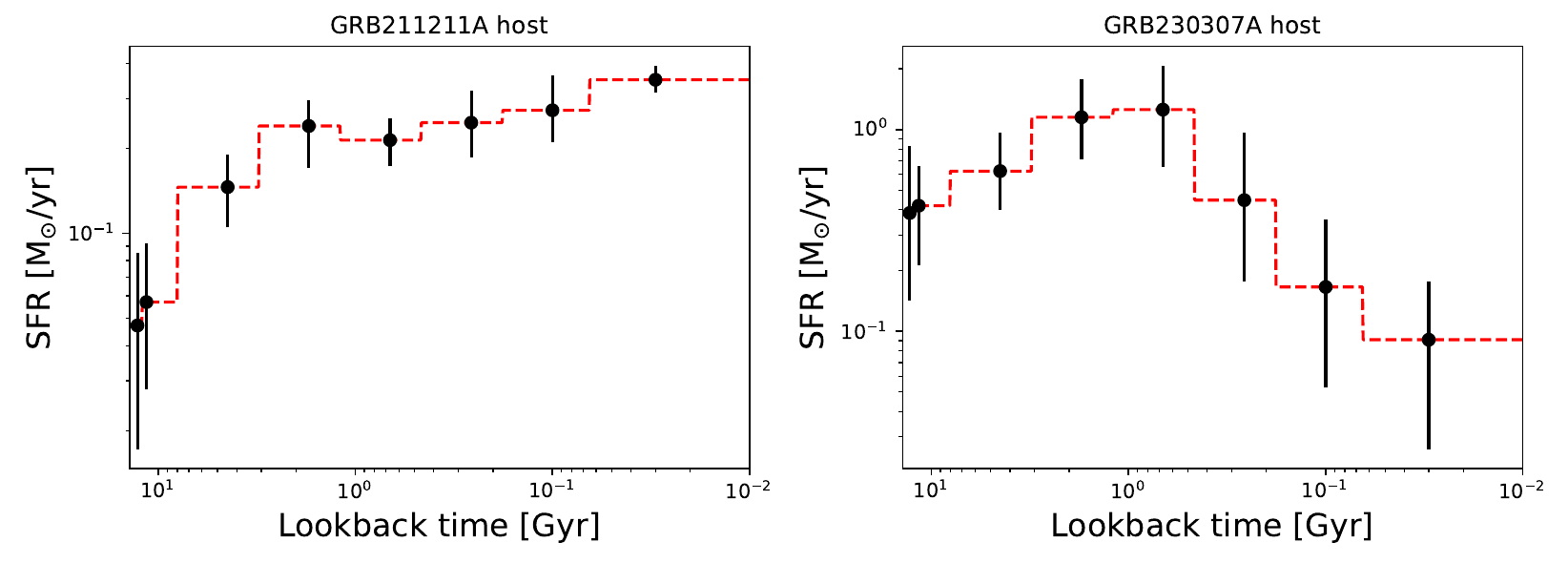}
\caption{Star formation histories, determined with {\sc prospector}, for the host galaxies of GRBs\,211211A and 230307A \citep[from][]{2025ApJ...982..144N}. Errorbars are given on the SFR estimates at each time-step, while the red dashed lines are the constructed histrograms which we use to weight the model seeding in time. The populations of both galaxies are dominated by star-formation within the last 10\,Gyr. This biases the merging WDNS binaries in these galaxies to the WD1-NS2 pathway (see Fig. \ref{fig:vsystdelay}).}
\label{fig:sfh}
\end{figure*}

We employ non-parametric star-formation histories (SFHs) for the host galaxies of GRBs\,211211A and 230307A, derived from {\sc prospector} SED-fitting \citep{2019ApJ...876....3L,2021ApJS..254...22J} with {\sc MIST} stellar populations by \citet{2025ApJ...982..144N}. The star-formation histories are shown in Figure \ref{fig:sfh}, with characteristic ages given in Table \ref{tab:host_galaxy}. The initial binary stellar models (at ZAMS) are seeded in time in proportion to the SFR at each time step. The total delay time is then the stellar evolution timescale of the longer-lived star, plus the gravitational wave in-spiral time. The dominant contribution to the delay time depends on the evolutionary pathway, as discussed in Sections \ref{sec:pathways} and \ref{sec:delays}. Although the ages of these galaxies are quite far down the delay time distribution for WDNSs (see Figure \ref{fig:vsystdelay}), which might suggest that these are not the likeliest hosts for WDNS mergers, we note that the same is true for BNSs.

\subsubsection{Stellar light profiles}
We model the stellar light of the host galaxies of GRBs\,211211A and 230307A as infinitely thin discs described by S{\'e}rsic profiles \citep{1963BAAA....6...41S}. For the host of GRB\,211211A we adopt the morphological fit parameters of \citet{2022Natur.612..228T}, based on \textit{HST}/F184W imaging. These are a half-light radius $r_e$ of 1.58\,kpc (this is the projected physical half-light radius at $z=0.076$) and a S{\'e}rsic index $n=1$. For the host of GRB\,230307A, a single S{\'e}rsic produces a poor fit, so we instead use two components \citep{2024Natur.626..737L}. The first has effective radius $r_e$=0.03\,kpc with n=0.50 and an effective surface brightness at $r_e$ of $\mu_e$=0.47\,MJy/sr, the second has r$_{e}$=2\,kpc, n=1.05 and $\mu_e$=0.35\,MJy/sr (where $\mu_e$ is used for the relative scaling of the two components). The stellar models are then seeded in proportion to the surface brightness at each location in the disc.

\subsubsection{Potentials}
We next require a gravitational potential to accurately follow the trajectory of the kicked binaries. We model the potential as having two components, the stellar disc and a dark matter (DM) halo. The potential of the disc is modelled as a double exponential with scale height related to the scale length by a factor $\gamma=0.2$ \citep[][and references therein]{2025arXiv250404825G}. For the dark matter halo we use the methodology of \citet[][and references therein]{2011MNRAS.413.2004C}. Succinctly, the DM halo mass depends on the $B$-band absolute magnitude $M_B$ of the galaxy, its morphologic type and its effective radius $r_e$ \citep{2009ApJ...691..770T}. For the host of GRB\,211211A we use  $M_B$ from \citet{2025arXiv250404825G}. For the host of GRB\,230307A we use the best-fit intrinsic (non-extincted) host spectrum from \citet{2024Natur.626..737L} and apply a $B$-band filter response function \citep{2020sea..confE.182R,2012ivoa.rept.1015R} to obtain $M_B$. Following \citet{2025arXiv250404825G} we assume that the potentials are static: this is a limitation of the approach, but is hard to address without cosmological simulations  \citep{2010ApJ...725L..91K,2018ApJ...865...27W} which apply to populations at a statistical level, rather than to individual, observed host galaxies. However, we note that the hosts of GRBs\,211211A and 230307A are star-forming with discy/spiral morphologies, suggesting that they have not undergone major mergers \citep[e.g.][]{2009MNRAS.399..966S}.

\subsection{Offset predictions}
\begin{figure*}
\centering
\includegraphics[width=\textwidth]{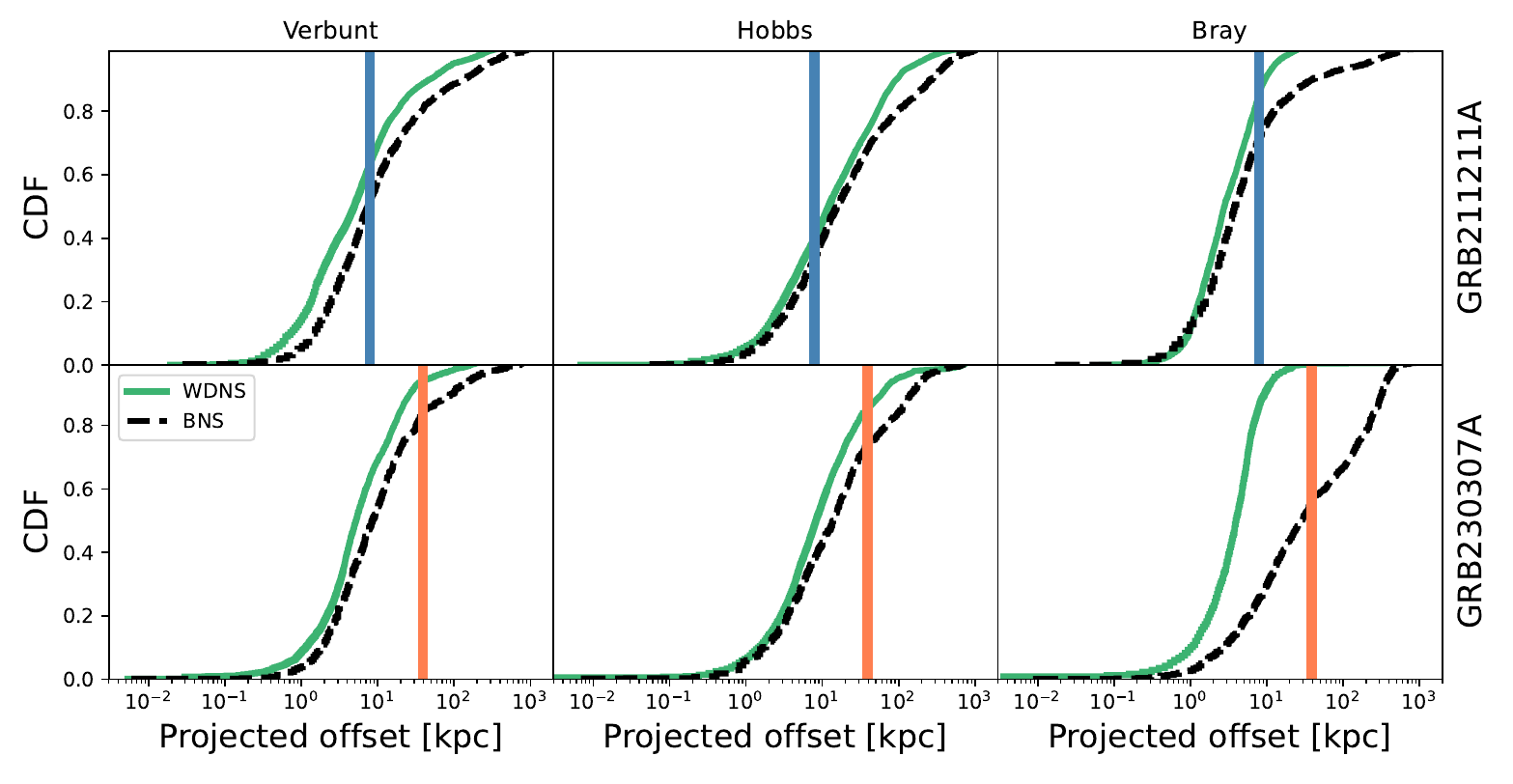}
\caption{WDNS and BNS offsets for the host galaxies of GRB\,211211A (top row) and GRB\,230307A (bottom row). The offsets are projected assuming random, isotropic viewing angles. The observed projected offsets of GRB\,211211A (7.92\,kpc) and GRB\,230307A (38.9\,kpc) are marked on each panel by blue and orange vertical lines respectively.}
\label{fig:results}
\end{figure*}

Given binary stellar populations, stellar light distributions and star-formation histories (for seeding the populations in space and time), and potentials (to trace the trajectories of binaries which stay bound after supernovae), we can now model the merger offset distribution of compact binaries around the host galaxies of GRBs\,211211A and 230307A. The binaries are initialised on circular velocities defined by the galactocentric radius and enclosed mass, and given random orbital plane orientations with respect to the galactic plane \citep{2024A&A...692A..21G,2025arXiv250404825G}. Tracking the trajectories of binaries post SN1 and SN2 (in the BNS case), we follow the binaries through the potential until the point of merger and record the position. The final offset distributions are projected assuming random viewing angles. These final distributions are weighted by both the {\sc bpass} weightings, and a weighting arising from the SFH. They can be considered as probability distributions for where we expect to see a merger today. The results are shown in Figure \ref{fig:results} for both host galaxies and for the three neutron star natal kick distributions. Although only a few hundred distinct (WDNS) models contribute to mergers at each metallicity, the large number of random starting positions and kick velocities has the effect of smoothing this granularity and producing the smooth, continuous offset distributions shown. In table \ref{tab:offsets} we list the fraction of mergers, for all combinations of progenitor, kick and host galaxy, which occur at or below the observed projected offset of the GRB. In nearly every case, the observed offset is consistent with either a BNS or a WDNS progenitor. The offset of GRB\,230307A is at the upper end of the predicted range however, with a probability of offsets this high or higher of $\sim$10\%.

\section{Volumetric rates}\label{sec:rates}
\begin{figure}
\centering
\includegraphics[width=\columnwidth]{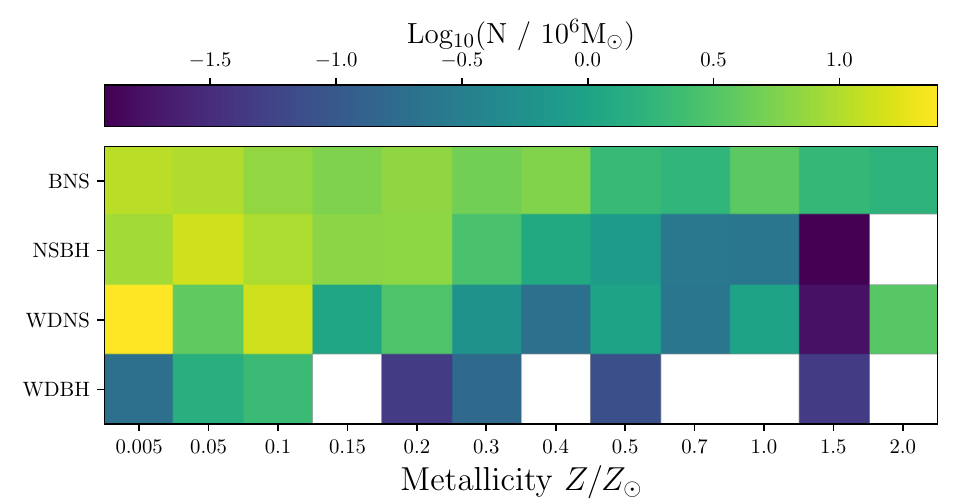}
\caption{The total number of BNS, NSBH, WDNS and WDBH binaries merging within a Hubble time, per 10$^{6}$\,M$_{\odot}$ of stars formed, as a function of metallicity.}
\label{fig:zbias}
\end{figure}

Another observational constraint on the progenitors of long-duration compact merger GRBs is their volumetric event rate. We calculate, at every {\sc bpass} metallicity, the number of compact binary merger progenitors born per 10$^{6}$\,M$_{\odot}$ of star formation, and the total delay times (from ZAMS to merger) associated with each binary. The total number of BNS, NSBH, WDNS and WDBH mergers within a Hubble time, per 10$^{6}$\,M$_{\odot}$ of stars formed, is shown in Figure \ref{fig:zbias} as a function of metallicity. The apparently stochastic nature of these results as a function of $Z$ is due to the relatively small number of relevant models in the model grid, and the complexities introduced by binary evolution, which results in `randomness' on top of broader trends. In general, compact binaries containing a NS or BH are rarer at higher metallicities, where mass-loss rates are higher. This effect is most pronounced for BH formation and weaker for NSs \citep{2025ApJ...979..209V}. We can also see a trend for WDNS mergers to strongly prefer low $Z$, this arises because (i) nuclear timescales are shorter at lower $Z$ and (ii) stars are more compact at lower $Z$, allowing them have tighter orbits \citep{2014ApJ...791..127J}. Both effects increase the number of systems merging within a Hubble time. 

\begin{figure}
\centering
\includegraphics[width=\columnwidth]{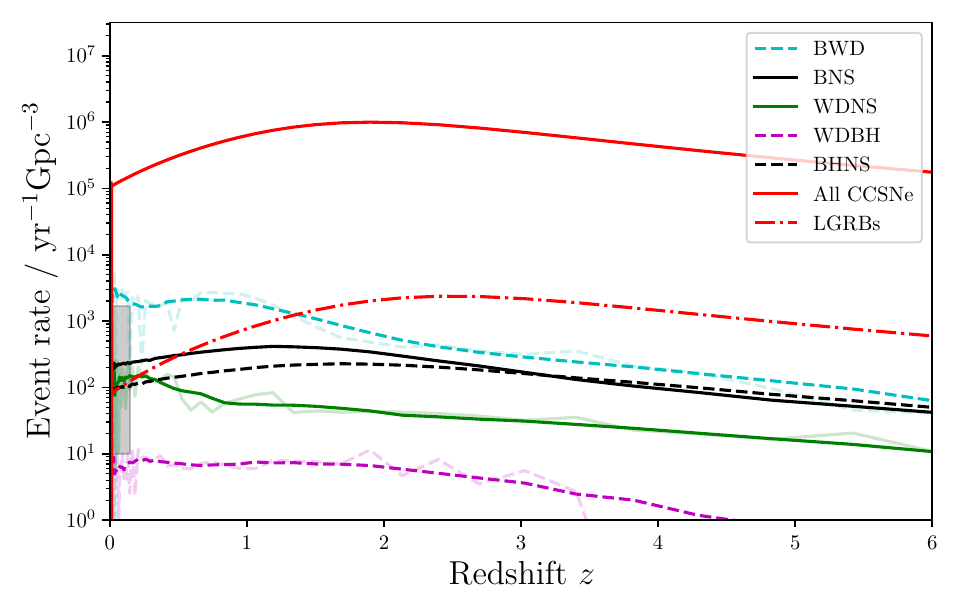}
\caption{Volumetric merger rates as a function of redshift. The cosmic SFH history of \citet{2006ApJ...638L..63L} is adopted. Also shown is the core-collapse supernova rate. The shaded grey region on the left indicates the range of possible BNS merger rates in the local Universe ($z=0$), constrained solely from LIGO/VIRGO/KAGRA gravitational wave observations \citep{2023PhRvX..13a1048A}. Short GRB volumetric rate estimates are consistent with this \citep[][]{2023A&A...680A..45S}. In addition to the rolling average of the WDNS, WDBH and BWD rates (see text), the un-averaged rates are shown in paler shades. The dash-dot red line is the derived intrinsic rate of LGRBs from \citet{2022ApJ...932...10G}. }
\label{fig:volrate}
\end{figure}

We next apply a metallicity-dependent cosmic star formation history (CSFH) prescription to determine volumetric merger rates as a function of redshift, as shown in Figure \ref{fig:volrate}. We adopt the star formation history of \citet{2014ARA&A..52..415M}, with the metallicity dependence of \citep{2006ApJ...638L..63L}. The CSFH defines the volumetric star formation rate, at each of the 12 {\sc bpass} metallicities, as a function of redshift. We refer the reader to \citet{2019MNRAS.488.5300C} and \citet{2022MNRAS.514.1315B} for discussion of how different CSFHs impact transient rate predictions. In addition to WDNS, WDBH and BNS systems, we include BHNS and binary white dwarf (BWD) mergers, as well as the core-collapse supernova rate \citep[for other {\sc bpass} transient rate predictions see][]{2019MNRAS.482..870E,2022MNRAS.514.1315B,2023MNRAS.520.5724B}. The conditions for core-collapse versus WD formation are as described in Section \ref{sec:popsynth}. Due to the relatively small number of models producing mergers involving a WD within a Hubble time, the rates are stochastic, with a scatter of around an order of magnitude at low redshifts. The effect is particularly noticeable at low redshifts, where equal time steps represent smaller redshift steps, a small number of highly-weighted long-delay time models start to contribute, and the overall number of models contributing to the merger rate drops due to increasing metallicity over cosmic time (Fig \ref{fig:zbias}). In the interests of visual clarity, we show a rolling average of the WDNS, WDBH and BWD rates, in addition to the un-averaged, more stochastic rates. We also show in Figure \ref{fig:volrate} the local Universe ($z=0$) BNS merger rate, determined from LIGO/VIRGO/KAGRA gravitational wave observations \citep{2023PhRvX..13a1048A}, and the volumetric rate of long-duration GRBs \citep[under certain assumptions for the luminosity function and jet opening angles,][]{2022ApJ...932...10G}. 

While on the topic of the CSFH, we might expect given the strong low-$Z$ preference in Figure \ref{fig:zbias} to find WDNS and WDBH mergers in low-$Z$ galaxies. However, GRBs 211211A and 230307A were at low redshifts, where very low $Z$ star-formation is rare, which should be accounted for. Furthermore, for long delay times, the metallicity of the host may change substantially during the in-spiral of the binary. Expectations for the host metallicities are therefore more complex that what we see in Figure \ref{fig:zbias}, and in any case difficult to assess at present with such a small sample size.

The inferred volumetric rates of GRBs as a function of redshift is highly sensitive to the energy band of the detectors, biases in follow-up observations (required to determine the redshift), as well as assumptions about the GRB luminosity function and beaming angles. Further more, compact binary merger rate predictions from population synthesis codes are affected by an array of factors, from assumptions about the initial binary parameter distributions, to how the various phases of binary stellar evolution are modelled. These all play into large (order of magnitude) systematic uncertainties on the rates \citep[e.g.][]{2014A&A...563A..83C,2022MNRAS.516.5737B}. Perhaps the best way to quantify these uncertainties is to compare rate predictions from different population synthesis codes, and rates inferred using different methods (e.g. Galactic binary population observations). In Table \ref{tab:ratecompare} of Appendix \ref{app:B} we list our WDNS and WDBH merger rates - in the form of predicted Galactic merger rates - alongside other estimates from the literature. This demonstrates the order of magnitude uncertainty on the rates, not only between population synthesis codes, but also between different (direct) observational constraints.

Given these caveats and proceeding under the assumptions outlined throughout this paper, we find that at $z<0.5$, both BNS and WDNS mergers occur at a similar rate to long GRBs. This leaves open the possibility of large fraction of long GRBs arising these channels, dependent on the detailed criteria for successful merger and GRB launching as we will discuss in the next section. WDBH mergers are a factor of 10 rarer, but we again stress that the order of magnitude observational and theoretical rate uncertainties (see Table \ref{tab:ratecompare}) preclude definitive statements at this time.

\begin{table}
	\centering
	\caption{We list here the fraction of simulated merger offsets below the observed offset for GRB\,211211A and 230307A, for various combinations of compact binary and natal kick. If greater than 95\% of the population are merging within the observed offset, we disfavour the model, although this only occurs for GRB\,230307A and the WDNS$+$Bray combination.}
	\label{tab:offsets}
	\begin{tabular}{llcc} 
		\hline
		\hline
		    & & \multicolumn{2}{c}{$f(r_{\rm merge}<r_{\rm grb}$)} \\
		  Binaries & Kicks & GRB211211A & GRB230307A \\
		\hline
            WDNS & Verbunt & 0.63 & 0.94\\
            BNS & Verbunt & 0.52 & 0.84\\
		\hline
            WDNS & Hobbs & 0.39 & 0.85\\
            BNS & Hobbs & 0.34 & 0.74\\
		\hline
            WDNS & Bray & 0.85 & 0.99\\
            BNS & Bray & 0.71 & 0.57\\
		\hline
	\end{tabular}
\end{table}

\section{Discussion}
\subsection{Modelling assumptions}
The results presented in this paper depend on the assumptions inherent to binary stellar evolution modelling and population synthesis. For example, an implication of the high common envelope (ejection) efficiency in {\sc bpass} is that it predicts relatively few Thorne–${\rm \dot{Z}}$ytkow objects \citep[TZOs][]{1977ApJ...212..832T}, which are only possible if the common envelope phases survives long enough for the NS or BH to in-spiral and ultimately settle at the core of the other star. In {\sc bpass}, TZOs are rare since the common envelope is quickly ejected, leaving the NS/BH in a (tighter) orbit around the low mass star, which ultimately continues to evolve into a WD. Observational constraints on Galactic WDNS binaries, both with electromagnetic and GW observations with LISA \citep{2017arXiv170200786A,2024arXiv240207571C}, will be a important observational anchor point against which population synthesis can be calibrated going forwards \citep[e.g.][]{2024MNRAS.530..844K}, helping to resolve the current merger rate discrepancies (Table \ref{tab:ratecompare}). 

While we have adopted three NS natal kick distributions, we have not explored variation in BH kicks in a similar way. This is due to the lack of well-defined observational distributions for this population, but the real distribution is likely between the one adopted in {\sc BPASS} and one closer to that of pulsars \citep{2017MNRAS.467..298R}. Our adopted black hole kicks are therefore at the low end of literature estimates, and we note that increasing them would decrease the WDBH merger rates even further (as more of these binaries would unbind up BH formation).

\subsection{Merger, GRB and kilonova criteria}\label{sec:criteria}
It is worth considering that not all WDNS binaries which come into contact will merge, which has been the implicit assumption made thus far in this paper. If the donor white dwarf is low-mass \citep[see][who suggest less than 0.2--0.3\.M$_{\odot}$]{2012A&A...537A.104V,2017MNRAS.467.3556B,2023ApJ...956...71K}, then the in-spiral can be arrested and a stable ultra-compact X-ray binary can form. Similar results have been found for WDBH systems \citep{2017ApJ...851L...4C}. However, such low mass WDs are rare, constituting only a few percent of the population both observationally \citep[e.g.][]{2020MNRAS.499.1890M} and in the {\sc bpass} models. Therefore our assumption that all WDNS in-spirals lead to merger is, for the purposes of this work, a reasonable approximation. 

Other considerations include whether accretion rates in WDNS mergers are high enough to launch a GRB, or if the merger conditions are suitable for producing a kilonova. GRB launch is likely only plausible for mergers with less asymmetric mass ratios $q$ (defined as M$_{\rm WD}$/M$_{\rm NS}$), and hence large disc masses \citep[up to a few tenths of a Solar mass,][]{2023ApJ...956...71K}. Similarly, kilonovae require accretion rates (and hence disc densities) high enough to neutronise the gas, which is considered unlikely in WDNS mergers \citep[e.g.][and references therein]{2023ApJ...956...71K,2025ApJ...984...77G}. 

To explore the impact of these additional criteria, for half-Solar metallicity and \citet{2017A&A...608A..57V} natal kicks, we select WDNS models with $q>0.5$ and which in-spiral within a Hubble time. These binaries are expected to merge and produce relatively large disc masses, making them more likely GRB (even if not kilonova) progenitors. We find that they have systemic velocities, delay times and hence merger offsets which are not substantially different from the overall population of WDNS mergers: a median of V$_{\rm sys}=101^{+42}_{-40}$,km\,s$^{-1}$ versus V$_{\rm sys}=104^{+64}_{-37}$,km\,s$^{-1}$ for all WDNS mergers, where the uncertainties represent the central 68\% of the distribution (i.e. they bound the 16th to the 84th percentiles). For the ages, we find medians of log$_{10}$(age/yr)$=10.1^{+0.0}_{-1.9}$ versus log$_{10}$(age/yr)$=10.0^{+0.1}_{-1.3}$. These values are calculated from the simulated populations shown in Figure \ref{fig:delay_sep}, with each contributing model weighted as described in Sec \ref{sec:popsynth}). However, binaries with $q>0.5$ constitute only 30\% of the merging WDNS population in our models, and the fraction declines rapidly as the mass ratio approaches one, as the WD mass distribution peaks around 0.5-0.6\,M$_{\odot}$. Therefore, strict constraints on $q_{\rm min}$ result in a smaller fraction of long duration GRBs which can plausibly be explained by WDNS mergers. Events with massive white dwarfs, exceeding $q_{\rm min}>0.9$, are predicted to be as rare as WDBH mergers. In this case, WDNS mergers could contribute no more than $\sim$10\% of the LGRB rate, modulo the order of magnitude rate uncertainties discussed in Section \ref{sec:rates}.

\subsection{Offsets as progenitor probes}
We have shown that the galactocentric merger offsets of different compact binaries are expected to have different distributions \citep[see also][]{2025arXiv250404825G}, however, these differences can be subtle. Indeed, it is clear from Figure \ref{fig:results} that the offset alone is not a strong constraint on the specific binary progenitor of any given merger event. However, if a large sample can be constructed, such constraints may be possible on a population level. To quantify this, we draw 100 samples of size $N$ from the WDNS and BNS offsets distributions for GRBs\,211211A and 230307A, with \citet{2017A&A...608A..57V} kicks, and perform a Kolmogorov–Smirnov (KS) consistency test, noting the mean $p$-value for each $N$. We decrease $N$ incrementally until the KS-test can no longer reject the null hypothesis that the distributions have the same parent distribution, using a (mean) $p$-value of 0.05 as the threshold. In both cases, we find that for sample sizes $N \gtrapprox 100$, our predicted WDNS and BNS distributions can be distinguished at $2\sigma$ significance. In the scenario that WDNS mergers are solely responsible for long GRBs with kilonovae, and BNS mergers are solely responsible for short GRBs, then this could be tested observationally, but the real situation is likely far more complex. BNS mergers might also contribute to long GRBs with KNe, NSBH mergers may play a role in both populations, and WDNS mergers may not even produce GRBs and KNe. Despite these uncertainties, we have demonstrated that we cannot make strong statements about any given transient having a WDNS progenitor based on its offset alone. However, low-mass host galaxies provide much better constraints on both the natal kicks being imparted, and the nature of the compact binary \citep{2025arXiv250404825G}, since the binaries are more likely to reach escape velocity, and the potential has less influence on the eventual merger offset distribution. Furthermore, when comparing offsets from different galaxies, the host-normalised offset should be used \citep[e.g.][]{2022ApJ...940...56F}. 

An important caveat when using offsets as a constraint is that host galaxy associations are not always robust \citep[e.g.][]{2007MNRAS.378.1439L,2014MNRAS.437.1495T}. $P_{\rm chance}$ arguments naturally favour bright hosts, unless the offset from a faint galaxy is very small \citep[e.g.][]{2024ApJ...962....5N} and the localisation is good \citep[i.e. a sub-arcsecond localisation from an optical counterpart][]{2007MNRAS.378.1439L,2010ApJ...722.1946B}. Galaxies with higher $P_{\rm chance}$ values should not be ruled out if we are to avoid a bias against high-offsets in our samples \citep{2025arXiv250404825G}. Of the GRBs studied in this work, GRB\,230307A in particular lies at a very large offset, but the association is made not only by $P_{\rm chance}$ but also energetics and spectroscopic arguments - placing it at the distance of the putative host yields a similar kilonova luminosity to other events, and with spectroscopic features (e.g. Te) at consistent wavelengths \citep{2024Natur.626..737L}. GRB\,211211A meanwhile is much closer in projection to its putative host, with a $P_{\rm chance}$ value of just 1.4\%, and no underlying alternative host even in deep {\it Hubble Space Telescope} imaging \citep{2022Natur.612..223R}. Therefore, the host associations for these two GRBs are considered secure. In both cases, the kilonovae display remarkable photometric and temporal similarity with previously confirmed events \citep[e.g. AT2017gfo,][]{2017ApJ...848L..27T}, as well as similar spectroscopic features \citep[][]{2023arXiv230800633G}.

Given that GRBs\,211211A and 230307A are at low redshifts, it is also worth noting that we predict WDNS mergers in particular to become more common with respect to BNS/NSBH mergers at lower redshifts. The current lack of a redshift distribution for SN-less long GRBs / long GRBs with kilonovae specifically limits the use of rates as a tool for inferring an evolving contribution from non-core-collapse progenitors \citep[but see][and Levan et al. prep]{2024MNRAS.535.2800L,2024ApJ...963L..12P}. Other than offsets and volumetric rates, kilonova properties \citep{2018ApJ...860...62G,2023MNRAS.526.4585G,2024ApJ...963..156W,2025ApJ...979..190R,2025ApJ...984...77G} may also provide insight into the specific progenitors of long-duration merger GRBs. 

Alternative ideas for long-duration SN-less GRBs include other kinds of compact binary merger \citep[e.g. NSBHs,][]{2025ApJ...984...77G} or failed supernova \citep[core-collapse events with no or very faint supernovae,][]{1993ApJ...405..273W,1999ApJ...526..152F}. The offsets of SN-less long GRBs, regardless of whether a kilonova is detected or ruled out, offer crucial insight: large host offsets strongly favour a compact binary progenitor of some variety.

\subsection{Alternative progenitor pathways involving a white dwarf}
An alternative route is white dwarf accretion (or merger) induced collapse (AIC). Some models suggest that AIC events can produce GRBs and/or kilonovae \citep{2006MNRAS.368L...1L,2007ApJ...669..585D,2019MNRAS.484..698R,2025ApJ...984..197B,2025ApJ...978L..38C}, but this is far from certain \citep[e.g.][]{1999ApJ...516..892F,2006ApJ...644.1063D,2025ApJ...984...77G}. Since white dwarfs do not receive substantial natal kicks, it is hard to eject them from their host galaxies. A possible solution is to have such events occurring in extended halos or globular clusters \citep{2021MNRAS.503.5997P}, but events such as GRB\,230307A - at an offset of 40\,kpc - are likely beyond the halo and globular cluster distribution \citep[e.g.][]{2008MNRAS.385..361S,2022MNRAS.509..180L}. Dynamical ejections of binaries from globular clusters, may, however, extend the offset distribution further in this scenario \citep[e.g.][]{2006ApJ...646..464S,2023ApJ...946..104W}. AIC may also play a role in producing WDNS and WDBH binaries by producing the NS or BH component from a WD or NS respectively \citep[e.g.][]{2023ApJ...951...91C}.

Finally, we recognise that this work focuses solely on field binaries. Although globular clusters likely see an elevated WDNS formation rate due to dynamical interactions \citep[e.g.][and references therein]{2018PhRvL.120s1103K,2024MNRAS.530..844K}, only a small fraction of the total stellar mass in any given galaxy is in globular clusters. Following from this, and based on the results of simulations and constraints from type Ia supernovae, the rate of dynamically-formed mergers is unlikely to be competitive with mergers in the field \citep[e.g.][]{2008ApJ...676.1162S,2013ApJ...762....1W,2020ApJ...888L..10Y}. Nonetheless, future work should consider the contributions from the AIC and dynamical formation scenarios.

\subsection{Other transients from WDNS and WDBH mergers}
Even in the case that WDNS and WDBH mergers do not produce GRBs and/or kilonovae, our predictions for the rates, systemic velocities and merger times of these systems still hold. Possibilities for alternative transients from NSWD mergers in particular include sub-luminous type Ia and/or type Iax supernovae \citep{2012MNRAS.419..827M,2014MNRAS.444.2157L,2022MNRAS.510.3758B}, rapidly fading bright transients \citep{2019ApJ...885L..23M,2020MNRAS.497..246G}, or fast and faint optical transients \citep{2016MNRAS.461.1154M,2019MNRAS.486.1805Z,2020MNRAS.493.3956Z}. In any case, WDNS and WDBH mergers should produce transients of some kind. They may occur at a comparable rate to BNS and NSBH mergers (Figure \ref{fig:volrate}), and could be found well outside their host galaxies.

\section{Conclusions}
In this paper, we performed population synthesis of WDNS and WDBH binaries using the {\sc bpass} code to investigate their evolutionary pathways and determine volumetric merger rates. We simulate the merger offset distribution of BNS and WDNS binaries around the host galaxies of long-duration GRBs from compact binary mergers for the first time, following suggestions in the literature that WDNS mergers can explain these events. Our conclusions are as follows,
\begin{itemize}
    \item Within our host galaxy and fiducial stellar population models, WDNS mergers are consistent with the observed projected offsets of GRBs 211211A and 230307A.   
    \item Binary neutron star merger offsets are also consistent with GRBs\,211211A and 230307A. WDNS binaries merge at lower offsets than BNSs, however, we predict that the disparity will only become apparent for large sample sizes of order $\sim$100. Therefore, offset alone is not a effective way to distinguish progenitors on a case by case basis. The picture is complicated by the possibility of other compact binary progenitors for SN-less long GRBs with kilonovae, such as BHNS mergers, and the fact that WDNS mergers may produce different transients altogether.
    \item We find that the rate of WDNS mergers at low redshift ($z<0.5$) is similar to the BNS merger and intrinsic long GRB rate. If low mass asymmetry is required for a successful GRB, the rate drops by as much as another factor of ten (if mass ratios close to one are imposed), similar to the WDBH merger rate. In this case, WDNS mergers can explain at most $\sim$10\% of long GRBs at low redshift. However, large (order of magnitude) observational and theoretical uncertainties preclude definitive statements based on volumetric rates at this stage. 
\end{itemize}

\noindent In summary, we have shown that WDNS mergers are relatively frequent events (with respect to BNS mergers), and are capable of occurring at large offsets from their host galaxies, which we have demonstrated with the host galaxies of two long duration GRBs with kilonovae for the first time. These results corroborate previous work in this area with different population synthesis codes \citep[e.g.][]{2018AandA...619A..53T}. In future, a reduction in GRB rate uncertainties - particularly in the rate of SN-less long-duration GRBs and those with kilonovae -  along with a reduction in binary stellar evolution modelling uncertainties - will enable stronger statements on their origins based on events rates. A large sample of SN-less long GRB offsets will also allow us to investigate if there is a difference with respect to short GRBs, but only if statistically significant sample sizes can be reached. The strongest evidence, however, for a merger involving a white dwarf would be a LIGO/VIRGO/KAGRA gravitational wave non-detection of a nearby SN-less long-duration GRB, given that mergers involving a WD lie well below the frequency range of these detectors \citep[e.g.][]{2024A&A...681A..41M}.

\begin{acknowledgements} 
We thank the anonymous referee for their careful consideration of this manuscript. AAC acknowledges support through the European Space Agency (ESA) research fellowship programme. NG acknowledges studentship support from the Dutch Research Council (NWO) under the project number 680.92.18.02. BPG acknowledges support from STFC grant No. ST/Y002253/1 and The Leverhulme Trust grant No. RPG-2024-117.

This work made use of v2.2.1 of the Binary Population and Spectral Synthesis ({\sc bpass}) models as described in \citet{2017PASA...34...58E} and \citet{2018MNRAS.479...75S}. This work has made use of {\sc ipython} \citep{2007CSE.....9c..21P}, {\sc numpy} \citep{2020Natur.585..357H}, {\sc scipy} \citep{2020NatMe..17..261V}; {\sc matplotlib} \citep{2007CSE.....9...90H}, Seaborn packages \citep{Waskom2021} and {\sc astropy},(\url{https://www.astropy.org}) a community-developed core Python package for Astronomy \citep{astropy:2013, astropy:2018}. This research has made use of the SVO Filter Profile Service "Carlos Rodrigo", funded by MCIN/AEI/10.13039/501100011033/ through grant PID2020-112949GB-I00.

\end{acknowledgements}

%
\bibliographystyle{aa} 
\bibliography{NSWD_long_GRBs.bib} 

\begin{thebibliography}{213}
\expandafter\ifx\csname natexlab\endcsname\relax\def\natexlab#1{#1}\fi

\bibitem[{{Abbott} {et~al.}(2017{\natexlab{a}}){Abbott}, {Abbott}, {Abbott},
  {Acernese}, {Ackley}, {Adams}, {Adams}, {Addesso}, {Adhikari}, {Adya},
  {Affeldt}, {Afrough}, {Agarwal}, {Agathos}, {Agatsuma}, {Aggarwal}, {Aguiar},
  {Aiello}, {Ain}, {Ajith}, {Allen}, {Allen}, {Allocca}, {Aloy}, {Altin},
  {Amato}, {Ananyeva}, {Anderson}, {Anderson}, {Angelova}, {Antier}, {Appert},
  {Arai}, {Araya}, {Areeda}, {Arnaud}, {Arun}, {Ascenzi}, {Ashton}, {Ast},
  {Aston}, {Astone}, {Atallah}, {Aufmuth}, {Aulbert}, {AultONeal}, {Austin},
  {Avila-Alvarez}, {Babak}, {Bacon}, {Bader}, {Bae}, {Baker}, {Baldaccini},
  {Ballardin}, {Ballmer}, {Banagiri}, {Barayoga}, {Barclay}, {Barish},
  {Barker}, {Barkett}, {Barone}, {Barr}, {Barsotti}, {Barsuglia}, {Barta},
  {Bartlett}, {Bartos}, {Bassiri}, {Basti}, {Batch}, {Bawaj}, {Bayley},
  {Bazzan}, {B{\'e}csy}, {Beer}, {Bejger}, {Belahcene}, {Bell}, {Berger},
  {Bergmann}, {Bero}, {Berry}, {Bersanetti}, {Bertolini}, {Betzwieser},
  {Bhagwat}, {Bhandare}, {Bilenko}, {Billingsley}, {Billman}, {Birch},
  {Birney}, {Birnholtz}, {Biscans}, {Biscoveanu}, {Bisht}, {Bitossi}, {Biwer},
  {Bizouard}, {Blackburn}, {Blackman}, {Blair}, {Blair}, {Blair}, {Bloemen},
  {Bock}, {Bode}, {Boer}, {Bogaert}, {Bohe}, {Bondu}, {Bonilla}, {Bonnand},
  {Boom}, {Bork}, {Boschi}, {Bose}, {Bossie}, {Bouffanais}, {Bozzi},
  {Bradaschia}, {Brady}, {Branchesi}, {Brau}, {Briant}, {Brillet}, {Brinkmann},
  {Brisson}, {Brockill}, {Broida}, {Brooks}, {Brown}, {Brown}, {Brunett},
  {Buchanan}, {Buikema}, {Bulik}, {Bulten}, {Buonanno}, {Buskulic}, {Buy},
  {Byer}, {Cabero}, {Cadonati}, {Cagnoli}, {Cahillane}, {Calder{\'o}n
  Bustillo}, {Callister}, {Calloni}, {Camp}, {Canepa}, {Canizares}, {Cannon},
  {Cao}, {Cao}, {Capano}, {Capocasa}, {Carbognani}, {Caride}, {Carney},
  {Casanueva Diaz}, {Casentini}, {Caudill}, {Cavagli{\`a}}, {Cavalier},
  {Cavalieri}, {Cella}, {Cepeda}, {Cerd{\'a}-Dur{\'a}n}, {Cerretani},
  {Cesarini}, {Chamberlin}, {Chan}, {Chao}, {Charlton}, {Chase},
  {Chassande-Mottin}, {Chatterjee}, {Chatziioannou}, {Cheeseboro}, {Chen},
  {Chen}, {Chen}, {Cheng}, {Chia}, {Chincarini}, {Chiummo}, {Chmiel}, {Cho},
  {Cho}, {Chow}, {Christensen}, {Chu}, {Chua}, {Chua}, {Chung}, {Chung},
  {Ciani}, {Ciolfi}, {Cirelli}, {Cirone}, {Clara}, {Clark}, {Clearwater},
  {Cleva}, {Cocchieri}, {Coccia}, {Cohadon}, {Cohen}, {Colla}, {Collette},
  {Cominsky}, {Constancio}, {Conti}, {Cooper}, {Corban}, {Corbitt},
  {Cordero-Carri{\'o}n}, {Corley}, {Cornish}, {Corsi}, {Cortese}, {Costa},
  {Coughlin}, {Coughlin}, {Coulon}, {Countryman}, {Couvares}, {Covas}, {Cowan},
  {Coward}, {Cowart}, {Coyne}, {Coyne}, {Creighton}, {Creighton}, {Cripe},
  {Crowder}, {Cullen}, {Cumming}, {Cunningham}, {Cuoco}, {Dal Canton},
  {D{\'a}lya}, {Danilishin}, {D'Antonio}, {Danzmann}, {Dasgupta}, {Da Silva
  Costa}, {Dattilo}, {Dave}, {Davier}, {Davis}, {Daw}, {Day}, {De}, {DeBra},
  {Degallaix}, {De Laurentis}, {Del{\'e}glise}, {Del Pozzo}, {Demos}, {Denker},
  {Dent}, {De Pietri}, {Dergachev}, {De Rosa}, {DeRosa}, {De Rossi}, {DeSalvo},
  {de Varona}, {Devenson}, {Dhurandhar}, {D{\'\i}az}, {Di Fiore}, {Di
  Giovanni}, {Di Girolamo}, {Di Lieto}, {Di Pace}, {Di Palma}, {Di Renzo},
  {Doctor}, {Dolique}, {Donovan}, {Dooley}, {Doravari}, {Dorrington},
  {Douglas}, {Dovale {\'A}lvarez}, {Downes}, {Drago}, {Dreissigacker},
  {Driggers}, {Du}, {Ducrot}, {Dupej}, {Dwyer}, {Edo}, {Edwards}, {Effler},
  {Eggenstein}, {Ehrens}, {Eichholz}, {Eikenberry}, {Eisenstein}, {Essick},
  {Estevez}, {Etienne}, {Etzel}, {Evans}, {Evans}, {Factourovich}, {Fafone},
  {Fair}, {Fairhurst}, {Fan}, {Farinon}, {Farr}, {Farr}, {Fauchon-Jones},
  {Favata}, {Fays}, {Fee}, {Fehrmann}, {Feicht}, {Fejer}, {Fernandez-Galiana},
  {Ferrante}, {Ferreira}, {Ferrini}, {Fidecaro}, {Finstad}, {Fiori},
  {Fiorucci}, {Fishbach}, {Fisher}, {Fitz-Axen}, {Flaminio}, {Fletcher},
  {Fong}, {Font}, {Forsyth}, {Forsyth}, {Fournier}, {Frasca}, {Frasconi},
  {Frei}, {Freise}, {Frey}, {Frey}, {Fries}, {Fritschel}, {Frolov}, {Fulda},
  {Fyffe}, {Gabbard}, {Gadre}, {Gaebel}, {Gair}, {Gammaitoni}, {Ganija},
  {Gaonkar}, {Garcia-Quiros}, {Garufi}, {Gateley}, {Gaudio}, {Gaur},
  {Gayathri}, {Gehrels}, {Gemme}, {Genin}, {Gennai}, {George}, {George},
  {Gergely}, {Germain}, {Ghonge}, {Ghosh}, {Ghosh}, {Ghosh}, {Giaime},
  {Giardina}, {Giazotto}, {Gill}, {Glover}, {Goetz}, {Goetz}, {Gomes},
  {Goncharov}, {Gonz{\'a}lez}, {Gonzalez Castro}, {Gopakumar}, {Gorodetsky},
  {Gossan}, {Gosselin}, {Gouaty}, {Grado}, {Graef}, {Granata}, {Grant}, {Gras},
  {Gray}, {Greco}, {Green}, {Gretarsson}, {Groot}, {Grote}, {Grunewald},
  {Gruning}, {Guidi}, {Guo}, {Gupta}, {Gupta}, {Gushwa}, {Gustafson},
  {Gustafson}, {Halim}, {Hall}, {Hall}, {Hamilton}, {Hammond}, {Haney},
  {Hanke}, {Hanks}, {Hanna}, {Hannam}, {Hannuksela}, {Hanson}, {Hardwick},
  {Harms}, {Harry}, {Harry}, {Hart}, {Haster}, {Haughian}, {Healy}, {Heidmann},
  {Heintze}, {Heitmann}, {Hello}, {Hemming}, {Hendry}, {Heng}, {Hennig},
  {Heptonstall}, {Heurs}, {Hild}, {Hinderer}, {Hoak}, {Hofman}, {Holt}, {Holz},
  {Hopkins}, {Horst}, {Hough}, {Houston}, {Howell}, {Hreibi}, {Hu}, {Huerta},
  {Huet}, {Hughey}, {Husa}, {Huttner}, {Huynh-Dinh}, {Indik}, {Inta}, {Intini},
  {Isa}, {Isac}, {Isi}, {Iyer}, {Izumi}, {Jacqmin}, {Jani}, {Jaranowski},
  {Jawahar}, {Jim{\'e}nez-Forteza}, {Johnson}, {Johnson-McDaniel}, {Jones},
  {Jones}, {Jonker}, {Ju}, {Junker}, {Kalaghatgi}, {Kalogera}, {Kamai},
  {Kandhasamy}, {Kang}, {Kanner}, {Kapadia}, {Karki}, {Karvinen}, {Kasprzack},
  {Kastaun}, {Katolik}, {Katsavounidis}, {Katzman}, {Kaufer}, {Kawabe},
  {K{\'e}f{\'e}lian}, {Keitel}, {Kemball}, {Kennedy}, {Kent}, {Key}, {Khalili},
  {Khan}, {Khan}, {Khan}, {Khazanov}, {Kijbunchoo}, {Kim}, {Kim}, {Kim}, {Kim},
  {Kim}, {Kim}, {Kimbrell}, {King}, {King}, {Kinley-Hanlon}, {Kirchhoff},
  {Kissel}, {Kleybolte}, {Klimenko}, {Knowles}, {Koch}, {Koehlenbeck}, {Koley},
  {Kondrashov}, {Kontos}, {Korobko}, {Korth}, {Kowalska}, {Kozak},
  {Kr{\"a}mer}, {Kringel}, {Krishnan}, {Kr{\'o}lak}, {Kuehn}, {Kumar}, {Kumar},
  {Kumar}, {Kuo}, {Kutynia}, {Kwang}, {Lackey}, {Lai}, {Landry}, {Lang},
  {Lange}, {Lantz}, {Lanza}, {Lartaux-Vollard}, {Lasky}, {Laxen}, {Lazzarini},
  {Lazzaro}, {Leaci}, {Leavey}, {Lee}, {Lee}, {Lee}, {Lee}, {Lee}, {Lehmann},
  {Lenon}, {Leonardi}, {Leroy}, {Letendre}, {Levin}, {Li}, {Linker},
  {Littenberg}, {Liu}, {Lo}, {Lockerbie}, {London}, {Lord}, {Lorenzini},
  {Loriette}, {Lormand}, {Losurdo}, {Lough}, {Lousto}, {Lovelace}, {L{\"u}ck},
  {Lumaca}, {Lundgren}, {Lynch}, {Ma}, {Macas}, {Macfoy}, {Machenschalk},
  {MacInnis}, {Macleod}, {Maga{\~n}a Hernandez}, {Maga{\~n}a-Sandoval},
  {Maga{\~n}a Zertuche}, {Magee}, {Majorana}, {Maksimovic}, {Man}, {Mandic},
  {Mangano}, {Mansell}, {Manske}, {Mantovani}, {Marchesoni}, {Marion},
  {M{\'a}rka}, {M{\'a}rka}, {Markakis}, {Markosyan}, {Markowitz}, {Maros},
  {Marquina}, {Martelli}, {Martellini}, {Martin}, {Martin}, {Martynov},
  {Mason}, {Massera}, {Masserot}, {Massinger}, {Masso-Reid}, {Mastrogiovanni},
  {Matas}, {Matichard}, {Matone}, {Mavalvala}, {Mazumder}, {McCarthy},
  {McClelland}, {McCormick}, {McCuller}, {McGuire}, {McIntyre}, {McIver},
  {McManus}, {McNeill}, {McRae}, {McWilliams}, {Meacher}, {Meadors}, {Mehmet},
  {Meidam}, {Mejuto-Villa}, {Melatos}, {Mendell}, {Mercer}, {Merilh},
  {Merzougui}, {Meshkov}, {Messenger}, {Messick}, {Metzdorff}, {Meyers},
  {Miao}, {Michel}, {Middleton}, {Mikhailov}, {Milano}, {Miller}, {Miller},
  {Miller}, {Millhouse}, {Milovich-Goff}, {Minazzoli}, {Minenkov}, {Ming},
  {Mishra}, {Mitra}, {Mitrofanov}, {Mitselmakher}, {Mittleman}, {Moffa},
  {Moggi}, {Mogushi}, {Mohan}, {Mohapatra}, {Montani}, {Moore}, {Moraru},
  {Moreno}, {Morriss}, {Mours}, {Mow-Lowry}, {Mueller}, {Muir}, {Mukherjee},
  {Mukherjee}, {Mukherjee}, {Mukund}, {Mullavey}, {Munch}, {Mu{\~n}iz},
  {Muratore}, {Murray}, {Napier}, {Nardecchia}, {Naticchioni}, {Nayak},
  {Neilson}, {Nelemans}, {Nelson}, {Nery}, {Neunzert}, {Nevin}, {Newport},
  {Newton}, {Ng}, {Nguyen}, {Nichols}, {Nielsen}, {Nissanke}, {Nitz}, {Noack},
  {Nocera}, {Nolting}, {North}, {Nuttall}, {Oberling}, {O'Dea}, {Ogin}, {Oh},
  {Oh}, {Ohme}, {Okada}, {Oliver}, {Oppermann}, {Oram}, {O'Reilly}, {Ormiston},
  {Ortega}, {O'Shaughnessy}, {Ossokine}, {Ottaway}, {Overmier}, {Owen}, {Pace},
  {Page}, {Page}, {Pai}, {Pai}, {Palamos}, {Palashov}, {Palomba}, {Pal-Singh},
  {Pan}, {Pan}, {Pang}, {Pang}, {Pankow}, {Pannarale}, {Pant}, {Paoletti},
  {Paoli}, {Papa}, {Parida}, {Parker}, {Pascucci}, {Pasqualetti},
  {Passaquieti}, {Passuello}, {Patil}, {Patricelli}, {Pearlstone}, {Pedraza},
  {Pedurand}, {Pekowsky}, {Pele}, {Penn}, {Perez}, {Perreca}, {Perri},
  {Pfeiffer}, {Phelps}, {Piccinni}, {Pichot}, {Piergiovanni}, {Pierro},
  {Pillant}, {Pinard}, {Pinto}, {Pirello}, {Pitkin}, {Poe}, {Poggiani},
  {Popolizio}, {Porter}, {Post}, {Powell}, {Prasad}, {Pratt}, {Pratten},
  {Predoi}, {Prestegard}, {Prijatelj}, {Principe}, {Privitera}, {Prodi},
  {Prokhorov}, {Puncken}, {Punturo}, {Puppo}, {P{\"u}rrer}, {Qi}, {Quetschke},
  {Quintero}, {Quitzow-James}, {Raab}, {Rabeling}, {Radkins}, {Raffai}, {Raja},
  {Rajan}, {Rajbhandari}, {Rakhmanov}, {Ramirez}, {Ramos-Buades}, {Rapagnani},
  {Raymond}, {Razzano}, {Read}, {Regimbau}, {Rei}, {Reid}, {Reitze}, {Ren},
  {Reyes}, {Ricci}, {Ricker}, {Rieger}, {Riles}, {Rizzo}, {Robertson}, {Robie},
  {Robinet}, {Rocchi}, {Rolland}, {Rollins}, {Roma}, {Romano}, {Romel},
  {Romie}, {Rosi{\'n}ska}, {Ross}, {Rowan}, {R{\"u}diger}, {Ruggi}, {Rutins},
  {Ryan}, {Sachdev}, {Sadecki}, {Sadeghian}, {Sakellariadou}, {Salconi},
  {Saleem}, {Salemi}, {Samajdar}, {Sammut}, {Sampson}, {Sanchez}, {Sanchez},
  {Sanchis-Gual}, {Sandberg}, {Sanders}, {Sassolas}, {Sathyaprakash},
  {Saulson}, {Sauter}, {Savage}, {Sawadsky}, {Schale}, {Scheel}, {Scheuer},
  {Schmidt}, {Schmidt}, {Schnabel}, {Schofield}, {Sch{\"o}nbeck}, {Schreiber},
  {Schuette}, {Schulte}, {Schutz}, {Schwalbe}, {Scott}, {Scott}, {Seidel},
  {Sellers}, {Sengupta}, {Sentenac}, {Sequino}, {Sergeev}, {Shaddock},
  {Shaffer}, {Shah}, {Shahriar}, {Shaner}, {Shao}, {Shapiro}, {Shawhan},
  {Sheperd}, {Shoemaker}, {Shoemaker}, {Siellez}, {Siemens}, {Sieniawska},
  {Sigg}, {Silva}, {Singer}, {Singh}, {Singhal}, {Sintes}, {Slagmolen},
  {Smith}, {Smith}, {Smith}, {Somala}, {Son}, {Sonnenberg}, {Sorazu},
  {Sorrentino}, {Souradeep}, {Spencer}, {Srivastava}, {Staats}, {Staley},
  {Steinke}, {Steinlechner}, {Steinlechner}, {Steinmeyer}, {Stevenson},
  {Stone}, {Stops}, {Strain}, {Stratta}, {Strigin}, {Strunk}, {Sturani},
  {Stuver}, {Summerscales}, {Sun}, {Sunil}, {Suresh}, {Sutton}, {Swinkels},
  {Szczepa{\'n}czyk}, {Tacca}, {Tait}, {Talbot}, {Talukder}, {Tanner},
  {T{\'a}pai}, {Taracchini}, {Tasson}, {Taylor}, {Taylor}, {Tewari}, {Theeg},
  {Thies}, {Thomas}, {Thomas}, {Thomas}, {Thorne}, {Thorne}, {Thrane},
  {Tiwari}, {Tiwari}, {Tokmakov}, {Toland}, {Tonelli}, {Tornasi},
  {Torres-Forn{\'e}}, {Torrie}, {T{\"o}yr{\"a}}, {Travasso}, {Traylor},
  {Trinastic}, {Tringali}, {Trozzo}, {Tsang}, {Tse}, {Tso}, {Tsukada}, {Tsuna},
  {Tuyenbayev}, {Ueno}, {Ugolini}, {Unnikrishnan}, {Urban}, {Usman},
  {Vahlbruch}, {Vajente}, {Valdes}, {van Bakel}, {van Beuzekom}, {van den
  Brand}, {Van Den Broeck}, {Vander-Hyde}, {van der Schaaf}, {van Heijningen},
  {van Veggel}, {Vardaro}, {Varma}, {Vass}, {Vas{\'u}th}, {Vecchio},
  {Vedovato}, {Veitch}, {Veitch}, {Venkateswara}, {Venugopalan}, {Verkindt},
  {Vetrano}, {Vicer{\'e}}, {Viets}, {Vinciguerra}, {Vine}, {Vinet}, {Vitale},
  {Vo}, {Vocca}, {Vorvick}, {Vyatchanin}, {Wade}, {Wade}, {Wade}, {Walet},
  {Walker}, {Wallace}, {Walsh}, {Wang}, {Wang}, {Wang}, {Wang}, {Wang}, {Ward},
  {Warner}, {Was}, {Watchi}, {Weaver}, {Wei}, {Weinert}, {Weinstein}, {Weiss},
  {Wen}, {Wessel}, {We{\ss}els}, {Westerweck}, {Westphal}, {Wette}, {Whelan},
  {Whitcomb}, {Whiting}, {Whittle}, {Wilken}, {Williams}, {Williams},
  {Williamson}, {Willis}, {Willke}, {Wimmer}, {Winkler}, {Wipf}, {Wittel},
  {Woan}, {Woehler}, {Wofford}, {Wong}, {Worden}, {Wright}, {Wu}, {Wysocki},
  {Xiao}, {Yamamoto}, {Yancey}, {Yang}, {Yap}, {Yazback}, {Yu}, {Yu}, {Yvert},
  {Zadro{\.z}ny}, {Zanolin}, {Zelenova}, {Zendri}, {Zevin}, {Zhang}, {Zhang},
  {Zhang}, {Zhang}, {Zhao}, {Zhou}, {Zhou}, {Zhu}, {Zhu}, {Zimmerman},
  {Zucker}, {Zweizig}, {(LIGO Scientific Collaboration}, {Virgo Collaboration},
  {Burns}, {Veres}, {Kocevski}, {Racusin}, {Goldstein}, {Connaughton},
  {Briggs}, {Blackburn}, {Hamburg}, {Hui}, {von Kienlin}, {McEnery}, {Preece},
  {Wilson-Hodge}, {Bissaldi}, {Cleveland}, {Gibby}, {Giles}, {Kippen},
  {McBreen}, {Meegan}, {Paciesas}, {Poolakkil}, {Roberts}, {Stanbro},
  {Gamma-ray Burst Monitor}, {Savchenko}, {Ferrigno}, {Kuulkers}, {Bazzano},
  {Bozzo}, {Brandt}, {Chenevez}, {Courvoisier}, {Diehl}, {Domingo}, {Hanlon},
  {Jourdain}, {Laurent}, {Lebrun}, {Lutovinov}, {Mereghetti}, {Natalucci},
  {Rodi}, {Roques}, {Sunyaev}, {Ubertini}, \&
  {(INTEGRAL}}]{2017ApJ...848L..13A}
{Abbott}, B.~P., {Abbott}, R., {Abbott}, T.~D., {et~al.} 2017{\natexlab{a}},
  \apjl, 848, L13

\bibitem[{{Abbott} {et~al.}(2017{\natexlab{b}}){Abbott}, {Abbott}, {Abbott},
  {Acernese}, {Ackley}, {Adams}, {Adams}, {Addesso}, {Adhikari}, {Adya},
  {Affeldt}, {Afrough}, {Agarwal}, {Agathos}, {Agatsuma}, {Aggarwal}, {Aguiar},
  {Aiello}, {Ain}, {Ajith}, {Allen}, {Allen}, {Allocca}, {Altin}, {Amato},
  {Ananyeva}, {Anderson}, {Anderson}, {Angelova}, {Antier}, {Appert}, {Arai},
  {Araya}, {Areeda}, {Arnaud}, {Arun}, {Ascenzi}, {Ashton}, {Ast}, {Aston},
  {Astone}, {Atallah}, {Aufmuth}, {Aulbert}, {AultONeal}, {Austin},
  {Avila-Alvarez}, {Babak}, {Bacon}, {Bader}, {Bae}, {Bailes}, {Baker},
  {Baldaccini}, {Ballardin}, {Ballmer}, {Banagiri}, {Barayoga}, {Barclay},
  {Barish}, {Barker}, {Barkett}, {Barone}, {Barr}, {Barsotti}, {Barsuglia},
  {Barta}, {Barthelmy}, {Bartlett}, {Bartos}, {Bassiri}, {Basti}, {Batch},
  {Bawaj}, {Bayley}, {Bazzan}, {B{\'e}csy}, {Beer}, {Bejger}, {Belahcene},
  {Bell}, {Berger}, {Bergmann}, {Bernuzzi}, {Bero}, {Berry}, {Bersanetti},
  {Bertolini}, {Betzwieser}, {Bhagwat}, {Bhandare}, {Bilenko}, {Billingsley},
  {Billman}, {Birch}, {Birney}, {Birnholtz}, {Biscans}, {Biscoveanu}, {Bisht},
  {Bitossi}, {Biwer}, {Bizouard}, {Blackburn}, {Blackman}, {Blair}, {Blair},
  {Blair}, {Bloemen}, {Bock}, {Bode}, {Boer}, {Bogaert}, {Bohe}, {Bondu},
  {Bonilla}, {Bonnand}, {Boom}, {Bork}, {Boschi}, {Bose}, {Bossie},
  {Bouffanais}, {Bozzi}, {Bradaschia}, {Brady}, {Branchesi}, {Brau}, {Briant},
  {Brillet}, {Brinkmann}, {Brisson}, {Brockill}, {Broida}, {Brooks}, {Brown},
  {Brown}, {Brunett}, {Buchanan}, {Buikema}, {Bulik}, {Bulten}, {Buonanno},
  {Buskulic}, {Buy}, {Byer}, {Cabero}, {Cadonati}, {Cagnoli}, {Cahillane},
  {Calder{\'o}n Bustillo}, {Callister}, {Calloni}, {Camp}, {Canepa},
  {Canizares}, {Cannon}, {Cao}, {Cao}, {Capano}, {Capocasa}, {Carbognani},
  {Caride}, {Carney}, {Carullo}, {Casanueva Diaz}, {Casentini}, {Caudill},
  {Cavagli{\`a}}, {Cavalier}, {Cavalieri}, {Cella}, {Cepeda},
  {Cerd{\'a}-Dur{\'a}n}, {Cerretani}, {Cesarini}, {Chamberlin}, {Chan}, {Chao},
  {Charlton}, {Chase}, {Chassande-Mottin}, {Chatterjee}, {Chatziioannou},
  {Cheeseboro}, {Chen}, {Chen}, {Chen}, {Cheng}, {Chia}, {Chincarini},
  {Chiummo}, {Chmiel}, {Cho}, {Cho}, {Chow}, {Christensen}, {Chu}, {Chua},
  {Chua}, {Chung}, {Chung}, {Ciani}, {Ciolfi}, {Cirelli}, {Cirone}, {Clara},
  {Clark}, {Clearwater}, {Cleva}, {Cocchieri}, {Coccia}, {Cohadon}, {Cohen},
  {Colla}, {Collette}, {Cominsky}, {Constancio}, {Conti}, {Cooper}, {Corban},
  {Corbitt}, {Cordero-Carri{\'o}n}, {Corley}, {Cornish}, {Corsi}, {Cortese},
  {Costa}, {Coughlin}, {Coughlin}, {Coulon}, {Countryman}, {Couvares}, {Covas},
  {Cowan}, {Coward}, {Cowart}, {Coyne}, {Coyne}, {Creighton}, {Creighton},
  {Cripe}, {Crowder}, {Cullen}, {Cumming}, {Cunningham}, {Cuoco}, {Dal Canton},
  {D{\'a}lya}, {Danilishin}, {D'Antonio}, {Danzmann}, {Dasgupta}, {Da Silva
  Costa}, {Dattilo}, {Dave}, {Davier}, {Davis}, {Daw}, {Day}, {De}, {DeBra},
  {Degallaix}, {De Laurentis}, {Del{\'e}glise}, {Del Pozzo}, {Demos}, {Denker},
  {Dent}, {De Pietri}, {Dergachev}, {De Rosa}, {DeRosa}, {De Rossi}, {DeSalvo},
  {de Varona}, {Devenson}, {Dhurandhar}, {D{\'\i}az}, {Dietrich}, {Di Fiore},
  {Di Giovanni}, {Di Girolamo}, {Di Lieto}, {Di Pace}, {Di Palma}, {Di Renzo},
  {Doctor}, {Dolique}, {Donovan}, {Dooley}, {Doravari}, {Dorrington},
  {Douglas}, {Dovale {\'A}lvarez}, {Downes}, {Drago}, {Dreissigacker},
  {Driggers}, {Du}, {Ducrot}, {Dudi}, {Dupej}, {Dwyer}, {Edo}, {Edwards},
  {Effler}, {Eggenstein}, {Ehrens}, {Eichholz}, {Eikenberry}, {Eisenstein},
  {Essick}, {Estevez}, {Etienne}, {Etzel}, {Evans}, {Evans}, {Factourovich},
  {Fafone}, {Fair}, {Fairhurst}, {Fan}, {Farinon}, {Farr}, {Farr},
  {Fauchon-Jones}, {Favata}, {Fays}, {Fee}, {Fehrmann}, {Feicht}, {Fejer},
  {Fernandez-Galiana}, {Ferrante}, {Ferreira}, {Ferrini}, {Fidecaro},
  {Finstad}, {Fiori}, {Fiorucci}, {Fishbach}, {Fisher}, {Fitz-Axen},
  {Flaminio}, {Fletcher}, {Fong}, {Font}, {Forsyth}, {Forsyth}, {Fournier},
  {Frasca}, {Frasconi}, {Frei}, {Freise}, {Frey}, {Frey}, {Fries}, {Fritschel},
  {Frolov}, {Fulda}, {Fyffe}, {Gabbard}, {Gadre}, {Gaebel}, {Gair},
  {Gammaitoni}, {Ganija}, {Gaonkar}, {Garcia-Quiros}, {Garufi}, {Gateley},
  {Gaudio}, {Gaur}, {Gayathri}, {Gehrels}, {Gemme}, {Genin}, {Gennai},
  {George}, {George}, {Gergely}, {Germain}, {Ghonge}, {Ghosh}, {Ghosh},
  {Ghosh}, {Giaime}, {Giardina}, {Giazotto}, {Gill}, {Glover}, {Goetz},
  {Goetz}, {Gomes}, {Goncharov}, {Gonz{\'a}lez}, {Gonzalez Castro},
  {Gopakumar}, {Gorodetsky}, {Gossan}, {Gosselin}, {Gouaty}, {Grado}, {Graef},
  {Granata}, {Grant}, {Gras}, {Gray}, {Greco}, {Green}, {Gretarsson}, {Groot},
  {Grote}, {Grunewald}, {Gruning}, {Guidi}, {Guo}, {Gupta}, {Gupta}, {Gushwa},
  {Gustafson}, {Gustafson}, {Halim}, {Hall}, {Hall}, {Hamilton}, {Hammond},
  {Haney}, {Hanke}, {Hanks}, {Hanna}, {Hannam}, {Hannuksela}, {Hanson},
  {Hardwick}, {Harms}, {Harry}, {Harry}, {Hart}, {Haster}, {Haughian}, {Healy},
  {Heidmann}, {Heintze}, {Heitmann}, {Hello}, {Hemming}, {Hendry}, {Heng},
  {Hennig}, {Heptonstall}, {Heurs}, {Hild}, {Hinderer}, {Ho}, {Hoak}, {Hofman},
  {Holt}, {Holz}, {Hopkins}, {Horst}, {Hough}, {Houston}, {Howell}, {Hreibi},
  {Hu}, {Huerta}, {Huet}, {Hughey}, {Husa}, {Huttner}, {Huynh-Dinh}, {Indik},
  {Inta}, {Intini}, {Isa}, {Isac}, {Isi}, {Iyer}, {Izumi}, {Jacqmin}, {Jani},
  {Jaranowski}, {Jawahar}, {Jim{\'e}nez-Forteza}, {Johnson},
  {Johnson-McDaniel}, {Jones}, {Jones}, {Jonker}, {Ju}, {Junker}, {Kalaghatgi},
  {Kalogera}, {Kamai}, {Kandhasamy}, {Kang}, {Kanner}, {Kapadia}, {Karki},
  {Karvinen}, {Kasprzack}, {Kastaun}, {Katolik}, {Katsavounidis}, {Katzman},
  {Kaufer}, {Kawabe}, {K{\'e}f{\'e}lian}, {Keitel}, {Kemball}, {Kennedy},
  {Kent}, {Key}, {Khalili}, {Khan}, {Khan}, {Khan}, {Khazanov}, {Kijbunchoo},
  {Kim}, {Kim}, {Kim}, {Kim}, {Kim}, {Kim}, {Kimbrell}, {King}, {King},
  {Kinley-Hanlon}, {Kirchhoff}, {Kissel}, {Kleybolte}, {Klimenko}, {Knowles},
  {Koch}, {Koehlenbeck}, {Koley}, {Kondrashov}, {Kontos}, {Korobko}, {Korth},
  {Kowalska}, {Kozak}, {Kr{\"a}mer}, {Kringel}, {Krishnan}, {Kr{\'o}lak},
  {Kuehn}, {Kumar}, {Kumar}, {Kumar}, {Kuo}, {Kutynia}, {Kwang}, {Lackey},
  {Lai}, {Landry}, {Lang}, {Lange}, {Lantz}, {Lanza}, {Larson},
  {Lartaux-Vollard}, {Lasky}, {Laxen}, {Lazzarini}, {Lazzaro}, {Leaci},
  {Leavey}, {Lee}, {Lee}, {Lee}, {Lee}, {Lee}, {Lehmann}, {Lenon}, {Leon},
  {Leonardi}, {Leroy}, {Letendre}, {Levin}, {Li}, {Linker}, {Littenberg},
  {Liu}, {Liu}, {Lo}, {Lockerbie}, {London}, {Lord}, {Lorenzini}, {Loriette},
  {Lormand}, {Losurdo}, {Lough}, {Lousto}, {Lovelace}, {L{\"u}ck}, {Lumaca},
  {Lundgren}, {Lynch}, {Ma}, {Macas}, {Macfoy}, {Machenschalk}, {MacInnis},
  {Macleod}, {Maga{\~n}a Hernandez}, {Maga{\~n}a-Sandoval}, {Maga{\~n}a
  Zertuche}, {Magee}, {Majorana}, {Maksimovic}, {Man}, {Mandic}, {Mangano},
  {Mansell}, {Manske}, {Mantovani}, {Marchesoni}, {Marion}, {M{\'a}rka},
  {M{\'a}rka}, {Markakis}, {Markosyan}, {Markowitz}, {Maros}, {Marquina},
  {Marsh}, {Martelli}, {Martellini}, {Martin}, {Martin}, {Martynov}, {Marx},
  {Mason}, {Massera}, {Masserot}, {Massinger}, {Masso-Reid}, {Mastrogiovanni},
  {Matas}, {Matichard}, {Matone}, {Mavalvala}, {Mazumder}, {McCarthy},
  {McClelland}, {McCormick}, {McCuller}, {McGuire}, {McIntyre}, {McIver},
  {McManus}, {McNeill}, {McRae}, {McWilliams}, {Meacher}, {Meadors}, {Mehmet},
  {Meidam}, {Mejuto-Villa}, {Melatos}, {Mendell}, {Mercer}, {Merilh},
  {Merzougui}, {Meshkov}, {Messenger}, {Messick}, {Metzdorff}, {Meyers},
  {Miao}, {Michel}, {Middleton}, {Mikhailov}, {Milano}, {Miller}, {Miller},
  {Miller}, {Millhouse}, {Milovich-Goff}, {Minazzoli}, {Minenkov}, {Ming},
  {Mishra}, {Mitra}, {Mitrofanov}, {Mitselmakher}, {Mittleman}, {Moffa},
  {Moggi}, {Mogushi}, {Mohan}, {Mohapatra}, {Molina}, {Montani}, {Moore},
  {Moraru}, {Moreno}, {Morisaki}, {Morriss}, {Mours}, {Mow-Lowry}, {Mueller},
  {Muir}, {Mukherjee}, {Mukherjee}, {Mukherjee}, {Mukund}, {Mullavey}, {Munch},
  {Mu{\~n}iz}, {Muratore}, {Murray}, {Nagar}, {Napier}, {Nardecchia},
  {Naticchioni}, {Nayak}, {Neilson}, {Nelemans}, {Nelson}, {Nery}, {Neunzert},
  {Nevin}, {Newport}, {Newton}, {Ng}, {Nguyen}, {Nguyen}, {Nichols}, {Nielsen},
  {Nissanke}, {Nitz}, {Noack}, {Nocera}, {Nolting}, {North}, {Nuttall},
  {Oberling}, {O'Dea}, {Ogin}, {Oh}, {Oh}, {Ohme}, {Okada}, {Oliver},
  {Oppermann}, {Oram}, {O'Reilly}, {Ormiston}, {Ortega}, {O'Shaughnessy},
  {Ossokine}, {Ottaway}, {Overmier}, {Owen}, {Pace}, {Page}, {Page}, {Pai},
  {Pai}, {Palamos}, {Palashov}, {Palomba}, {Pal-Singh}, {Pan}, {Pan}, {Pang},
  {Pang}, {Pankow}, {Pannarale}, {Pant}, {Paoletti}, {Paoli}, {Papa}, {Parida},
  {Parker}, {Pascucci}, {Pasqualetti}, {Passaquieti}, {Passuello}, {Patil},
  {Patricelli}, {Pearlstone}, {Pedraza}, {Pedurand}, {Pekowsky}, {Pele},
  {Penn}, {Perez}, {Perreca}, {Perri}, {Pfeiffer}, {Phelps}, {Piccinni},
  {Pichot}, {Piergiovanni}, {Pierro}, {Pillant}, {Pinard}, {Pinto}, {Pirello},
  {Pitkin}, {Poe}, {Poggiani}, {Popolizio}, {Porter}, {Post}, {Powell},
  {Prasad}, {Pratt}, {Pratten}, {Predoi}, {Prestegard}, {Prijatelj},
  {Principe}, {Privitera}, {Prix}, {Prodi}, {Prokhorov}, {Puncken}, {Punturo},
  {Puppo}, {P{\"u}rrer}, {Qi}, {Quetschke}, {Quintero}, {Quitzow-James},
  {Raab}, {Rabeling}, {Radkins}, {Raffai}, {Raja}, {Rajan}, {Rajbhandari},
  {Rakhmanov}, {Ramirez}, {Ramos-Buades}, {Rapagnani}, {Raymond}, {Razzano},
  {Read}, {Regimbau}, {Rei}, {Reid}, {Reitze}, {Ren}, {Reyes}, {Ricci},
  {Ricker}, {Rieger}, {Riles}, {Rizzo}, {Robertson}, {Robie}, {Robinet},
  {Rocchi}, {Rolland}, {Rollins}, {Roma}, {Romano}, {Romano}, {Romel}, {Romie},
  {Rosi{\'n}ska}, {Ross}, {Rowan}, {R{\"u}diger}, {Ruggi}, {Rutins}, {Ryan},
  {Sachdev}, {Sadecki}, {Sadeghian}, {Sakellariadou}, {Salconi}, {Saleem},
  {Salemi}, {Samajdar}, {Sammut}, {Sampson}, {Sanchez}, {Sanchez},
  {Sanchis-Gual}, {Sandberg}, {Sanders}, {Sassolas}, {Sathyaprakash},
  {Saulson}, {Sauter}, {Savage}, {Sawadsky}, {Schale}, {Scheel}, {Scheuer},
  {Schmidt}, {Schmidt}, {Schnabel}, {Schofield}, {Sch{\"o}nbeck}, {Schreiber},
  {Schuette}, {Schulte}, {Schutz}, {Schwalbe}, {Scott}, {Scott}, {Seidel},
  {Sellers}, {Sengupta}, {Sentenac}, {Sequino}, {Sergeev}, {Shaddock},
  {Shaffer}, {Shah}, {Shahriar}, {Shaner}, {Shao}, {Shapiro}, {Shawhan},
  {Sheperd}, {Shoemaker}, {Shoemaker}, {Siellez}, {Siemens}, {Sieniawska},
  {Sigg}, {Silva}, {Singer}, {Singh}, {Singhal}, {Sintes}, {Slagmolen},
  {Smith}, {Smith}, {Smith}, {Somala}, {Son}, {Sonnenberg}, {Sorazu},
  {Sorrentino}, {Souradeep}, {Spencer}, {Srivastava}, {Staats}, {Staley},
  {Steinke}, {Steinlechner}, {Steinlechner}, {Steinmeyer}, {Stevenson},
  {Stone}, {Stops}, {Strain}, {Stratta}, {Strigin}, {Strunk}, {Sturani},
  {Stuver}, {Summerscales}, {Sun}, {Sunil}, {Suresh}, {Sutton}, {Swinkels},
  {Szczepa{\'n}czyk}, {Tacca}, {Tait}, {Talbot}, {Talukder}, {Tanner},
  {T{\'a}pai}, {Taracchini}, {Tasson}, {Taylor}, {Taylor}, {Tewari}, {Theeg},
  {Thies}, {Thomas}, {Thomas}, {Thomas}, {Thorne}, {Thorne}, {Thrane},
  {Tiwari}, {Tiwari}, {Tokmakov}, {Toland}, {Tonelli}, {Tornasi},
  {Torres-Forn{\'e}}, {Torrie}, {T{\"o}yr{\"a}}, {Travasso}, {Traylor},
  {Trinastic}, {Tringali}, {Trozzo}, {Tsang}, {Tse}, {Tso}, {Tsukada}, {Tsuna},
  {Tuyenbayev}, {Ueno}, {Ugolini}, {Unnikrishnan}, {Urban}, {Usman},
  {Vahlbruch}, {Vajente}, {Valdes}, {Vallisneri}, {van Bakel}, {van Beuzekom},
  {van den Brand}, {Van Den Broeck}, {Vander-Hyde}, {van der Schaaf}, {van
  Heijningen}, {van Veggel}, {Vardaro}, {Varma}, {Vass}, {Vas{\'u}th},
  {Vecchio}, {Vedovato}, {Veitch}, {Veitch}, {Venkateswara}, {Venugopalan},
  {Verkindt}, {Vetrano}, {Vicer{\'e}}, {Viets}, {Vinciguerra}, {Vine}, {Vinet},
  {Vitale}, {Vo}, {Vocca}, {Vorvick}, {Vyatchanin}, {Wade}, {Wade}, {Wade},
  {Walet}, {Walker}, {Wallace}, {Walsh}, {Wang}, {Wang}, {Wang}, {Wang},
  {Wang}, {Ward}, {Warner}, {Was}, {Watchi}, {Weaver}, {Wei}, {Weinert},
  {Weinstein}, {Weiss}, {Wen}, {Wessel}, {We{\ss}els}, {Westerweck},
  {Westphal}, {Wette}, {Whelan}, {Whitcomb}, {Whiting}, {Whittle}, {Wilken},
  {Williams}, {Williams}, {Williamson}, {Willis}, {Willke}, {Wimmer},
  {Winkler}, {Wipf}, {Wittel}, {Woan}, {Woehler}, {Wofford}, {Wong}, {Worden},
  {Wright}, {Wu}, {Wysocki}, {Xiao}, {Yamamoto}, {Yancey}, {Yang}, {Yap},
  {Yazback}, {Yu}, {Yu}, {Yvert}, {Zadro{\.z}ny}, {Zanolin}, {Zelenova},
  {Zendri}, {Zevin}, {Zhang}, {Zhang}, {Zhang}, {Zhang}, {Zhao}, {Zhou},
  {Zhou}, {Zhu}, {Zhu}, {Zimmerman}, {Zucker}, {Zweizig}, {LIGO Scientific
  Collaboration}, \& {Virgo Collaboration}}]{2017PhRvL.119p1101A}
{Abbott}, B.~P., {Abbott}, R., {Abbott}, T.~D., {et~al.} 2017{\natexlab{b}},
  \prl, 119, 161101

\bibitem[{{Abbott} {et~al.}(2017{\natexlab{c}}){Abbott}, {Abbott}, {Abbott},
  {Acernese}, {Ackley}, {Adams}, {Adams}, {Addesso}, {Adhikari}, {Adya},
  {Affeldt}, {Afrough}, {Agarwal}, {Agathos}, {Agatsuma}, {Aggarwal}, {Aguiar},
  {Aiello}, {Ain}, {Ajith}, {Allen}, {Allen}, {Allocca}, {Altin}, {Amato},
  {Ananyeva}, {Anderson}, {Anderson}, {Angelova}, {Antier}, {Appert}, {Arai},
  {Araya}, {Areeda}, {Arnaud}, {Arun}, {Ascenzi}, {Ashton}, {Ast}, {Aston},
  {Astone}, {Atallah}, {Aufmuth}, {Aulbert}, {AultONeal}, {Austin},
  {Avila-Alvarez}, {Babak}, {Bacon}, {Bader}, {Bae}, {Baker}, {Baldaccini},
  {Ballardin}, {Ballmer}, {Banagiri}, {Barayoga}, {Barclay}, {Barish},
  {Barker}, {Barkett}, {Barone}, {Barr}, {Barsotti}, {Barsuglia}, {Barta},
  {Barthelmy}, {Bartlett}, {Bartos}, {Bassiri}, {Basti}, {Batch}, {Bawaj},
  {Bayley}, {Bazzan}, {B{\'e}csy}, {Beer}, {Bejger}, {Belahcene}, {Bell},
  {Berger}, {Bergmann}, {Bero}, {Berry}, {Bersanetti}, {Bertolini},
  {Betzwieser}, {Bhagwat}, {Bhandare}, {Bilenko}, {Billingsley}, {Billman},
  {Birch}, {Birney}, {Birnholtz}, {Biscans}, {Biscoveanu}, {Bisht}, {Bitossi},
  {Biwer}, {Bizouard}, {Blackburn}, {Blackman}, {Blair}, {Blair}, {Blair},
  {Bloemen}, {Bock}, {Bode}, {Boer}, {Bogaert}, {Bohe}, {Bondu}, {Bonilla},
  {Bonnand}, {Boom}, {Bork}, {Boschi}, {Bose}, {Bossie}, {Bouffanais}, {Bozzi},
  {Bradaschia}, {Brady}, {Branchesi}, {Brau}, {Briant}, {Brillet}, {Brinkmann},
  {Brisson}, {Brockill}, {Broida}, {Brooks}, {Brown}, {Brown}, {Brunett},
  {Buchanan}, {Buikema}, {Bulik}, {Bulten}, {Buonanno}, {Buskulic}, {Buy},
  {Byer}, {Cabero}, {Cadonati}, {Cagnoli}, {Cahillane}, {Calder{\'o}n
  Bustillo}, {Callister}, {Calloni}, {Camp}, {Canepa}, {Canizares}, {Cannon},
  {Cao}, {Cao}, {Capano}, {Capocasa}, {Carbognani}, {Caride}, {Carney},
  {Casanueva Diaz}, {Casentini}, {Caudill}, {Cavagli{\`a}}, {Cavalier},
  {Cavalieri}, {Cella}, {Cepeda}, {Cerd{\'a}-Dur{\'a}n}, {Cerretani},
  {Cesarini}, {Chamberlin}, {Chan}, {Chao}, {Charlton}, {Chase},
  {Chassande-Mottin}, {Chatterjee}, {Chatziioannou}, {Cheeseboro}, {Chen},
  {Chen}, {Chen}, {Cheng}, {Chia}, {Chincarini}, {Chiummo}, {Chmiel}, {Cho},
  {Cho}, {Chow}, {Christensen}, {Chu}, {Chua}, {Chua}, {Chung}, {Chung},
  {Ciani}, {Ciolfi}, {Cirelli}, {Cirone}, {Clara}, {Clark}, {Clearwater},
  {Cleva}, {Cocchieri}, {Coccia}, {Cohadon}, {Cohen}, {Colla}, {Collette},
  {Cominsky}, {Constancio}, {Conti}, {Cooper}, {Corban}, {Corbitt},
  {Cordero-Carri{\'o}n}, {Corley}, {Cornish}, {Corsi}, {Cortese}, {Costa},
  {Coughlin}, {Coughlin}, {Coulon}, {Countryman}, {Couvares}, {Covas}, {Cowan},
  {Coward}, {Cowart}, {Coyne}, {Coyne}, {Creighton}, {Creighton}, {Cripe},
  {Crowder}, {Cullen}, {Cumming}, {Cunningham}, {Cuoco}, {Dal Canton},
  {D{\'a}lya}, {Danilishin}, {D'Antonio}, {Danzmann}, {Dasgupta}, {Da Silva
  Costa}, {Dattilo}, {Dave}, {Davier}, {Davis}, {Daw}, {Day}, {De}, {DeBra},
  {Degallaix}, {De Laurentis}, {Del{\'e}glise}, {Del Pozzo}, {Demos}, {Denker},
  {Dent}, {De Pietri}, {Dergachev}, {De Rosa}, {DeRosa}, {De Rossi}, {DeSalvo},
  {de Varona}, {Devenson}, {Dhurandhar}, {D{\'\i}az}, {Di Fiore}, {Di
  Giovanni}, {Di Girolamo}, {Di Lieto}, {Di Pace}, {Di Palma}, {Di Renzo},
  {Doctor}, {Dolique}, {Donovan}, {Dooley}, {Doravari}, {Dorrington},
  {Douglas}, {Dovale {\'A}lvarez}, {Downes}, {Drago}, {Dreissigacker},
  {Driggers}, {Du}, {Ducrot}, {Dupej}, {Dwyer}, {Edo}, {Edwards}, {Effler},
  {Ehrens}, {Eichholz}, {Eikenberry}, {Eisenstein}, {Essick}, {Estevez},
  {Etienne}, {Etzel}, {Evans}, {Evans}, {Factourovich}, {Fafone}, {Fair},
  {Fairhurst}, {Fan}, {Farinon}, {Farr}, {Farr}, {Fauchon-Jones}, {Favata},
  {Fays}, {Fee}, {Fehrmann}, {Feicht}, {Fejer}, {Fernandez-Galiana},
  {Ferrante}, {Ferreira}, {Ferrini}, {Fidecaro}, {Finstad}, {Fiori},
  {Fiorucci}, {Fishbach}, {Fisher}, {Fitz-Axen}, {Flaminio}, {Fletcher},
  {Fong}, {Font}, {Forsyth}, {Forsyth}, {Fournier}, {Frasca}, {Frasconi},
  {Frei}, {Freise}, {Frey}, {Frey}, {Fries}, {Fritschel}, {Frolov}, {Fulda},
  {Fyffe}, {Gabbard}, {Gadre}, {Gaebel}, {Gair}, {Gammaitoni}, {Ganija},
  {Gaonkar}, {Garcia-Quiros}, {Garufi}, {Gateley}, {Gaudio}, {Gaur},
  {Gayathri}, {Gehrels}, {Gemme}, {Genin}, {Gennai}, {George}, {George},
  {Gergely}, {Germain}, {Ghonge}, {Ghosh}, {Ghosh}, {Ghosh}, {Giaime},
  {Giardina}, {Giazotto}, {Gill}, {Glover}, {Goetz}, {Goetz}, {Gomes},
  {Goncharov}, {Gonz{\'a}lez}, {Gonzalez Castro}, {Gopakumar}, {Gorodetsky},
  {Gossan}, {Gosselin}, {Gouaty}, {Grado}, {Graef}, {Granata}, {Grant}, {Gras},
  {Gray}, {Greco}, {Green}, {Gretarsson}, {Griswold}, {Groot}, {Grote},
  {Grunewald}, {Gruning}, {Guidi}, {Guo}, {Gupta}, {Gupta}, {Gushwa},
  {Gustafson}, {Gustafson}, {Halim}, {Hall}, {Hall}, {Hamilton}, {Hammond},
  {Haney}, {Hanke}, {Hanks}, {Hanna}, {Hannam}, {Hannuksela}, {Hanson},
  {Hardwick}, {Harms}, {Harry}, {Harry}, {Hart}, {Haster}, {Haughian}, {Healy},
  {Heidmann}, {Heintze}, {Heitmann}, {Hello}, {Hemming}, {Hendry}, {Heng},
  {Hennig}, {Heptonstall}, {Heurs}, {Hild}, {Hinderer}, {Hoak}, {Hofman},
  {Holt}, {Holz}, {Hopkins}, {Horst}, {Hough}, {Houston}, {Howell}, {Hreibi},
  {Hu}, {Huerta}, {Huet}, {Hughey}, {Husa}, {Huttner}, {Huynh-Dinh}, {Indik},
  {Inta}, {Intini}, {Isa}, {Isac}, {Isi}, {Iyer}, {Izumi}, {Jacqmin}, {Jani},
  {Jaranowski}, {Jawahar}, {Jim{\'e}nez-Forteza}, {Johnson}, {Jones}, {Jones},
  {Jonker}, {Ju}, {Junker}, {Kalaghatgi}, {Kalogera}, {Kamai}, {Kandhasamy},
  {Kang}, {Kanner}, {Kapadia}, {Karki}, {Karvinen}, {Kasprzack}, {Katolik},
  {Katsavounidis}, {Katzman}, {Kaufer}, {Kawabe}, {K{\'e}f{\'e}lian}, {Keitel},
  {Kemball}, {Kennedy}, {Kent}, {Key}, {Khalili}, {Khan}, {Khan}, {Khan},
  {Khazanov}, {Kijbunchoo}, {Kim}, {Kim}, {Kim}, {Kim}, {Kim}, {Kim},
  {Kimbrell}, {King}, {King}, {Kinley-Hanlon}, {Kirchhoff}, {Kissel},
  {Kleybolte}, {Klimenko}, {Knowles}, {Koch}, {Koehlenbeck}, {Koley},
  {Kondrashov}, {Kontos}, {Korobko}, {Korth}, {Kowalska}, {Kozak},
  {Kr{\"a}mer}, {Kringel}, {Krishnan}, {Kr{\'o}lak}, {Kuehn}, {Kumar}, {Kumar},
  {Kumar}, {Kuo}, {Kutynia}, {Kwang}, {Lackey}, {Lai}, {Landry}, {Lang},
  {Lange}, {Lantz}, {Lanza}, {Larson}, {Lartaux-Vollard}, {Lasky}, {Laxen},
  {Lazzarini}, {Lazzaro}, {Leaci}, {Leavey}, {Lee}, {Lee}, {Lee}, {Lee}, {Lee},
  {Lehmann}, {Lenon}, {Leonardi}, {Leroy}, {Letendre}, {Levin}, {Li}, {Linker},
  {Littenberg}, {Liu}, {Lo}, {Lockerbie}, {London}, {Lord}, {Lorenzini},
  {Loriette}, {Lormand}, {Losurdo}, {Lough}, {Lousto}, {Lovelace}, {L{\"u}ck},
  {Lumaca}, {Lundgren}, {Lynch}, {Ma}, {Macas}, {Macfoy}, {Machenschalk},
  {MacInnis}, {Macleod}, {Maga{\~n}a Hernandez}, {Maga{\~n}a-Sandoval},
  {Maga{\~n}a Zertuche}, {Magee}, {Majorana}, {Maksimovic}, {Man}, {Mandic},
  {Mangano}, {Mansell}, {Manske}, {Mantovani}, {Marchesoni}, {Marion},
  {M{\'a}rka}, {M{\'a}rka}, {Markakis}, {Markosyan}, {Markowitz}, {Maros},
  {Marquina}, {Marsh}, {Martelli}, {Martellini}, {Martin}, {Martin},
  {Martynov}, {Mason}, {Massera}, {Masserot}, {Massinger}, {Masso-Reid},
  {Mastrogiovanni}, {Matas}, {Matichard}, {Matone}, {Mavalvala}, {Mazumder},
  {McCarthy}, {McClelland}, {McCormick}, {McCuller}, {McGuire}, {McIntyre},
  {McIver}, {McManus}, {McNeill}, {McRae}, {McWilliams}, {Meacher}, {Meadors},
  {Mehmet}, {Meidam}, {Mejuto-Villa}, {Melatos}, {Mendell}, {Mercer}, {Merilh},
  {Merzougui}, {Meshkov}, {Messenger}, {Messick}, {Metzdorff}, {Meyers},
  {Miao}, {Michel}, {Middleton}, {Mikhailov}, {Milano}, {Miller}, {Miller},
  {Miller}, {Millhouse}, {Milovich-Goff}, {Minazzoli}, {Minenkov}, {Ming},
  {Mishra}, {Mitra}, {Mitrofanov}, {Mitselmakher}, {Mittleman}, {Moffa},
  {Moggi}, {Mogushi}, {Mohan}, {Mohapatra}, {Montani}, {Moore}, {Moraru},
  {Moreno}, {Morriss}, {Mours}, {Mow-Lowry}, {Mueller}, {Muir}, {Mukherjee},
  {Mukherjee}, {Mukherjee}, {Mukund}, {Mullavey}, {Munch}, {Mu{\~n}iz},
  {Muratore}, {Murray}, {Napier}, {Nardecchia}, {Naticchioni}, {Nayak},
  {Neilson}, {Nelemans}, {Nelson}, {Nery}, {Neunzert}, {Nevin}, {Newport},
  {Newton}, {Ng}, {Nguyen}, {Nguyen}, {Nichols}, {Nielsen}, {Nissanke}, {Nitz},
  {Noack}, {Nocera}, {Nolting}, {North}, {Nuttall}, {Oberling}, {O'Dea},
  {Ogin}, {Oh}, {Oh}, {Ohme}, {Okada}, {Oliver}, {Oppermann}, {Oram},
  {O'Reilly}, {Ormiston}, {Ortega}, {O'Shaughnessy}, {Ossokine}, {Ottaway},
  {Overmier}, {Owen}, {Pace}, {Page}, {Page}, {Pai}, {Pai}, {Palamos},
  {Palashov}, {Palomba}, {Pal-Singh}, {Pan}, {Pan}, {Pang}, {Pang}, {Pankow},
  {Pannarale}, {Pant}, {Paoletti}, {Paoli}, {Papa}, {Parida}, {Parker},
  {Pascucci}, {Pasqualetti}, {Passaquieti}, {Passuello}, {Patil}, {Patricelli},
  {Pearlstone}, {Pedraza}, {Pedurand}, {Pekowsky}, {Pele}, {Penn}, {Perez},
  {Perreca}, {Perri}, {Pfeiffer}, {Phelps}, {Piccinni}, {Pichot},
  {Piergiovanni}, {Pierro}, {Pillant}, {Pinard}, {Pinto}, {Pirello}, {Pitkin},
  {Poe}, {Poggiani}, {Popolizio}, {Porter}, {Post}, {Powell}, {Prasad},
  {Pratt}, {Pratten}, {Predoi}, {Prestegard}, {Price}, {Prijatelj}, {Principe},
  {Privitera}, {Prodi}, {Prokhorov}, {Puncken}, {Punturo}, {Puppo},
  {P{\"u}rrer}, {Qi}, {Quetschke}, {Quintero}, {Quitzow-James}, {Raab},
  {Rabeling}, {Radkins}, {Raffai}, {Raja}, {Rajan}, {Rajbhandari}, {Rakhmanov},
  {Ramirez}, {Ramos-Buades}, {Rapagnani}, {Raymond}, {Razzano}, {Read},
  {Regimbau}, {Rei}, {Reid}, {Reitze}, {Ren}, {Reyes}, {Ricci}, {Ricker},
  {Rieger}, {Riles}, {Rizzo}, {Robertson}, {Robie}, {Robinet}, {Rocchi},
  {Rolland}, {Rollins}, {Roma}, {Romano}, {Romel}, {Romie}, {Rosi{\'n}ska},
  {Ross}, {Rowan}, {R{\"u}diger}, {Ruggi}, {Rutins}, {Ryan}, {Sachdev},
  {Sadecki}, {Sadeghian}, {Sakellariadou}, {Salconi}, {Saleem}, {Salemi},
  {Samajdar}, {Sammut}, {Sampson}, {Sanchez}, {Sanchez}, {Sanchis-Gual},
  {Sandberg}, {Sanders}, {Sassolas}, {Sathyaprakash}, {Saulson}, {Sauter},
  {Savage}, {Sawadsky}, {Schale}, {Scheel}, {Scheuer}, {Schmidt}, {Schmidt},
  {Schnabel}, {Schofield}, {Sch{\"o}nbeck}, {Schreiber}, {Schuette}, {Schulte},
  {Schutz}, {Schwalbe}, {Scott}, {Scott}, {Seidel}, {Sellers}, {Sengupta},
  {Sentenac}, {Sequino}, {Sergeev}, {Shaddock}, {Shaffer}, {Shah}, {Shahriar},
  {Shaner}, {Shao}, {Shapiro}, {Shawhan}, {Sheperd}, {Shoemaker}, {Shoemaker},
  {Siellez}, {Siemens}, {Sieniawska}, {Sigg}, {Silva}, {Singer}, {Singh},
  {Singhal}, {Sintes}, {Slagmolen}, {Smith}, {Smith}, {Smith}, {Somala}, {Son},
  {Sonnenberg}, {Sorazu}, {Sorrentino}, {Souradeep}, {Spencer}, {Srivastava},
  {Staats}, {Staley}, {Steinke}, {Steinlechner}, {Steinlechner}, {Steinmeyer},
  {Stevenson}, {Stone}, {Stops}, {Strain}, {Stratta}, {Strigin}, {Strunk},
  {Sturani}, {Stuver}, {Summerscales}, {Sun}, {Sunil}, {Suresh}, {Sutton},
  {Swinkels}, {Szczepa{\'n}czyk}, {Tacca}, {Tait}, {Talbot}, {Talukder},
  {Tanner}, {T{\'a}pai}, {Taracchini}, {Tasson}, {Taylor}, {Taylor}, {Tewari},
  {Theeg}, {Thies}, {Thomas}, {Thomas}, {Thomas}, {Thorne}, {Thorne}, {Thrane},
  {Tiwari}, {Tiwari}, {Tokmakov}, {Toland}, {Tonelli}, {Tornasi},
  {Torres-Forn{\'e}}, {Torrie}, {T{\"o}yr{\"a}}, {Travasso}, {Traylor},
  {Trinastic}, {Tringali}, {Trozzo}, {Tsang}, {Tse}, {Tso}, {Tsukada}, {Tsuna},
  {Tuyenbayev}, {Ueno}, {Ugolini}, {Unnikrishnan}, {Urban}, {Usman},
  {Vahlbruch}, {Vajente}, {Valdes}, {van Bakel}, {van Beuzekom}, {van den
  Brand}, {Van Den Broeck}, {Vander-Hyde}, {van der Schaaf}, {van Heijningen},
  {van Veggel}, {Vardaro}, {Varma}, {Vass}, {Vas{\'u}th}, {Vecchio},
  {Vedovato}, {Veitch}, {Veitch}, {Venkateswara}, {Venugopalan}, {Verkindt},
  {Vetrano}, {Vicer{\'e}}, {Viets}, {Vinciguerra}, {Vine}, {Vinet}, {Vitale},
  {Vo}, {Vocca}, {Vorvick}, {Vyatchanin}, {Wade}, {Wade}, {Wade}, {Walet},
  {Walker}, {Wallace}, {Walsh}, {Wang}, {Wang}, {Wang}, {Wang}, {Wang}, {Ward},
  {Warner}, {Was}, {Watchi}, {Weaver}, {Wei}, {Weinert}, {Weinstein}, {Weiss},
  {Wen}, {Wessel}, {Wessels}, {Westerweck}, {Westphal}, {Wette}, {Whelan},
  {Whitcomb}, {Whiting}, {Whittle}, {Wilken}, {Williams}, {Williams},
  {Williamson}, {Willis}, {Willke}, {Wimmer}, {Winkler}, {Wipf}, {Wittel},
  {Woan}, {Woehler}, {Wofford}, {Wong}, {Worden}, {Wright}, {Wu}, {Wysocki},
  {Xiao}, {Yamamoto}, {Yancey}, {Yang}, {Yap}, {Yazback}, {Yu}, {Yu}, {Yvert},
  {Zadro{\.z}ny}, {Zanolin}, {Zelenova}, {Zendri}, {Zevin}, {Zhang}, {Zhang},
  {Zhang}, {Zhang}, {Zhao}, {Zhou}, {Zhou}, {Zhu}, {Zhu}, {Zimmerman},
  {Zucker}, {Zweizig}, {LIGO Scientific Collaboration}, {Virgo Collaboration},
  {Wilson-Hodge}, {Bissaldi}, {Blackburn}, {Briggs}, {Burns}, {Cleveland},
  {Connaughton}, {Gibby}, {Giles}, {Goldstein}, {Hamburg}, {Jenke}, {Hui},
  {Kippen}, {Kocevski}, {McBreen}, {Meegan}, {Paciesas}, {Poolakkil}, {Preece},
  {Racusin}, {Roberts}, {Stanbro}, {Veres}, {von Kienlin}, {GBM}, {Savchenko},
  {Ferrigno}, {Kuulkers}, {Bazzano}, {Bozzo}, {Brandt}, {Chenevez},
  {Courvoisier}, {Diehl}, {Domingo}, {Hanlon}, {Jourdain}, {Laurent}, {Lebrun},
  {Lutovinov}, {Martin-Carrillo}, {Mereghetti}, {Natalucci}, {Rodi}, {Roques},
  {Sunyaev}, {Ubertini}, {INTEGRAL}, {Aartsen}, {Ackermann}, {Adams},
  {Aguilar}, {Ahlers}, {Ahrens}, {Samarai}, {Altmann}, {Andeen}, {Anderson},
  {Ansseau}, {Anton}, {Arg{\"u}elles}, {Auffenberg}, {Axani}, {Bagherpour},
  {Bai}, {Barron}, {Barwick}, {Baum}, {Bay}, {Beatty}, {Becker Tjus},
  {Bernardini}, {Besson}, {Binder}, {Bindig}, {Blaufuss}, {Blot}, {Bohm},
  {B{\"o}rner}, {Bos}, {Bose}, {B{\"o}ser}, {Botner}, {Bourbeau}, {Bourbeau},
  {Bradascio}, {Braun}, {Brayeur}, {Brenzke}, {Bretz}, {Bron},
  {Brostean-Kaiser}, {Burgman}, {Carver}, {Casey}, {Casier}, {Cheung},
  {Chirkin}, {Christov}, {Clark}, {Classen}, {Coenders}, {Collin}, {Conrad},
  {Cowen}, {Cross}, {Day}, {de Andr{\'e}}, {De Clercq}, {DeLaunay},
  {Dembinski}, {De Ridder}, {Desiati}, {de Vries}, {de Wasseige}, {de With},
  {DeYoung}, {D{\'\i}az-V{\'e}lez}, {di Lorenzo}, {Dujmovic}, {Dumm},
  {Dunkman}, {Dvorak}, {Eberhardt}, {Ehrhardt}, {Eichmann}, {Eller}, {Evenson},
  {Fahey}, {Fazely}, {Felde}, {Filimonov}, {Finley}, {Flis}, {Franckowiak},
  {Friedman}, {Fuchs}, {Gaisser}, {Gallagher}, {Gerhardt}, {Ghorbani}, {Giang},
  {Glauch}, {Gl{\"u}senkamp}, {Goldschmidt}, {Gonzalez}, {Grant}, {Griffith},
  {Haack}, {Hallgren}, {Halzen}, {Hanson}, {Hebecker}, {Heereman}, {Helbing},
  {Hellauer}, {Hickford}, {Hignight}, {Hill}, {Hoffman}, {Hoffmann},
  {Hokanson-Fasig}, {Hoshina}, {Huang}, {Huber}, {Hultqvist}, {H{\"u}nnefeld},
  {In}, {Ishihara}, {Jacobi}, {Japaridze}, {Jeong}, {Jero}, {Jones},
  {Kalaczynski}, {Kang}, {Kappes}, {Karg}, {Karle}, {Kauer}, {Keivani},
  {Kelley}, {Kheirandish}, {Kim}, {Kim}, {Kintscher}, {Kiryluk}, {Kittler},
  {Klein}, {Kohnen}, {Koirala}, {Kolanoski}, {K{\"o}pke}, {Kopper}, {Kopper},
  {Koschinsky}, {Koskinen}, {Kowalski}, {Krings}, {Kroll}, {Kr{\"u}ckl},
  {Kunnen}, {Kunwar}, {Kurahashi}, {Kuwabara}, {Kyriacou}, {Labare},
  {Lanfranchi}, {Larson}, {Lauber}, {Lesiak-Bzdak}, {Leuermann}, {Liu}, {Lu},
  {L{\"u}nemann}, {Luszczak}, {Madsen}, {Maggi}, {Mahn}, {Mancina}, {Maruyama},
  {Mase}, {Maunu}, {McNally}, {Meagher}, {Medici}, {Meier}, {Menne}, {Merino},
  {Meures}, {Miarecki}, {Micallef}, {Moment{\'e}}, {Montaruli}, {Moore},
  {Moulai}, {Nahnhauer}, {Nakarmi}, {Naumann}, {Neer}, {Niederhausen},
  {Nowicki}, {Nygren}, {Obertacke Pollmann}, {Olivas}, {O'Murchadha},
  {Palczewski}, {Pandya}, {Pankova}, {Peiffer}, {Pepper}, {P{\'e}rez de los
  Heros}, {Pieloth}, {Pinat}, {Price}, {Przybylski}, {Raab}, {R{\"a}del},
  {Rameez}, {Rawlins}, {Rea}, {Reimann}, {Relethford}, {Relich}, {Resconi},
  {Rhode}, {Richman}, {Robertson}, {Rongen}, {Rott}, {Ruhe}, {Ryckbosch},
  {Rysewyk}, {S{\"a}lzer}, {Sanchez Herrera}, {Sandrock}, {Sandroos},
  {Santander}, {Sarkar}, {Sarkar}, {Satalecka}, {Schlunder}, {Schmidt},
  {Schneider}, {Schoenen}, {Sch{\"o}neberg}, {Schumacher}, {Seckel},
  {Seunarine}, {Soedingrekso}, {Soldin}, {Song}, {Spiczak}, {Spiering},
  {Stachurska}, {Stamatikos}, {Stanev}, {Stasik}, {Stettner}, {Steuer},
  {Stezelberger}, {Stokstad}, {St{\"o}ssl}, {Strotjohann}, {Stuttard},
  {Sullivan}, {Sutherland}, {Taboada}, {Tatar}, {Tenholt}, {Ter-Antonyan},
  {Terliuk}, {Te{\v{s}}i{\'c}}, {Tilav}, {Toale}, {Tobin}, {Toscano}, {Tosi},
  {Tselengidou}, {Tung}, {Turcati}, {Turley}, {Ty}, {Unger}, {Usner},
  {Vandenbroucke}, {Van Driessche}, {van Eijndhoven}, {Vanheule}, {van Santen},
  {Vehring}, {Vogel}, {Vraeghe}, {Walck}, {Wallace}, {Wallraff}, {Wandler},
  {Wandkowsky}, {Waza}, {Weaver}, {Weiss}, {Wendt}, {Werthebach}, {Whelan},
  {Wiebe}, {Wiebusch}, {Wille}, {Williams}, {Wills}, {Wolf}, {Wood}, {Woolsey},
  {Woschnagg}, {Xu}, {Xu}, {Xu}, {Yanez}, {Yodh}, {Yoshida}, {Yuan}, {Zoll},
  {IceCube Collaboration}, {Balasubramanian}, {Mate}, {Bhalerao},
  {Bhattacharya}, {Vibhute}, {Dewangan}, {Rao}, {Vadawale}, {AstroSat Cadmium
  Zinc Telluride Imager Team}, {Svinkin}, {Hurley}, {Aptekar}, {Frederiks},
  {Golenetskii}, {Kozlova}, {Lysenko}, {Oleynik}, {Tsvetkova}, {Ulanov},
  {Cline}, {IPN Collaboration}, {Li}, {Xiong}, {Zhang}, {Lu}, {Song}, {Cao},
  {Chang}, {Chen}, {Chen}, {Chen}, {Chen}, {Chen}, {Chen}, {Cui}, {Cui},
  {Deng}, {Dong}, {Du}, {Fu}, {Gao}, {Gao}, {Gao}, {Ge}, {Gu}, {Guan}, {Guo},
  {Han}, {Hu}, {Huang}, {Huo}, {Jia}, {Jiang}, {Jiang}, {Jin}, {Jin}, {Li},
  {Li}, {Li}, {Li}, {Li}, {Li}, {Li}, {Li}, {Li}, {Li}, {Li}, {Liang}, {Liao},
  {Liu}, {Liu}, {Liu}, {Liu}, {Liu}, {Liu}, {Liu}, {Lu}, {Lu}, {Luo}, {Ma},
  {Meng}, {Nang}, {Nie}, {Ou}, {Qu}, {Sai}, {Sun}, {Tan}, {Tao}, {Tao}, {Tuo},
  {Wang}, {Wang}, {Wang}, {Wang}, {Wang}, {Wen}, {Wu}, {Wu}, {Xiao}, {Xu},
  {Xu}, {Yan}, {Yang}, {Yang}, {Yang}, {Zhang}, {Zhang}, {Zhang}, {Zhang},
  {Zhang}, {Zhang}, {Zhang}, {Zhang}, {Zhang}, {Zhang}, {Zhang}, {Zhang},
  {Zhang}, {Zhang}, {Zhang}, {Zhang}, {Zhang}, {Zhang}, {Zhao}, {Zhao}, {Zhao},
  {Zheng}, {Zhu}, {Zhu}, {Zou}, {Insight-HXMT Collaboration}, {Albert},
  {Andr{\'e}}, {Anghinolfi}, {Ardid}, {Aubert}, {Aublin}, {Avgitas}, {Baret},
  {Barrios-Mart{\'\i}}, {Basa}, {Belhorma}, {Bertin}, {Biagi}, {Bormuth},
  {Bourret}, {Bouwhuis}, {Br{\^a}nza{\c{s}}}, {Bruijn}, {Brunner}, {Busto},
  {Capone}, {Caramete}, {Carr}, {Celli}, {Cherkaoui El Moursli}, {Chiarusi},
  {Circella}, {Coelho}, {Coleiro}, {Coniglione}, {Costantini}, {Coyle},
  {Creusot}, {D{\'\i}az}, {Deschamps}, {De Bonis}, {Distefano}, {Di Palma},
  {Domi}, {Donzaud}, {Dornic}, {Drouhin}, {Eberl}, {El Bojaddaini}, {El
  Khayati}, {Els{\"a}sser}, {Enzenh{\"o}fer}, {Ettahiri}, {Fassi}, {Felis},
  {Fusco}, {Gay}, {Giordano}, {Glotin}, {Gr{\'e}goire}, {Ruiz}, {Graf},
  {Hallmann}, {van Haren}, {Heijboer}, {Hello}, {Hern{\'a}ndez-Rey},
  {H{\"o}ssl}, {Hofest{\"a}dt}, {Hugon}, {Illuminati}, {James}, {de Jong},
  {Jongen}, {Kadler}, {Kalekin}, {Katz}, {Kiessling}, {Kouchner}, {Kreter},
  {Kreykenbohm}, {Kulikovskiy}, {Lachaud}, {Lahmann}, {Lef{\`e}vre}, {Leonora},
  {Lotze}, {Loucatos}, {Marcelin}, {Margiotta}, {Marinelli},
  {Mart{\'\i}nez-Mora}, {Mele}, {Melis}, {Michael}, {Migliozzi}, {Moussa},
  {Navas}, {Nezri}, {Organokov}, {P{\u{a}}v{\u{a}}la{\c{s}}}, {Pellegrino},
  {Perrina}, {Piattelli}, {Popa}, {Pradier}, {Quinn}, {Racca}, {Riccobene},
  {S{\'a}nchez-Losa}, {Salda{\~n}a}, {Salvadori}, {Samtleben}, {Sanguineti},
  {Sapienza}, {Sieger}, {Spurio}, {Stolarczyk}, {Taiuti}, {Tayalati},
  {Trovato}, {Turpin}, {T{\"o}nnis}, {Vallage}, {Van Elewyck}, {Versari},
  {Vivolo}, {Vizzoca}, {Wilms}, {Zornoza}, {Z{\'u}{\~n}iga}, {ANTARES
  Collaboration}, {Beardmore}, {Breeveld}, {Burrows}, {Cenko}, {Cusumano},
  {D'A{\`\i}}, {de Pasquale}, {Emery}, {Evans}, {Giommi}, {Gronwall}, {Kennea},
  {Krimm}, {Kuin}, {Lien}, {Marshall}, {Melandri}, {Nousek}, {Oates},
  {Osborne}, {Pagani}, {Page}, {Palmer}, {Perri}, {Siegel}, {Sbarufatti},
  {Tagliaferri}, {Tohuvavohu}, {Swift Collaboration}, {Tavani}, {Verrecchia},
  {Bulgarelli}, {Evangelista}, {Pacciani}, {Feroci}, {Pittori}, {Giuliani},
  {Del Monte}, {Donnarumma}, {Argan}, {Trois}, {Ursi}, {Cardillo}, {Piano},
  {Longo}, {Lucarelli}, {Munar-Adrover}, {Fuschino}, {Labanti}, {Marisaldi},
  {Minervini}, {Fioretti}, {Parmiggiani}, {Gianotti}, {Trifoglio}, {Di Persio},
  {Antonelli}, {Barbiellini}, {Caraveo}, {Cattaneo}, {Costa}, {Colafrancesco},
  {D'Amico}, {Ferrari}, {Morselli}, {Paoletti}, {Picozza}, {Pilia}, {Rappoldi},
  {Soffitta}, {Vercellone}, {AGILE Team}, {Foley}, {Coulter}, {Kilpatrick},
  {Drout}, {Piro}, {Shappee}, {Siebert}, {Simon}, {Ulloa}, {Kasen}, {Madore},
  {Murguia-Berthier}, {Pan}, {Prochaska}, {Ramirez-Ruiz}, {Rest},
  {Rojas-Bravo}, {1M2H Team}, {Berger}, {Soares-Santos}, {Annis}, {Alexander},
  {Allam}, {Balbinot}, {Blanchard}, {Brout}, {Butler}, {Chornock}, {Cook},
  {Cowperthwaite}, {Diehl}, {Drlica-Wagner}, {Drout}, {Durret}, {Eftekhari},
  {Finley}, {Fong}, {Frieman}, {Fryer}, {Garc{\'\i}a-Bellido}, {Gruendl},
  {Hartley}, {Herner}, {Kessler}, {Lin}, {Lopes}, {Louren{\c{c}}o}, {Margutti},
  {Marshall}, {Matheson}, {Medina}, {Metzger}, {Mu{\~n}oz}, {Muir}, {Nicholl},
  {Nugent}, {Palmese}, {Paz-Chinch{\'o}n}, {Quataert}, {Sako}, {Sauseda},
  {Schlegel}, {Scolnic}, {Secco}, {Smith}, {Sobreira}, {Villar}, {Vivas},
  {Wester}, {Williams}, {Yanny}, {Zenteno}, {Zhang}, {Abbott}, {Banerji},
  {Bechtol}, {Benoit-L{\'e}vy}, {Bertin}, {Brooks}, {Buckley-Geer}, {Burke},
  {Capozzi}, {Carnero Rosell}, {Carrasco Kind}, {Castander}, {Crocce}, {Cunha},
  {D'Andrea}, {da Costa}, {Davis}, {DePoy}, {Desai}, {Dietrich}, {Eifler},
  {Fernandez}, {Flaugher}, {Fosalba}, {Gaztanaga}, {Gerdes}, {Giannantonio},
  {Goldstein}, {Gruen}, {Gschwend}, {Gutierrez}, {Honscheid}, {James},
  {Jeltema}, {Johnson}, {Johnson}, {Kent}, {Krause}, {Kron}, {Kuehn}, {Lahav},
  {Lima}, {Maia}, {March}, {Martini}, {McMahon}, {Menanteau}, {Miller},
  {Miquel}, {Mohr}, {Nichol}, {Ogando}, {Plazas}, {Romer}, {Roodman}, {Rykoff},
  {Sanchez}, {Scarpine}, {Schindler}, {Schubnell}, {Sevilla-Noarbe}, {Sheldon},
  {Smith}, {Smith}, {Stebbins}, {Suchyta}, {Swanson}, {Tarle}, {Thomas},
  {Troxel}, {Tucker}, {Vikram}, {Walker}, {Wechsler}, {Weller}, {Carlin},
  {Gill}, {Li}, {Marriner}, {Neilsen}, {Dark Energy Camera GW-EM
  Collaboration}, {DES Collaboration}, {Haislip}, {Kouprianov}, {Reichart},
  {Sand}, {Tartaglia}, {Valenti}, {Yang}, {DLT40 Collaboration}, {Benetti},
  {Brocato}, {Campana}, {Cappellaro}, {Covino}, {D'Avanzo}, {D'Elia}, {Getman},
  {Ghirlanda}, {Ghisellini}, {Limatola}, {Nicastro}, {Palazzi}, {Pian},
  {Piranomonte}, {Possenti}, {Rossi}, {Salafia}, {Tomasella}, {Amati},
  {Antonelli}, {Bernardini}, {Bufano}, {Capaccioli}, {Casella}, {Dadina}, {De
  Cesare}, {Di Paola}, {Giuffrida}, {Giunta}, {Israel}, {Lisi}, {Maiorano},
  {Mapelli}, {Masetti}, {Pescalli}, {Pulone}, {Salvaterra}, {Schipani},
  {Spera}, {Stamerra}, {Stella}, {Testa}, {Turatto}, {Vergani}, {Aresu},
  {Bachetti}, {Buffa}, {Burgay}, {Buttu}, {Caria}, {Carretti}, {Casasola},
  {Castangia}, {Carboni}, {Casu}, {Concu}, {Corongiu}, {Deiana}, {Egron},
  {Fara}, {Gaudiomonte}, {Gusai}, {Ladu}, {Loru}, {Leurini}, {Marongiu},
  {Melis}, {Melis}, {Migoni}, {Milia}, {Navarrini}, {Orlati}, {Ortu}, {Palmas},
  {Pellizzoni}, {Perrodin}, {Pisanu}, {Poppi}, {Righini}, {Saba}, {Serra},
  {Serrau}, {Stagni}, {Surcis}, {Vacca}, {Vargiu}, {Hunt}, {Jin}, {Klose},
  {Kouveliotou}, {Mazzali}, {M{\o}ller}, {Nava}, {Piran}, {Selsing}, {Vergani},
  {Wiersema}, {Toma}, {Higgins}, {Mundell}, {di Serego Alighieri}, {G{\'o}tz},
  {Gao}, {Gomboc}, {Kaper}, {Kobayashi}, {Kopac}, {Mao}, {Starling}, {Steele},
  {van der Horst}, {GRAWITA: GRAvitational Wave Inaf TeAm}, {Acero}, {Atwood},
  {Baldini}, {Barbiellini}, {Bastieri}, {Berenji}, {Bellazzini}, {Bissaldi},
  {Blandford}, {Bloom}, {Bonino}, {Bottacini}, {Bregeon}, {Buehler}, {Buson},
  {Cameron}, {Caputo}, {Caraveo}, {Cavazzuti}, {Chekhtman}, {Cheung}, {Chiang},
  {Ciprini}, {Cohen-Tanugi}, {Cominsky}, {Costantin}, {Cuoco}, {D'Ammando}, {de
  Palma}, {Digel}, {Di Lalla}, {Di Mauro}, {Di Venere}, {Dubois}, {Fegan},
  {Focke}, {Franckowiak}, {Fukazawa}, {Funk}, {Fusco}, {Gargano}, {Gasparrini},
  {Giglietto}, {Giordano}, {Giroletti}, {Glanzman}, {Green}, {Grondin},
  {Guillemot}, {Guiriec}, {Harding}, {Horan}, {J{\'o}hannesson}, {Kamae},
  {Kensei}, {Kuss}, {La Mura}, {Latronico}, {Lemoine-Goumard}, {Longo},
  {Loparco}, {Lovellette}, {Lubrano}, {Magill}, {Maldera}, {Manfreda},
  {Mazziotta}, {McEnery}, {Meyer}, {Michelson}, {Mirabal}, {Monzani},
  {Moretti}, {Morselli}, {Moskalenko}, {Negro}, {Nuss}, {Ojha}, {Omodei},
  {Orienti}, {Orlando}, {Palatiello}, {Paliya}, {Paneque}, {Pesce-Rollins},
  {Piron}, {Porter}, {Principe}, {Rain{\`o}}, {Rando}, {Razzano}, {Razzaque},
  {Reimer}, {Reimer}, {Reposeur}, {Rochester}, {Saz Parkinson}, {Sgr{\`o}},
  {Siskind}, {Spada}, {Spandre}, {Suson}, {Takahashi}, {Tanaka}, {Thayer},
  {Thayer}, {Thompson}, {Tibaldo}, {Torres}, {Torresi}, {Troja}, {Venters},
  {Vianello}, {Zaharijas}, {Fermi Large Area Telescope Collaboration},
  {Allison}, {Bannister}, {Dobie}, {Kaplan}, {Lenc}, {Lynch}, {Murphy},
  {Sadler}, {Australia Telescope Compact Array}, {Hotan}, {James}, {Oslowski},
  {Raja}, {Shannon}, {Whiting}, {Australian SKA Pathfinder}, {Arcavi},
  {Howell}, {McCully}, {Hosseinzadeh}, {Hiramatsu}, {Poznanski}, {Barnes},
  {Zaltzman}, {Vasylyev}, {Maoz}, {Las Cumbres Observatory Group}, {Cooke},
  {Bailes}, {Wolf}, {Deller}, {Lidman}, {Wang}, {Gendre}, {Andreoni}, {Ackley},
  {Pritchard}, {Bessell}, {Chang}, {M{\"o}ller}, {Onken}, {Scalzo},
  {Ridden-Harper}, {Sharp}, {Tucker}, {Farrell}, {Elmer}, {Johnston},
  {Venkatraman Krishnan}, {Keane}, {Green}, {Jameson}, {Hu}, {Ma}, {Sun}, {Wu},
  {Wang}, {Shang}, {Hu}, {Ashley}, {Yuan}, {Li}, {Tao}, {Zhu}, {Zhang},
  {Suntzeff}, {Zhou}, {Yang}, {Orange}, {Morris}, {Cucchiara}, {Giblin},
  {Klotz}, {Staff}, {Thierry}, {Schmidt}, {OzGrav}, {(Deeper}, {Wider},
  {program}, {AST3}, {CAASTRO Collaborations}, {Tanvir}, {Levan}, {Cano}, {de
  Ugarte-Postigo}, {Gonz{\'a}lez-Fern{\'a}ndez}, {Greiner}, {Hjorth}, {Irwin},
  {Kr{\"u}hler}, {Mandel}, {Milvang-Jensen}, {O'Brien}, {Rol}, {Rosetti},
  {Rosswog}, {Rowlinson}, {Steeghs}, {Th{\"o}ne}, {Ulaczyk}, {Watson}, {Bruun},
  {Cutter}, {Figuera Jaimes}, {Fujii}, {Fruchter}, {Gompertz}, {Jakobsson},
  {Hodosan}, {J{\`e}rgensen}, {Kangas}, {Kann}, {Rabus}, {Schr{\o}der},
  {Stanway}, {Wijers}, {VINROUGE Collaboration}, {Lipunov}, {Gorbovskoy},
  {Kornilov}, {Tyurina}, {Balanutsa}, {Kuznetsov}, {Vlasenko}, {Podesta},
  {Lopez}, {Podesta}, {Levato}, {Saffe}, {Mallamaci}, {Budnev}, {Gress},
  {Kuvshinov}, {Gorbunov}, {Vladimirov}, {Zimnukhov}, {Gabovich}, {Yurkov},
  {Sergienko}, {Rebolo}, {Serra-Ricart}, {Tlatov}, {Ishmuhametova}, {MASTER
  Collaboration}, {Abe}, {Aoki}, {Aoki}, {Asakura}, {Baar}, {Barway}, {Bond},
  {Doi}, {Finet}, {Fujiyoshi}, {Furusawa}, {Honda}, {Itoh}, {Kanda},
  {Kawabata}, {Kawabata}, {Kim}, {Koshida}, {Kuroda}, {Lee}, {Liu},
  {Matsubayashi}, {Miyazaki}, {Morihana}, {Morokuma}, {Motohara}, {Murata},
  {Nagai}, {Nagashima}, {Nagayama}, {Nakaoka}, {Nakata}, {Ohsawa}, {Ohshima},
  {Ohta}, {Okita}, {Saito}, {Saito}, {Sako}, {Sekiguchi}, {Sumi}, {Tajitsu},
  {Takahashi}, {Takayama}, {Tamura}, {Tanaka}, {Tanaka}, {Terai}, {Tominaga},
  {Tristram}, {Uemura}, {Utsumi}, {Yamaguchi}, {Yasuda}, {Yoshida}, {Zenko},
  {J-GEM}, {Adams}, {Anupama}, {Bally}, {Barway}, {Bellm}, {Blagorodnova},
  {Cannella}, {Chandra}, {Chatterjee}, {Clarke}, {Cobb}, {Cook}, {Copperwheat},
  {De}, {Emery}, {Feindt}, {Foster}, {Fox}, {Frail}, {Fremling}, {Frohmaier},
  {Garcia}, {Ghosh}, {Giacintucci}, {Goobar}, {Gottlieb}, {Grefenstette},
  {Hallinan}, {Harrison}, {Heida}, {Helou}, {Ho}, {Horesh}, {Hotokezaka}, {Ip},
  {Itoh}, {Jacobs}, {Jencson}, {Kasen}, {Kasliwal}, {Kassim}, {Kim}, {Kiran},
  {Kuin}, {Kulkarni}, {Kupfer}, {Lau}, {Madsen}, {Mazzali}, {Miller},
  {Miyasaka}, {Mooley}, {Myers}, {Nakar}, {Ngeow}, {Nugent}, {Ofek},
  {Palliyaguru}, {Pavana}, {Perley}, {Peters}, {Pike}, {Piran}, {Qi}, {Quimby},
  {Rana}, {Rosswog}, {Rusu}, {Sadler}, {Van Sistine}, {Sollerman}, {Xu}, {Yan},
  {Yatsu}, {Yu}, {Zhang}, {Zhao}, {GROWTH}, {JAGWAR}, {Caltech-NRAO},
  {TTU-NRAO}, {NuSTAR Collaborations}, {Chambers}, {Huber}, {Schultz},
  {Bulger}, {Flewelling}, {Magnier}, {Lowe}, {Wainscoat}, {Waters}, {Willman},
  {Pan-STARRS}, {Ebisawa}, {Hanyu}, {Harita}, {Hashimoto}, {Hidaka}, {Hori},
  {Ishikawa}, {Isobe}, {Iwakiri}, {Kawai}, {Kawai}, {Kawamuro}, {Kawase},
  {Kitaoka}, {Makishima}, {Matsuoka}, {Mihara}, {Morita}, {Morita}, {Nakahira},
  {Nakajima}, {Nakamura}, {Negoro}, {Oda}, {Sakamaki}, {Sasaki}, {Serino},
  {Shidatsu}, {Shimomukai}, {Sugawara}, {Sugita}, {Sugizaki}, {Tachibana},
  {Takao}, {Tanimoto}, {Tomida}, {Tsuboi}, {Tsunemi}, {Ueda}, {Ueno}, {Yamada},
  {Yamaoka}, {Yamauchi}, {Yatabe}, {Yoneyama}, {Yoshii}, {MAXI Team}, {Coward},
  {Crisp}, {Macpherson}, {Andreoni}, {Laugier}, {Noysena}, {Klotz}, {Gendre},
  {Thierry}, {Turpin}, {Consortium}, {Im}, {Choi}, {Kim}, {Yoon}, {Lim}, {Lee},
  {Lee}, {Kim}, {Ko}, {Joe}, {Kwon}, {Kim}, {Lim}, {Choi}, {KU Collaboration},
  {Fynbo}, {Malesani}, {Xu}, {Optical Telescope}, {Smartt}, {Jerkstrand},
  {Kankare}, {Sim}, {Fraser}, {Inserra}, {Maguire}, {Leloudas}, {Magee},
  {Shingles}, {Smith}, {Young}, {Kotak}, {Gal-Yam}, {Lyman}, {Homan},
  {Agliozzo}, {Anderson}, {Angus}, {Ashall}, {Barbarino}, {Bauer}, {Berton},
  {Botticella}, {Bulla}, {Cannizzaro}, {Cartier}, {Cikota}, {Clark}, {De Cia},
  {Della Valle}, {Dennefeld}, {Dessart}, {Dimitriadis}, {Elias-Rosa}, {Firth},
  {Fl{\"o}rs}, {Frohmaier}, {Galbany}, {Gonz{\'a}lez-Gait{\'a}n}, {Gromadzki},
  {Guti{\'e}rrez}, {Hamanowicz}, {Harmanen}, {Heintz}, {Hernandez}, {Hodgkin},
  {Hook}, {Izzo}, {James}, {Jonker}, {Kerzendorf}, {Kostrzewa-Rutkowska},
  {Kromer}, {Kuncarayakti}, {Lawrence}, {Manulis}, {Mattila}, {McBrien},
  {M{\"u}ller}, {Nordin}, {O'Neill}, {Onori}, {Palmerio}, {Pastorello},
  {Patat}, {Pignata}, {Podsiadlowski}, {Razza}, {Reynolds}, {Roy}, {Ruiter},
  {Rybicki}, {Salmon}, {Pumo}, {Prentice}, {Seitenzahl}, {Smith}, {Sollerman},
  {Sullivan}, {Szegedi}, {Taddia}, {Taubenberger}, {Terreran}, {Van Soelen},
  {Vos}, {Walton}, {Wright}, {Wyrzykowski}, {Yaron}, {pre=''(''>ePESSTO},
  {Chen}, {Kr{\"u}hler}, {Schady}, {Wiseman}, {Greiner}, {Rau}, {Schweyer},
  {Klose}, {Nicuesa Guelbenzu}, {GROND}, {Palliyaguru}, {Tech University},
  {Shara}, {Williams}, {Vaisanen}, {Potter}, {Romero Colmenero}, {Crawford},
  {Buckley}, {Mao}, {SALT Group}, {D{\'\i}az}, {Macri}, {Garc{\'\i}a Lambas},
  {Mendes de Oliveira}, {Nilo Castell{\'o}n}, {Ribeiro}, {S{\'a}nchez},
  {Schoenell}, {Abramo}, {Akras}, {Alcaniz}, {Artola}, {Beroiz}, {Bonoli},
  {Cabral}, {Camuccio}, {Chavushyan}, {Coelho}, {Colazo}, {Costa-Duarte},
  {Cuevas Larenas}, {Dom{\'\i}nguez Romero}, {Dultzin}, {Fern{\'a}ndez},
  {Garc{\'\i}a}, {Girardini}, {Gon{\c{c}}alves}, {Gon{\c{c}}alves}, {Gurovich},
  {Jim{\'e}nez-Teja}, {Kanaan}, {Lares}, {Lopes de Oliveira}, {L{\'o}pez-Cruz},
  {Melia}, {Molino}, {Padilla}, {Pe{\~n}uela}, {Placco}, {Qui{\~n}ones},
  {Ram{\'\i}rez Rivera}, {Renzi}, {Riguccini}, {R{\'\i}os-L{\'o}pez},
  {Rodriguez}, {Sampedro}, {Schneiter}, {Sodr{\'e}}, {Starck}, {Torres-Flores},
  {Tornatore}, {Zadro{\.z}ny}, {Castillo}, {TOROS: Transient Robotic
  Observatory of South Collaboration}, {Castro-Tirado}, {Tello}, {Hu}, {Zhang},
  {Cunniffe}, {Castell{\'o}n}, {Hiriart}, {Caballero-Garc{\'\i}a},
  {Jel{\'\i}nek}, {Kub{\'a}nek}, {P{\'e}rez del Pulgar}, {Park}, {Jeong},
  {Castro Cer{\'o}n}, {Pandey}, {Yock}, {Querel}, {Fan}, {Wang}, {BOOTES
  Collaboration}, {Beardsley}, {Brown}, {Crosse}, {Emrich}, {Franzen},
  {Gaensler}, {Horsley}, {Johnston-Hollitt}, {Kenney}, {Morales}, {Pallot},
  {Sokolowski}, {Steele}, {Tingay}, {Trott}, {Walker}, {Wayth}, {Williams},
  {Wu}, {Murchison Widefield Array}, {Yoshida}, {Sakamoto}, {Kawakubo},
  {Yamaoka}, {Takahashi}, {Asaoka}, {Ozawa}, {Torii}, {Shimizu}, {Tamura},
  {Ishizaki}, {Cherry}, {Ricciarini}, {Penacchioni}, {Marrocchesi}, {CALET
  Collaboration}, {Pozanenko}, {Volnova}, {Mazaeva}, {Minaev}, {Krugov},
  {Kusakin}, {Reva}, {Moskvitin}, {Rumyantsev}, {Inasaridze}, {Klunko},
  {Tungalag}, {Schmalz}, {Burhonov}, {IKI-GW Follow-up Collaboration},
  {Abdalla}, {Abramowski}, {Aharonian}, {Ait Benkhali}, {Ang{\"u}ner},
  {Arakawa}, {Arrieta}, {Aubert}, {Backes}, {Balzer}, {Barnard}, {Becherini},
  {Becker Tjus}, {Berge}, {Bernhard}, {Bernl{\"o}hr}, {Blackwell},
  {B{\"o}ttcher}, {Boisson}, {Bolmont}, {Bonnefoy}, {Bordas}, {Bregeon},
  {Brun}, {Brun}, {Bryan}, {B{\"u}chele}, {Bulik}, {Capasso}, {Caroff},
  {Carosi}, {Casanova}, {Cerruti}, {Chakraborty}, {Chaves}, {Chen},
  {Chevalier}, {Colafrancesco}, {Condon}, {Conrad}, {Davids}, {Decock}, {Deil},
  {Devin}, {deWilt}, {Dirson}, {Djannati-Ata{\"\i}}, {Donath}, {O'C. Drury},
  {Dutson}, {Dyks}, {Edwards}, {Egberts}, {Emery}, {Ernenwein}, {Eschbach},
  {Farnier}, {Fegan}, {Fernandes}, {Fiasson}, {Fontaine}, {Funk},
  {F{\"u}ssling}, {Gabici}, {Gallant}, {Garrigoux}, {Gat{\'e}}, {Giavitto},
  {Giebels}, {Glawion}, {Glicenstein}, {Gottschall}, {Grondin}, {Hahn},
  {Haupt}, {Hawkes}, {Heinzelmann}, {Henri}, {Hermann}, {Hinton}, {Hofmann},
  {Hoischen}, {Holch}, {Holler}, {Horns}, {Ivascenko}, {Iwasaki},
  {Jacholkowska}, {Jamrozy}, {Jankowsky}, {Jankowsky}, {Jingo}, {Jouvin},
  {Jung-Richardt}, {Kastendieck}, {Katarzy{\'n}ski}, {Katsuragawa},
  {Kerszberg}, {Khangulyan}, {Kh{\'e}lifi}, {King}, {Klepser}, {Klochkov},
  {Klu{\'z}niak}, {Komin}, {Kosack}, {Krakau}, {Kraus}, {Kr{\"u}ger}, {Laffon},
  {Lamanna}, {Lau}, {Lees}, {Lefaucheur}, {Lemi{\`e}re}, {Lemoine-Goumard},
  {Lenain}, {Leser}, {Lohse}, {Lorentz}, {Liu}, {Lypova}, {Malyshev},
  {Marandon}, {Marcowith}, {Mariaud}, {Marx}, {Maurin}, {Maxted}, {Mayer},
  {Meintjes}, {Meyer}, {Mitchell}, {Moderski}, {Mohamed}, {Mohrmann},
  {Mor{\r{a}}}, {Moulin}, {Murach}, {Nakashima}, {de Naurois}, {Ndiyavala},
  {Niederwanger}, {Niemiec}, {Oakes}, {O'Brien}, {Odaka}, {Ohm}, {Ostrowski},
  {Oya}, {Padovani}, {Panter}, {Parsons}, {Pekeur}, {Pelletier}, {Perennes},
  {Petrucci}, {Peyaud}, {Piel}, {Pita}, {Poireau}, {Poon}, {Prokhorov},
  {Prokoph}, {P{\"u}hlhofer}, {Punch}, {Quirrenbach}, {Raab}, {Rauth},
  {Reimer}, {Reimer}, {Renaud}, {de los Reyes}, {Rieger}, {Rinchiuso},
  {Romoli}, {Rowell}, {Rudak}, {Rulten}, {Sahakian}, {Saito}, {Sanchez},
  {Santangelo}, {Sasaki}, {Schlickeiser}, {Sch{\"u}ssler}, {Schulz},
  {Schwanke}, {Schwemmer}, {Seglar-Arroyo}, {Settimo}, {Seyffert}, {Shafi},
  {Shilon}, {Shiningayamwe}, {Simoni}, {Sol}, {Spanier}, {Spir-Jacob},
  {Stawarz}, {Steenkamp}, {Stegmann}, {Steppa}, {Sushch}, {Takahashi},
  {Tavernet}, {Tavernier}, {Taylor}, {Terrier}, {Tibaldo}, {Tiziani},
  {Tluczykont}, {Trichard}, {Tsirou}, {Tsuji}, {Tuffs}, {Uchiyama}, {van der
  Walt}, {van Eldik}, {van Rensburg}, {van Soelen}, {Vasileiadis}, {Veh},
  {Venter}, {Viana}, {Vincent}, {Vink}, {Voisin}, {V{\"o}lk}, {Vuillaume},
  {Wadiasingh}, {Wagner}, {Wagner}, {Wagner}, {White}, {Wierzcholska},
  {Willmann}, {W{\"o}rnlein}, {Wouters}, {Yang}, {Zaborov}, {Zacharias},
  {Zanin}, {Zdziarski}, {Zech}, {Zefi}, {Ziegler}, {Zorn}, {{\.Z}ywucka},
  {H.~E.~S.~S. Collaboration}, {Fender}, {Broderick}, {Rowlinson}, {Wijers},
  {Stewart}, {ter Veen}, {Shulevski}, {LOFAR Collaboration}, {Kavic},
  {Simonetti}, {League}, {Tsai}, {Obenberger}, {Nathaniel}, {Taylor}, {Dowell},
  {Liebling}, {Estes}, {Lippert}, {Sharma}, {Vincent}, {Farella}, {Wavelength
  Array}, {Abeysekara}, {Albert}, {Alfaro}, {Alvarez}, {Arceo},
  {Arteaga-Vel{\'a}zquez}, {Avila Rojas}, {Ayala Solares}, {Barber}, {Becerra
  Gonzalez}, {Becerril}, {Belmont-Moreno}, {BenZvi}, {Berley}, {Bernal},
  {Braun}, {Brisbois}, {Caballero-Mora}, {Capistr{\'a}n}, {Carrami{\~n}ana},
  {Casanova}, {Castillo}, {Cotti}, {Cotzomi}, {Couti{\~n}o de Le{\'o}n}, {De
  Le{\'o}n}, {De la Fuente}, {Diaz Hernandez}, {Dichiara}, {Dingus},
  {DuVernois}, {D{\'\i}az-V{\'e}lez}, {Ellsworth}, {Engel},
  {Enr{\'\i}quez-Rivera}, {Fiorino}, {Fleischhack}, {Fraija},
  {Garc{\'\i}a-Gonz{\'a}lez}, {Garfias}, {Gerhardt}, {Gonz{\~o}lez Mu{\~n}oz},
  {Gonz{\'a}lez}, {Goodman}, {Hampel-Arias}, {Harding}, {Hernandez},
  {Hernandez-Almada}, {Hona}, {H{\"u}ntemeyer}, {Iriarte}, {Jardin-Blicq},
  {Joshi}, {Kaufmann}, {Kieda}, {Lara}, {Lauer}, {Lennarz}, {Le{\'o}n Vargas},
  {Linnemann}, {Longinotti}, {Raya}, {Luna-Garc{\'\i}a}, {L{\'o}pez-Coto},
  {Malone}, {Marinelli}, {Martinez}, {Martinez-Castellanos},
  {Mart{\'\i}nez-Castro}, {Mart{\'\i}nez-Huerta}, {Matthews},
  {Miranda-Romagnoli}, {Moreno}, {Mostaf{\'a}}, {Nellen}, {Newbold}, {Nisa},
  {Noriega-Papaqui}, {Pelayo}, {Pretz}, {P{\'e}rez-P{\'e}rez}, {Ren}, {Rho},
  {Rivi{\`e}re}, {Rosa-Gonz{\'a}lez}, {Rosenberg}, {Ruiz-Velasco}, {Salazar},
  {Salesa Greus}, {Sandoval}, {Schneider}, {Schoorlemmer}, {Sinnis}, {Smith},
  {Springer}, {Surajbali}, {Tibolla}, {Tollefson}, {Torres}, {Ukwatta},
  {Weisgarber}, {Westerhoff}, {Wisher}, {Wood}, {Yapici}, {Yodh}, {Younk},
  {Zhou}, {{\'A}lvarez}, {HAWC Collaboration}, {Aab}, {Abreu}, {Aglietta},
  {Albuquerque}, {Albury}, {Allekotte}, {Almela}, {Alvarez Castillo},
  {Alvarez-Mu{\~n}iz}, {Anastasi}, {Anchordoqui}, {Andrada}, {Andringa},
  {Aramo}, {Arsene}, {Asorey}, {Assis}, {Avila}, {Badescu}, {Balaceanu},
  {Barbato}, {Barreira Luz}, {Becker}, {Bellido}, {Berat}, {Bertaina},
  {Bertou}, {Biermann}, {Biteau}, {Blaess}, {Blanco}, {Blazek}, {Bleve},
  {Boh{\'a}{\v{c}}ov{\'a}}, {Bonifazi}, {Borodai}, {Botti}, {Brack}, {Brancus},
  {Bretz}, {Bridgeman}, {Briechle}, {Buchholz}, {Bueno}, {Buitink}, {Buscemi},
  {Caballero-Mora}, {Caccianiga}, {Cancio}, {Canfora}, {Caruso}, {Castellina},
  {Catalani}, {Cataldi}, {Cazon}, {Chavez}, {Chinellato}, {Chudoba}, {Clay},
  {Cobos Cerutti}, {Colalillo}, {Coleman}, {Collica}, {Coluccia},
  {Concei{\c{c}}{\~a}o}, {Consolati}, {Contreras}, {Cooper}, {Coutu},
  {Covault}, {Cronin}, {D'Amico}, {Daniel}, {Dasso}, {Daumiller}, {Dawson},
  {Day}, {de Almeida}, {de Jong}, {De Mauro}, {de Mello Neto}, {De Mitri}, {de
  Oliveira}, {de Souza}, {Debatin}, {Deligny}, {D{\'\i}az Castro}, {Diogo},
  {Dobrigkeit}, {D'Olivo}, {Dorosti}, {Dos Anjos}, {Dova}, {Dundovic}, {Ebr},
  {Engel}, {Erdmann}, {Erfani}, {Escobar}, {Espadanal}, {Etchegoyen}, {Falcke},
  {Farmer}, {Farrar}, {Fauth}, {Fazzini}, {Feldbusch}, {Fenu}, {Fick},
  {Figueira}, {Filip{\v{c}}i{\v{c}}}, {Freire}, {Fujii}, {Fuster},
  {Ga{\"\i}or}, {Garc{\'\i}a}, {Gat{\'e}}, {Gemmeke}, {Gherghel-Lascu}, {Ghia},
  {Giaccari}, {Giammarchi}, {Giller}, {G{\l}as}, {Glaser}, {Golup}, {G{\'o}mez
  Berisso}, {G{\'o}mez Vitale}, {Gonz{\'a}lez}, {Gorgi}, {Gottowik}, {Grillo},
  {Grubb}, {Guarino}, {Guedes}, {Halliday}, {Hampel}, {Hansen}, {Harari},
  {Harrison}, {Harvey}, {Haungs}, {Hebbeker}, {Heck}, {Heimann}, {Herve},
  {Hill}, {Hojvat}, {Holt}, {Homola}, {H{\"o}randel}, {Horvath},
  {Hrabovsk{\'y}}, {Huege}, {Hulsman}, {Insolia}, {Isar}, {Jandt}, {Johnsen},
  {Josebachuili}, {Jurysek}, {K{\"a}{\"a}p{\"a}}, {Kampert}, {Keilhauer},
  {Kemmerich}, {Kemp}, {Kieckhafer}, {Klages}, {Kleifges}, {Kleinfeller},
  {Krause}, {Krohm}, {Kuempel}, {Kukec Mezek}, {Kunka}, {Kuotb Awad}, {Lago},
  {LaHurd}, {Lang}, {Lauscher}, {Legumina}, {Leigui de Oliveira},
  {Letessier-Selvon}, {Lhenry-Yvon}, {Link}, {Lo Presti}, {Lopes}, {L{\'o}pez},
  {L{\'o}pez Casado}, {Lorek}, {Luce}, {Lucero}, {Malacari}, {Mallamaci},
  {Mandat}, {Mantsch}, {Mariazzi}, {Maris}, {Marsella}, {Martello}, {Martinez},
  {Mart{\'\i}nez Bravo}, {Mas{\'\i}as Meza}, {Mathes}, {Mathys}, {Matthews},
  {Matthiae}, {Mayotte}, {Mazur}, {Medina}, {Medina-Tanco}, {Melo},
  {Menshikov}, {Merenda}, {Michal}, {Micheletti}, {Middendorf}, {Miramonti},
  {Mitrica}, {Mockler}, {Mollerach}, {Montanet}, {Morello}, {Morlino},
  {M{\"u}ller}, {M{\"u}ller}, {Muller}, {M{\"u}ller}, {Mussa}, {Naranjo},
  {Nguyen}, {Niculescu-Oglinzanu}, {Niechciol}, {Niemietz}, {Niggemann},
  {Nitz}, {Nosek}, {Novotny}, {No{\v{z}}ka}, {N{\'u}{\~n}ez}, {Oikonomou},
  {Olinto}, {Palatka}, {Pallotta}, {Papenbreer}, {Parente}, {Parra}, {Paul},
  {Pech}, {Pedreira}, {P{\c{e}}kala}, {Pe{\~n}a-Rodriguez}, {Pereira},
  {Perlin}, {Perrone}, {Peters}, {Petrera}, {Phuntsok}, {Pierog}, {Pimenta},
  {Pirronello}, {Platino}, {Plum}, {Poh}, {Porowski}, {Prado}, {Privitera},
  {Prouza}, {Quel}, {Querchfeld}, {Quinn}, {Ramos-Pollan}, {Rautenberg},
  {Ravignani}, {Ridky}, {Riehn}, {Risse}, {Ristori}, {Rizi}, {Rodrigues de
  Carvalho}, {Rodriguez Fernandez}, {Rodriguez Rojo}, {Roncoroni}, {Roth},
  {Roulet}, {Rovero}, {Ruehl}, {Saffi}, {Saftoiu}, {Salamida}, {Salazar},
  {Saleh}, {Salina}, {S{\'a}nchez}, {Sanchez-Lucas}, {Santos}, {Santos},
  {Sarazin}, {Sarmento}, {Sarmiento-Cano}, {Sato}, {Schauer}, {Scherini},
  {Schieler}, {Schimp}, {Schmidt}, {Scholten}, {Schov{\'a}nek}, {Schr{\"o}der},
  {Schr{\"o}der}, {Schulz}, {Schumacher}, {Sciutto}, {Segreto}, {Shadkam},
  {Shellard}, {Sigl}, {Silli}, {{\v{S}}m{\'\i}da}, {Snow}, {Sommers},
  {Sonntag}, {Soriano}, {Squartini}, {Stanca}, {Stani{\v{c}}}, {Stasielak},
  {Stassi}, {Stolpovskiy}, {Strafella}, {Streich}, {Suarez},
  {Suarez-Dur{\'a}n}, {Sudholz}, {Suomij{\"a}rvi}, {Supanitsky},
  {{\v{S}}up{\'\i}k}, {Swain}, {Szadkowski}, {Taboada}, {Taborda},
  {Timmermans}, {Todero Peixoto}, {Tomankova}, {Tom{\'e}}, {Torralba Elipe},
  {Travnicek}, {Trini}, {Tueros}, {Ulrich}, {Unger}, {Urban}, {Vald{\'e}s
  Galicia}, {Vali{\~n}o}, {Valore}, {van Aar}, {van Bodegom}, {van den Berg},
  {van Vliet}, {Varela}, {Vargas C{\'a}rdenas}, {V{\'a}zquez}, {Veberi{\v{c}}},
  {Ventura}, {Vergara Quispe}, {Verzi}, {Vicha}, {Villase{\~n}or}, {Vorobiov},
  {Wahlberg}, {Wainberg}, {Walz}, {Watson}, {Weber}, {Weindl}, {Wiede{\'n}ski},
  {Wiencke}, {Wilczy{\'n}ski}, {Wirtz}, {Wittkowski}, {Wundheiler}, {Yang},
  {Yushkov}, {Zas}, {Zavrtanik}, {Zavrtanik}, {Zepeda}, {Zimmermann},
  {Ziolkowski}, {Zong}, {Zuccarello}, {Pierre Auger Collaboration}, {Kim},
  {Schulze}, {Bauer}, {Corral-Santana}, {de Gregorio-Monsalvo},
  {Gonz{\'a}lez-L{\'o}pez}, {Hartmann}, {Ishwara-Chandra}, {Mart{\'\i}n},
  {Mehner}, {Misra}, {Micha{\l}owski}, {Resmi}, {ALMA Collaboration}, {Paragi},
  {Agudo}, {An}, {Beswick}, {Casadio}, {Frey}, {Jonker}, {Kettenis}, {Marcote},
  {Moldon}, {Szomoru}, {van Langevelde}, {Yang}, {Euro VLBI Team}, {Cwiek},
  {Cwiok}, {Czyrkowski}, {Dabrowski}, {Kasprowicz}, {Mankiewicz}, {Nawrocki},
  {Opiela}, {Piotrowski}, {Wrochna}, {Zaremba}, {{\.Z}arnecki}, {Pi of Sky
  Collaboration}, {Haggard}, {Nynka}, {Ruan}, {Chandra Team at McGill
  University}, {Bland}, {Booler}, {Devillepoix}, {de Gois}, {Hancock}, {Howie},
  {Paxman}, {Sansom}, {Towner}, {Desert Fireball Network}, {Tonry}, {Coughlin},
  {Stubbs}, {Denneau}, {Heinze}, {Stalder}, {Weiland}, {ATLAS}, {Eatough},
  {Kramer}, {Kraus}, {Time Resolution Universe Survey}, {Troja}, {Piro},
  {Becerra Gonz{\'a}lez}, {Butler}, {Fox}, {Khandrika}, {Kutyrev}, {Lee},
  {Ricci}, {Ryan}, {S{\'a}nchez-Ram{\'\i}rez}, {Veilleux}, {Watson},
  {Wieringa}, {Burgess}, {van Eerten}, {Fontes}, {Fryer}, {Korobkin},
  {Wollaeger}, {RIMAS}, {RATIR}, {Camilo}, {Foley}, {Goedhart}, {Makhathini},
  {Oozeer}, {Smirnov}, {Fender}, {Woudt}, \& {South
  Africa/MeerKAT}}]{2017ApJ...848L..12A}
{Abbott}, B.~P., {Abbott}, R., {Abbott}, T.~D., {et~al.} 2017{\natexlab{c}},
  \apjl, 848, L12

\bibitem[{{Abbott} {et~al.}(2023){Abbott}, {Abbott}, {Acernese}, {Ackley},
  {Adams}, {Adhikari}, {Adhikari}, {Adya}, {Affeldt}, {Agarwal}, {Agathos},
  {Agatsuma}, {Aggarwal}, {Aguiar}, {Aiello}, {Ain}, {Ajith}, {Akutsu}, {de
  Alarc{\'o}n}, {Akcay}, {Albanesi}, {Allocca}, {Altin}, {Amato}, {Anand},
  {Anand}, {Ananyeva}, {Anderson}, {Anderson}, {Ando}, {Andrade}, {Andres},
  {Andri{\'c}}, {Angelova}, {Ansoldi}, {Antelis}, {Antier}, {Antonini},
  {Appert}, {Arai}, {Arai}, {Arai}, {Araki}, {Araya}, {Araya}, {Areeda},
  {Ar{\`e}ne}, {Aritomi}, {Arnaud}, {Arogeti}, {Aronson}, {Arun}, {Asada},
  {Asali}, {Ashton}, {Aso}, {Assiduo}, {Aston}, {Astone}, {Aubin}, {Austin},
  {Babak}, {Badaracco}, {Bader}, {Badger}, {Bae}, {Bae}, {Baer}, {Bagnasco},
  {Bai}, {Baiotti}, {Baird}, {Bajpai}, {Ball}, {Ballardin}, {Ballmer},
  {Balsamo}, {Baltus}, {Banagiri}, {Bankar}, {Barayoga}, {Barbieri}, {Barish},
  {Barker}, {Barneo}, {Barone}, {Barr}, {Barsotti}, {Barsuglia}, {Barta},
  {Bartlett}, {Barton}, {Bartos}, {Bassiri}, {Basti}, {Bawaj}, {Bayley},
  {Baylor}, {Bazzan}, {B{\'e}csy}, {Bedakihale}, {Bejger}, {Belahcene},
  {Benedetto}, {Beniwal}, {Bennett}, {Bentley}, {Benyaala}, {Bergamin},
  {Berger}, {Bernuzzi}, {Berry}, {Bersanetti}, {Bertolini}, {Betzwieser},
  {Beveridge}, {Bhandare}, {Bhardwaj}, {Bhattacharjee}, {Bhaumik}, {Bilenko},
  {Billingsley}, {Bini}, {Birney}, {Birnholtz}, {Biscans}, {Bischi},
  {Biscoveanu}, {Bisht}, {Biswas}, {Bitossi}, {Bizouard}, {Blackburn}, {Blair},
  {Blair}, {Blair}, {Bobba}, {Bode}, {Boer}, {Bogaert}, {Boldrini}, {Bonavena},
  {Bondu}, {Bonilla}, {Bonnand}, {Booker}, {Boom}, {Bork}, {Boschi}, {Bose},
  {Bose}, {Bossilkov}, {Boudart}, {Bouffanais}, {Bozzi}, {Bradaschia}, {Brady},
  {Bramley}, {Branch}, {Branchesi}, {Brandt}, {Brau}, {Breschi}, {Briant},
  {Briggs}, {Brillet}, {Brinkmann}, {Brockill}, {Brooks}, {Brooks}, {Brown},
  {Brunett}, {Bruno}, {Bruntz}, {Bryant}, {Bulik}, {Bulten}, {Buonanno},
  {Buscicchio}, {Buskulic}, {Buy}, {Byer}, {Cadonati}, {Cagnoli}, {Cahillane},
  {Bustillo}, {Callaghan}, {Callister}, {Calloni}, {Cameron}, {Camp}, {Canepa},
  {Canevarolo}, {Cannavacciuolo}, {Cannon}, {Cao}, {Cao}, {Capocasa}, {Capote},
  {Carapella}, {Carbognani}, {Carlin}, {Carney}, {Carpinelli}, {Carrillo},
  {Carullo}, {Carver}, {Diaz}, {Casentini}, {Castaldi}, {Caudill},
  {Cavagli{\`a}}, {Cavalier}, {Cavalieri}, {Ceasar}, {Cella},
  {Cerd{\'a}-Dur{\'a}n}, {Cesarini}, {Chaibi}, {Chakravarti}, {Subrahmanya},
  {Champion}, {Chan}, {Chan}, {Chan}, {Chan}, {Chan}, {Chandra}, {Chanial},
  {Chao}, {Chapman-Bird}, {Charlton}, {Chase}, {Chassande-Mottin},
  {Chatterjee}, {Chatterjee}, {Chatterjee}, {Chaturvedi}, {Chaty},
  {Chatziioannou}, {Chen}, {Chen}, {Chen}, {Chen}, {Chen}, {Chen}, {Chen},
  {Chen}, {Cheng}, {Cheong}, {Cheung}, {Chia}, {Chiadini}, {Chiang},
  {Chiarini}, {Chierici}, {Chincarini}, {Chiofalo}, {Chiummo}, {Cho}, {Cho},
  {Choudhary}, {Choudhary}, {Christensen}, {Chu}, {Chu}, {Chu}, {Chua},
  {Chung}, {Ciani}, {Ciecielag}, {Cie{\'s}lar}, {Cifaldi}, {Ciobanu}, {Ciolfi},
  {Cipriano}, {Cirone}, {Clara}, {Clark}, {Clark}, {Clarke}, {Clearwater},
  {Clesse}, {Cleva}, {Coccia}, {Codazzo}, {Cohadon}, {Cohen}, {Cohen},
  {Colleoni}, {Collette}, {Colombo}, {Colpi}, {Compton}, {Constancio}, {Conti},
  {Cooper}, {Corban}, {Corbitt}, {Cordero-Carri{\'o}n}, {Corezzi}, {Corley},
  {Cornish}, {Corre}, {Corsi}, {Cortese}, {Costa}, {Cotesta}, {Coughlin},
  {Coulon}, {Countryman}, {Cousins}, {Couvares}, {Coward}, {Cowart}, {Coyne},
  {Coyne}, {Creighton}, {Creighton}, {Criswell}, {Croquette}, {Crowder},
  {Cudell}, {Cullen}, {Cumming}, {Cummings}, {Cunningham}, {Cuoco},
  {Cury{\l}o}, {Dabadie}, {Canton}, {Dall'Osso}, {D{\'a}lya}, {Dana},
  {Daneshgaranbajastani}, {D'Angelo}, {Danila}, {Danilishin}, {D'Antonio},
  {Danzmann}, {Darsow-Fromm}, {Dasgupta}, {Datrier}, {Datta}, {Dattilo},
  {Dave}, {Davier}, {Davies}, {Davis}, {Davis}, {Daw}, {Dean}, {Debra},
  {Deenadayalan}, {Degallaix}, {de Laurentis}, {Del{\'e}glise}, {Del Favero},
  {de Lillo}, {de Lillo}, {Del Pozzo}, {Demarchi}, {de Matteis}, {D'Emilio},
  {Demos}, {Dent}, {Depasse}, {de Pietri}, {De Rosa}, {de Rossi}, {Desalvo},
  {de Simone}, {Dhurandhar}, {D{\'\i}az}, {Diaz-Ortiz}, {Didio}, {Dietrich},
  {di Fiore}, {di Fronzo}, {di Giorgio}, {di Giovanni}, {di Giovanni}, {di
  Girolamo}, {di Lieto}, {Ding}, {di Pace}, {di Palma}, {di Renzo},
  {Divakarla}, {Dmitriev}, {Doctor}, {D'Onofrio}, {Donovan}, {Dooley},
  {Doravari}, {Dorrington}, {Drago}, {Driggers}, {Drori}, {Ducoin}, {Dupej},
  {Durante}, {D'Urso}, {Duverne}, {Dwyer}, {Eassa}, {Easter}, {Ebersold},
  {Eckhardt}, {Eddolls}, {Edelman}, {Edo}, {Edy}, {Effler}, {Eguchi},
  {Eichholz}, {Eikenberry}, {Eisenmann}, {Eisenstein}, {Ejlli}, {Engelby},
  {Enomoto}, {Errico}, {Essick}, {Estell{\'e}s}, {Estevez}, {Etienne}, {Etzel},
  {Evans}, {Evans}, {Ewing}, {Fafone}, {Fair}, {Fairhurst}, {Farah}, {Farinon},
  {Farr}, {Farr}, {Farrow}, {Fauchon-Jones}, {Favaro}, {Favata}, {Fays},
  {Fazio}, {Feicht}, {Fejer}, {Fenyvesi}, {Ferguson}, {Fernandez-Galiana},
  {Ferrante}, {Ferreira}, {Fidecaro}, {Figura}, {Fiori}, {Fishbach}, {Fisher},
  {Fittipaldi}, {Fiumara}, {Flaminio}, {Floden}, {Fong}, {Font}, {Fornal},
  {Forsyth}, {Franke}, {Frasca}, {Frasconi}, {Frederick}, {Freed}, {Frei},
  {Freise}, {Frey}, {Fritschel}, {Frolov}, {Fronz{\'e}}, {Fujii}, {Fujikawa},
  {Fukunaga}, {Fukushima}, {Fulda}, {Fyffe}, {Gabbard}, {Gadre}, {Gair},
  {Gais}, {Galaudage}, {Gamba}, {Ganapathy}, {Ganguly}, {Gao}, {Gaonkar},
  {Garaventa}, {Garc{\'\i}a}, {Garc{\'\i}a-N{\'u}{\~n}ez},
  {Garc{\'\i}a-Quir{\'o}s}, {Garufi}, {Gateley}, {Gaudio}, {Gayathri}, {Ge},
  {Gemme}, {Gennai}, {George}, {George}, {Gerberding}, {Gergely}, {Gewecke},
  {Ghonge}, {Ghosh}, {Ghosh}, {Ghosh}, {Ghosh}, {Giacomazzo}, {Giacoppo},
  {Giaime}, {Giardina}, {Gibson}, {Gier}, {Giesler}, {Giri}, {Gissi},
  {Glanzer}, {Gleckl}, {Godwin}, {Golomb}, {Goetz}, {Goetz}, {Gohlke},
  {Goncharov}, {Gonz{\'a}lez}, {Gopakumar}, {Gosselin}, {Gouaty}, {Gould},
  {Grace}, {Grado}, {Granata}, {Granata}, {Grant}, {Gras}, {Grassia}, {Gray},
  {Gray}, {Greco}, {Green}, {Green}, {Gretarsson}, {Gretarsson}, {Griffith},
  {Griffiths}, {Griggs}, {Grignani}, {Grimaldi}, {Grimm}, {Grote}, {Grunewald},
  {Gruning}, {Guerra}, {Guidi}, {Guimaraes}, {Guix{\'e}}, {Gulati}, {Guo},
  {Guo}, {Gupta}, {Gupta}, {Gupta}, {Gustafson}, {Gustafson}, {Guzman}, {Ha},
  {Haegel}, {Hagiwara}, {Haino}, {Halim}, {Hall}, {Hamilton}, {Hammond}, {Han},
  {Haney}, {Hanks}, {Hanna}, {Hannam}, {Hannuksela}, {Hansen}, {Hansen},
  {Hanson}, {Harder}, {Hardwick}, {Haris}, {Harms}, {Harry}, {Harry},
  {Hartwig}, {Hasegawa}, {Haskell}, {Hasskew}, {Haster}, {Hattori}, {Haughian},
  {Hayakawa}, {Hayama}, {Hayes}, {Healy}, {Heidmann}, {Heidt}, {Heintze},
  {Heinze}, {Heinzel}, {Heitmann}, {Hellman}, {Hello}, {Helmling-Cornell},
  {Hemming}, {Hendry}, {Heng}, {Hennes}, {Hennig}, {Hennig}, {Hernandez},
  {Vivanco}, {Heurs}, {Hild}, {Hill}, {Himemoto}, {Hines}, {Hiranuma},
  {Hirata}, {Hirose}, {Hochheim}, {Hofman}, {Hohmann}, {Holcomb}, {Holland},
  {Hollows}, {Holmes}, {Holt}, {Holz}, {Hong}, {Hopkins}, {Hough}, {Hourihane},
  {Howell}, {Hoy}, {Hoyland}, {Hreibi}, {Hsieh}, {Hsu}, {Huang}, {Huang},
  {Huang}, {Huang}, {Huang}, {Huang}, {H{\"u}bner}, {Huddart}, {Hughey}, {Hui},
  {Hui}, {Husa}, {Huttner}, {Huxford}, {Huynh-Dinh}, {Ide}, {Idzkowski},
  {Iess}, {Ikenoue}, {Imam}, {Inayoshi}, {Ingram}, {Inoue}, {Ioka}, {Isi},
  {Isleif}, {Ito}, {Itoh}, {Iyer}, {Izumi}, {Jaberianhamedan}, {Jacqmin},
  {Jadhav}, {Jadhav}, {James}, {Jan}, {Jani}, {Janquart}, {Janssens},
  {Janthalur}, {Jaranowski}, {Jariwala}, {Jaume}, {Jenkins}, {Jenner}, {Jeon},
  {Jeunon}, {Jia}, {Jin}, {Johns}, {Jones}, {Jones}, {Jones}, {Jones}, {Jones},
  {Jonker}, {Ju}, {Jung}, {Jung}, {Junker}, {Juste}, {Kaihotsu}, {Kajita},
  {Kakizaki}, {Kalaghatgi}, {Kalogera}, {Kamai}, {Kamiizumi}, {Kanda},
  {Kandhasamy}, {Kang}, {Kanner}, {Kao}, {Kapadia}, {Kapasi}, {Karat},
  {Karathanasis}, {Karki}, {Kashyap}, {Kasprzack}, {Kastaun}, {Katsanevas},
  {Katsavounidis}, {Katzman}, {Kaur}, {Kawabe}, {Kawaguchi}, {Kawai},
  {Kawasaki}, {K{\'e}f{\'e}lian}, {Keitel}, {Key}, {Khadka}, {Khalili}, {Khan},
  {Khazanov}, {Khetan}, {Khursheed}, {Kijbunchoo}, {Kim}, {Kim}, {Kim}, {Kim},
  {Kim}, {Kim}, {Kimball}, {Kimura}, {Kinley-Hanlon}, {Kirchhoff}, {Kissel},
  {Kita}, {Kitazawa}, {Kleybolte}, {Klimenko}, {Knee}, {Knowles}, {Knyazev},
  {Koch}, {Koekoek}, {Kojima}, {Kokeyama}, {Koley}, {Kolitsidou}, {Kolstein},
  {Komori}, {Kondrashov}, {Kong}, {Kontos}, {Koper}, {Korobko}, {Kotake},
  {Kovalam}, {Kozak}, {Kozakai}, {Kozu}, {Kringel}, {Krishnendu}, {Kr{\'o}lak},
  {Kuehn}, {Kuei}, {Kuijer}, {Kulkarni}, {Kumar}, {Kumar}, {Kumar}, {Kumar},
  {Kume}, {Kuns}, {Kuo}, {Kuo}, {Kuromiya}, {Kuroyanagi}, {Kusayanagi},
  {Kuwahara}, {Kwak}, {Lagabbe}, {Laghi}, {Lalande}, {Lam}, {Lamberts},
  {Landry}, {Landry}, {Lane}, {Lang}, {Lange}, {Lantz}, {La Rosa},
  {Lartaux-Vollard}, {Lasky}, {Laxen}, {Lazzarini}, {Lazzaro}, {Leaci},
  {Leavey}, {Lecoeuche}, {Lee}, {Lee}, {Lee}, {Lee}, {Lee}, {Lee}, {Lehmann},
  {Lema{\^\i}tre}, {Leonardi}, {Leroy}, {Letendre}, {Levesque}, {Levin},
  {Leviton}, {Leyde}, {Li}, {Li}, {Li}, {Li}, {Li}, {Li}, {Lin}, {Lin}, {Lin},
  {Lin}, {Lin}, {Linde}, {Linker}, {Linley}, {Littenberg}, {Liu}, {Liu}, {Liu},
  {Liu}, {Llamas}, {Llorens-Monteagudo}, {Lo}, {Lockwood}, {Loh}, {London},
  {Longo}, {Lopez}, {Portilla}, {Lorenzini}, {Loriette}, {Lormand}, {Losurdo},
  {Lott}, {Lough}, {Lousto}, {Lovelace}, {Lucaccioni}, {L{\"u}ck}, {Lumaca},
  {Lundgren}, {Luo}, {Lynam}, {Macas}, {Macinnis}, {MacLeod}, {MacMillan},
  {Macquet}, {Hernandez}, {Magazz{\`u}}, {Magee}, {Maggiore}, {Magnozzi},
  {Mahesh}, {Majorana}, {Makarem}, {Maksimovic}, {Maliakal}, {Malik}, {Man},
  {Mandic}, {Mangano}, {Mango}, {Mansell}, {Manske}, {Mantovani}, {Mapelli},
  {Marchesoni}, {Marchio}, {Marion}, {Mark}, {M{\'a}rka}, {M{\'a}rka},
  {Markakis}, {Markosyan}, {Markowitz}, {Maros}, {Marquina}, {Marsat},
  {Martelli}, {Martin}, {Martin}, {Martinez}, {Martinez}, {Martinez},
  {Martinovic}, {Martynov}, {Marx}, {Masalehdan}, {Mason}, {Massera},
  {Masserot}, {Massinger}, {Masso-Reid}, {Mastrogiovanni}, {Matas},
  {Mateu-Lucena}, {Matichard}, {Matiushechkina}, {Mavalvala}, {McCann},
  {McCarthy}, {McClelland}, {McClincy}, {McCormick}, {McCuller}, {McGhee},
  {McGuire}, {McIsaac}, {McIver}, {McRae}, {McWilliams}, {Meacher}, {Mehmet},
  {Mehta}, {Meijer}, {Melatos}, {Melchor}, {Mendell}, {Menendez-Vazquez},
  {Menoni}, {Mercer}, {Mereni}, {Merfeld}, {Merilh}, {Merritt}, {Merzougui},
  {Meshkov}, {Messenger}, {Messick}, {Meyers}, {Meylahn}, {Mhaske}, {Miani},
  {Miao}, {Michaloliakos}, {Michel}, {Michimura}, {Middleton}, {Milano},
  {Miller}, {Miller}, {Miller}, {Miller}, {Millhouse}, {Mills}, {Milotti},
  {Minazzoli}, {Minenkov}, {Mio}, {Mir}, {Miravet-Ten{\'e}s}, {Mishra},
  {Mishra}, {Mistry}, {Mitra}, {Mitrofanov}, {Mitselmakher}, {Mittleman},
  {Miyakawa}, {Miyamoto}, {Miyazaki}, {Miyo}, {Miyoki}, {Mo}, {Modafferi},
  {Moguel}, {Mogushi}, {Mohapatra}, {Mohite}, {Molina}, {Molina-Ruiz},
  {Mondin}, {Montani}, {Moore}, {Moraru}, {Morawski}, {More}, {Moreno},
  {Moreno}, {Mori}, {Morisaki}, {Moriwaki}, {Morr{\'a}s}, {Mours}, {Mow-Lowry},
  {Mozzon}, {Muciaccia}, {Mukherjee}, {Mukherjee}, {Mukherjee}, {Mukherjee},
  {Mukherjee}, {Mukund}, {Mullavey}, {Munch}, {Mu{\~n}iz}, {Murray},
  {Musenich}, {Muusse}, {Nadji}, {Nagano}, {Nagano}, {Nagar}, {Nakamura},
  {Nakano}, {Nakano}, {Nakashima}, {Nakayama}, {Napolano}, {Nardecchia},
  {Narikawa}, {Naticchioni}, {Nayak}, {Nayak}, {Negishi}, {Neil}, {Neilson},
  {Nelemans}, {Nelson}, {Nery}, {Neubauer}, {Neunzert}, {Ng}, {Ng}, {Nguyen},
  {Nguyen}, {Nguyen}, {Quynh}, {Ni}, {Nichols}, {Nishizawa}, {Nissanke},
  {Nitoglia}, {Nocera}, {Norman}, {North}, {Nozaki}, {Siles}, {Nuttall},
  {Oberling}, {O'Brien}, {Obuchi}, {O'Dell}, {Oelker}, {Ogaki}, {Oganesyan},
  {Oh}, {Oh}, {Oh}, {Ohashi}, {Ohishi}, {Ohkawa}, {Ohme}, {Ohta}, {Okada},
  {Okutani}, {Okutomi}, {Olivetto}, {Oohara}, {Ooi}, {Oram}, {O'Reilly},
  {Ormiston}, {Ormsby}, {Ortega}, {O'Shaughnessy}, {O'Shea}, {Oshino},
  {Ossokine}, {Osthelder}, {Otabe}, {Ottaway}, {Overmier}, {Pace}, {Pagano},
  {Page}, {Pagliaroli}, {Pai}, {Pai}, {Palamos}, {Palashov}, {Palomba}, {Pan},
  {Pan}, {Panda}, {Pang}, {Pang}, {Pankow}, {Pannarale}, {Pant}, {Panther},
  {Paoletti}, {Paoli}, {Paolone}, {Parisi}, {Park}, {Park}, {Parker},
  {Pascucci}, {Pasqualetti}, {Passaquieti}, {Passuello}, {Patel}, {Pathak},
  {Patricelli}, {Patron}, {Paul}, {Payne}, {Pedraza}, {Pegoraro}, {Pele},
  {Arellano}, {Penn}, {Perego}, {Pereira}, {Pereira}, {Perez}, {P{\'e}rigois},
  {Perkins}, {Perreca}, {Perri{\`e}s}, {Petermann}, {Petterson}, {Pfeiffer},
  {Pham}, {Phukon}, {Piccinni}, {Pichot}, {Piendibene}, {Piergiovanni},
  {Pierini}, {Pierro}, {Pillant}, {Pillas}, {Pilo}, {Pinard}, {Pinto}, {Pinto},
  {Piotrzkowski}, {Piotrzkowski}, {Pirello}, {Pitkin}, {Placidi}, {Planas},
  {Plastino}, {Pluchar}, {Poggiani}, {Polini}, {Pong}, {Ponrathnam},
  {Popolizio}, {Porter}, {Poulton}, {Powell}, {Pracchia}, {Pradier},
  {Prajapati}, {Prasai}, {Prasanna}, {Pratten}, {Principe}, {Prodi},
  {Prokhorov}, {Prosposito}, {Prudenzi}, {Puecher}, {Punturo}, {Puosi},
  {Puppo}, {P{\"u}rrer}, {Qi}, {Quetschke}, {Quitzow-James}, {Raab},
  {Raaijmakers}, {Radkins}, {Radulesco}, {Raffai}, {Rail}, {Raja}, {Rajan},
  {Ramirez}, {Ramirez}, {Ramos-Buades}, {Rana}, {Rapagnani}, {Rapol}, {Ray},
  {Raymond}, {Raza}, {Razzano}, {Read}, {Rees}, {Regimbau}, {Rei}, {Reid},
  {Reid}, {Reitze}, {Relton}, {Renzini}, {Rettegno}, {Reza}, {Rezac}, {Ricci},
  {Richards}, {Richardson}, {Richardson}, {Riemenschneider}, {Riles},
  {Rinaldi}, {Rink}, {Rizzo}, {Robertson}, {Robie}, {Robinet}, {Rocchi},
  {Rodriguez}, {Rolland}, {Rollins}, {Romanelli}, {Romano}, {Romel},
  {Romero-Rodr{\'\i}guez}, {Romero-Shaw}, {Romie}, {Ronchini}, {Rosa}, {Rose},
  {Rosi{\'n}ska}, {Ross}, {Rowan}, {Rowlinson}, {Roy}, {Roy}, {Roy}, {Rozza},
  {Ruggi}, {Ryan}, {Sachdev}, {Sadecki}, {Sadiq}, {Sago}, {Saito}, {Saito},
  {Sakai}, {Sakai}, {Sakellariadou}, {Sakuno}, {Salafia}, {Salconi}, {Saleem},
  {Salemi}, {Samajdar}, {Sanchez}, {Sanchez}, {Sanchez}, {Sanchis-Gual},
  {Sanders}, {Sanuy}, {Saravanan}, {Sarin}, {Sassolas}, {Satari},
  {Sathyaprakash}, {Sato}, {Sato}, {Sauter}, {Savage}, {Sawada}, {Sawant},
  {Sawant}, {Sayah}, {Schaetzl}, {Scheel}, {Scheuer}, {Schiworski}, {Schmidt},
  {Schmidt}, {Schnabel}, {Schneewind}, {Schofield}, {Sch{\"o}nbeck}, {Schulte},
  {Schutz}, {Schwartz}, {Scott}, {Scott}, {Seglar-Arroyo}, {Sekiguchi},
  {Sekiguchi}, {Sellers}, {Sengupta}, {Sentenac}, {Seo}, {Sequino}, {Sergeev},
  {Setyawati}, {Shaffer}, {Shahriar}, {Shams}, {Shao}, {Sharma}, {Sharma},
  {Shawhan}, {Shcheblanov}, {Shibagaki}, {Shikauchi}, {Shimizu}, {Shimoda},
  {Shimode}, {Shinkai}, {Shishido}, {Shoda}, {Shoemaker}, {Shoemaker},
  {Shyamsundar}, {Sieniawska}, {Sigg}, {Singer}, {Singh}, {Singh}, {Singha},
  {Sintes}, {Sipala}, {Skliris}, {Slagmolen}, {Slaven-Blair}, {Smetana},
  {Smith}, {Smith}, {Soldateschi}, {Somala}, {Somiya}, {Son}, {Soni}, {Soni},
  {Sordini}, {Sorrentino}, {Sorrentino}, {Sotani}, {Soulard}, {Souradeep},
  {Sowell}, {Spagnuolo}, {Spencer}, {Spera}, {Srinivasan}, {Srivastava},
  {Srivastava}, {Staats}, {Stachie}, {Steer}, {Steinhoff}, {Steinlechner},
  {Steinlechner}, {Stevenson}, {Stops}, {Stover}, {Strain}, {Strang},
  {Stratta}, {Strunk}, {Sturani}, {Stuver}, {Sudhagar}, {Sudhir}, {Sugimoto},
  {Suh}, {Sullivan}, {Summerscales}, {Sun}, {Sun}, {Sunil}, {Sur}, {Suresh},
  {Sutton}, {Suzuki}, {Suzuki}, {Swinkels}, {Szczepa{\'n}czyk}, {Szewczyk},
  {Tacca}, {Tagoshi}, {Tait}, {Takahashi}, {Takahashi}, {Takamori}, {Takano},
  {Takeda}, {Takeda}, {Talbot}, {Talbot}, {Tanaka}, {Tanaka}, {Tanaka},
  {Tanaka}, {Tanaka}, {Tanasijczuk}, {Tanioka}, {Tanner}, {Tao}, {Tao},
  {Mart{\'\i}n}, {Taranto}, {Tasson}, {Telada}, {Tenorio}, {Terhune},
  {Terkowski}, {Thirugnanasambandam}, {Thomas}, {Thomas}, {Thomas}, {Thompson},
  {Thondapu}, {Thorne}, {Thrane}, {Tiwari}, {Tiwari}, {Tiwari}, {Toivonen},
  {Toland}, {Tolley}, {Tomaru}, {Tomigami}, {Tomura}, {Tonelli},
  {Torres-Forn{\'e}}, {Torrie}, {E Melo}, {T{\"o}yr{\"a}}, {Trapananti},
  {Travasso}, {Traylor}, {Trevor}, {Tringali}, {Tripathee}, {Troiano},
  {Trovato}, {Trozzo}, {Trudeau}, {Tsai}, {Tsai}, {Tsang}, {Tsang}, {Tsao},
  {Tse}, {Tso}, {Tsubono}, {Tsuchida}, {Tsukada}, {Tsuna}, {Tsutsui},
  {Tsuzuki}, {Turbang}, {Turconi}, {Tuyenbayev}, {Ubhi}, {Uchikata},
  {Uchiyama}, {Udall}, {Ueda}, {Uehara}, {Ueno}, {Ueshima}, {Unnikrishnan},
  {Uraguchi}, {Urban}, {Ushiba}, {Utina}, {Vahlbruch}, {Vajente}, {Vajpeyi},
  {Valdes}, {Valentini}, {Valsan}, {van Bakel}, {van Beuzekom}, {van den
  Brand}, {van den Broeck}, {Vander-Hyde}, {van der Schaaf}, {van Heijningen},
  {Vanosky}, {van Putten}, {van Remortel}, {Vardaro}, {Vargas}, {Varma},
  {Vas{\'u}th}, {Vecchio}, {Vedovato}, {Veitch}, {Veitch}, {Venneberg},
  {Venugopalan}, {Verkindt}, {Verma}, {Verma}, {Veske}, {Vetrano},
  {Vicer{\'e}}, {Vidyant}, {Viets}, {Vijaykumar}, {Villa-Ortega}, {Vinet},
  {Virtuoso}, {Vitale}, {Vo}, {Vocca}, {von Reis}, {von Wrangel}, {Vorvick},
  {Vyatchanin}, {Wade}, {Wade}, {Wagner}, {Walet}, {Walker}, {Wallace},
  {Wallace}, {Walsh}, {Wang}, {Wang}, {Wang}, {Ward}, {Warner}, {Was},
  {Washimi}, {Washington}, {Watchi}, {Weaver}, {Webster}, {Weinert},
  {Weinstein}, {Weiss}, {Weller}, {Wellmann}, {Wen}, {We{\ss}els}, {Wette},
  {Whelan}, {White}, {Whiting}, {Whittle}, {Wilken}, {Williams}, {Williams},
  {Williamson}, {Willis}, {Willke}, {Wilson}, {Winkler}, {Wipf}, {Wlodarczyk},
  {Woan}, {Woehler}, {Wofford}, {Wong}, {Wu}, {Wu}, {Wu}, {Wu}, {Wysocki},
  {Xiao}, {Xu}, {Yamada}, {Yamamoto}, {Yamamoto}, {Yamamoto}, {Yamamoto},
  {Yamashita}, {Yamazaki}, {Yang}, {Yang}, {Yang}, {Yang}, {Yang}, {Yap},
  {Yeeles}, {Yelikar}, {Ying}, {Yokogawa}, {Yokoyama}, {Yokozawa}, {Yoo},
  {Yoshioka}, {Yu}, {Yu}, {Yuzurihara}, {Zadro{\.z}ny}, {Zanolin}, {Zeidler},
  {Zelenova}, {Zendri}, {Zevin}, {Zhan}, {Zhang}, {Zhang}, {Zhang}, {Zhang},
  {Zhang}, {Zhao}, {Zhao}, {Zhao}, {Zhao}, {Zheng}, {Zhou}, {Zhou}, {Zhu},
  {Zhu}, {Zimmerman}, {Zlochower}, {Zucker}, {Zweizig}, {LIGO Scientific
  Collaboration}, {VIRGO Collaboration}, \& {KAGRA
  Collaboration}}]{2023PhRvX..13a1048A}
{Abbott}, R., {Abbott}, T.~D., {Acernese}, F., {et~al.} 2023, Physical Review
  X, 13, 011048

\bibitem[{{Ai} {et~al.}(2025){Ai}, {Gao}, \& {Zhang}}]{2025ApJ...978...52A}
{Ai}, S., {Gao}, H., \& {Zhang}, B. 2025, \apj, 978, 52

\bibitem[{{Ai} {et~al.}(2022){Ai}, {Zhang}, \& {Zhu}}]{2022MNRAS.516.2614A}
{Ai}, S., {Zhang}, B., \& {Zhu}, Z. 2022, \mnras, 516, 2614

\bibitem[{{Amaro-Seoane} {et~al.}(2017){Amaro-Seoane}, {Audley}, {Babak},
  {Baker}, {Barausse}, {Bender}, {Berti}, {Binetruy}, {Born}, {Bortoluzzi},
  {Camp}, {Caprini}, {Cardoso}, {Colpi}, {Conklin}, {Cornish}, {Cutler},
  {Danzmann}, {Dolesi}, {Ferraioli}, {Ferroni}, {Fitzsimons}, {Gair}, {Gesa
  Bote}, {Giardini}, {Gibert}, {Grimani}, {Halloin}, {Heinzel}, {Hertog},
  {Hewitson}, {Holley-Bockelmann}, {Hollington}, {Hueller}, {Inchauspe},
  {Jetzer}, {Karnesis}, {Killow}, {Klein}, {Klipstein}, {Korsakova}, {Larson},
  {Livas}, {Lloro}, {Man}, {Mance}, {Martino}, {Mateos}, {McKenzie},
  {McWilliams}, {Miller}, {Mueller}, {Nardini}, {Nelemans}, {Nofrarias},
  {Petiteau}, {Pivato}, {Plagnol}, {Porter}, {Reiche}, {Robertson},
  {Robertson}, {Rossi}, {Russano}, {Schutz}, {Sesana}, {Shoemaker}, {Slutsky},
  {Sopuerta}, {Sumner}, {Tamanini}, {Thorpe}, {Troebs}, {Vallisneri},
  {Vecchio}, {Vetrugno}, {Vitale}, {Volonteri}, {Wanner}, {Ward}, {Wass},
  {Weber}, {Ziemer}, \& {Zweifel}}]{2017arXiv170200786A}
{Amaro-Seoane}, P., {Audley}, H., {Babak}, S., {et~al.} 2017, arXiv e-prints,
  arXiv:1702.00786

\bibitem[{{Andrews} {et~al.}(2024){Andrews}, {Bavera}, {Briel}, {Chattaraj},
  {Dotter}, {Fragos}, {Gallegos-Garcia}, {Gossage}, {Kalogera}, {Kasdagli},
  {Katsaggelos}, {Kimball}, {Kovlakas}, {Kruckow}, {Liotine}, {Misra}, {Rocha},
  {Souropanis}, {Srivastava}, {Sun}, {Teng}, {Xing}, {Zapartas}, \&
  {Zevin}}]{2024arXiv241102376A}
{Andrews}, J.~J., {Bavera}, S.~S., {Briel}, M., {et~al.} 2024, arXiv e-prints,
  arXiv:2411.02376

\bibitem[{{Astropy Collaboration} {et~al.}(2013){Astropy Collaboration},
  {Robitaille}, {Tollerud}, {Greenfield}, {Droettboom}, {Bray}, {Aldcroft},
  {Davis}, {Ginsburg}, {Price-Whelan}, {Kerzendorf}, {Conley}, {Crighton},
  {Barbary}, {Muna}, {Ferguson}, {Grollier}, {Parikh}, {Nair}, {Unther},
  {Deil}, {Woillez}, {Conseil}, {Kramer}, {Turner}, {Singer}, {Fox}, {Weaver},
  {Zabalza}, {Edwards}, {Azalee Bostroem}, {Burke}, {Casey}, {Crawford},
  {Dencheva}, {Ely}, {Jenness}, {Labrie}, {Lim}, {Pierfederici}, {Pontzen},
  {Ptak}, {Refsdal}, {Servillat}, \& {Streicher}}]{astropy:2013}
{Astropy Collaboration}, {Robitaille}, T.~P., {Tollerud}, E.~J., {et~al.} 2013,
  \aap, 558, A33

\bibitem[{{Atri} {et~al.}(2019){Atri}, {Miller-Jones}, {Bahramian}, {Plotkin},
  {Jonker}, {Nelemans}, {Maccarone}, {Sivakoff}, {Deller}, {Chaty}, {Torres},
  {Horiuchi}, {McCallum}, {Natusch}, {Phillips}, {Stevens}, \&
  {Weston}}]{2019MNRAS.489.3116A}
{Atri}, P., {Miller-Jones}, J.~C.~A., {Bahramian}, A., {et~al.} 2019, \mnras,
  489, 3116

\bibitem[{{Batziou} {et~al.}(2025){Batziou}, {Glas}, {Janka}, {Ehring},
  {Abdikamalov}, \& {Just}}]{2025ApJ...984..197B}
{Batziou}, E., {Glas}, R., {Janka}, H.~T., {et~al.} 2025, \apj, 984, 197

\bibitem[{{Belczynski} {et~al.}(2010){Belczynski}, {Dominik}, {Bulik},
  {O'Shaughnessy}, {Fryer}, \& {Holz}}]{2010ApJ...715L.138B}
{Belczynski}, K., {Dominik}, M., {Bulik}, T., {et~al.} 2010, \apjl, 715, L138

\bibitem[{{Belczynski} {et~al.}(2008){Belczynski}, {Kalogera}, {Rasio}, {Taam},
  {Zezas}, {Bulik}, {Maccarone}, \& {Ivanova}}]{2008ApJS..174..223B}
{Belczynski}, K., {Kalogera}, V., {Rasio}, F.~A., {et~al.} 2008, \apjs, 174,
  223

\bibitem[{{Belczynski} {et~al.}(2006){Belczynski}, {Perna}, {Bulik},
  {Kalogera}, {Ivanova}, \& {Lamb}}]{2006ApJ...648.1110B}
{Belczynski}, K., {Perna}, R., {Bulik}, T., {et~al.} 2006, \apj, 648, 1110

\bibitem[{{Berger}(2010)}]{2010ApJ...722.1946B}
{Berger}, E. 2010, \apj, 722, 1946

\bibitem[{{Blaauw}(1961)}]{1961BAN....15..265B}
{Blaauw}, A. 1961, \bain, 15, 265

\bibitem[{{Blanchard} {et~al.}(2017){Blanchard}, {Berger}, {Fong}, {Nicholl},
  {Leja}, {Conroy}, {Alexander}, {Margutti}, {Williams}, {Doctor}, {Chornock},
  {Villar}, {Cowperthwaite}, {Annis}, {Brout}, {Brown}, {Chen}, {Eftekhari},
  {Frieman}, {Holz}, {Metzger}, {Rest}, {Sako}, \&
  {Soares-Santos}}]{2017ApJ...848L..22B}
{Blanchard}, P.~K., {Berger}, E., {Fong}, W., {et~al.} 2017, \apjl, 848, L22

\bibitem[{{Bloom} {et~al.}(2002){Bloom}, {Kulkarni}, \&
  {Djorgovski}}]{2002AJ....123.1111B}
{Bloom}, J.~S., {Kulkarni}, S.~R., \& {Djorgovski}, S.~G. 2002, \aj, 123, 1111

\bibitem[{{Bloom} {et~al.}(1999){Bloom}, {Sigurdsson}, \&
  {Pols}}]{1999MNRAS.305..763B}
{Bloom}, J.~S., {Sigurdsson}, S., \& {Pols}, O.~R. 1999, \mnras, 305, 763

\bibitem[{{Bobrick} {et~al.}(2017){Bobrick}, {Davies}, \&
  {Church}}]{2017MNRAS.467.3556B}
{Bobrick}, A., {Davies}, M.~B., \& {Church}, R.~P. 2017, \mnras, 467, 3556

\bibitem[{{Bobrick} {et~al.}(2022){Bobrick}, {Zenati}, {Perets}, {Davies}, \&
  {Church}}]{2022MNRAS.510.3758B}
{Bobrick}, A., {Zenati}, Y., {Perets}, H.~B., {Davies}, M.~B., \& {Church}, R.
  2022, \mnras, 510, 3758

\bibitem[{{Boersma}(1961)}]{1961BAN....15..291B}
{Boersma}, J. 1961, \bain, 15, 291

\bibitem[{{Bray} \& {Eldridge}(2016)}]{2016MNRAS.461.3747B}
{Bray}, J.~C. \& {Eldridge}, J.~J. 2016, \mnras, 461, 3747

\bibitem[{{Bray} \& {Eldridge}(2018)}]{2018MNRAS.480.5657B}
{Bray}, J.~C. \& {Eldridge}, J.~J. 2018, \mnras, 480, 5657

\bibitem[{{Breivik} {et~al.}(2020){Breivik}, {Coughlin}, {Zevin}, {Rodriguez},
  {Kremer}, {Ye}, {Andrews}, {Kurkowski}, {Digman}, {Larson}, \&
  {Rasio}}]{2020ApJ...898...71B}
{Breivik}, K., {Coughlin}, S., {Zevin}, M., {et~al.} 2020, \apj, 898, 71

\bibitem[{{Briel} {et~al.}(2022){Briel}, {Eldridge}, {Stanway}, {Stevance}, \&
  {Chrimes}}]{2022MNRAS.514.1315B}
{Briel}, M.~M., {Eldridge}, J.~J., {Stanway}, E.~R., {Stevance}, H.~F., \&
  {Chrimes}, A.~A. 2022, \mnras, 514, 1315

\bibitem[{{Briel} {et~al.}(2024){Briel}, {Metha}, {Eldridge}, {Moriya}, \&
  {Trenti}}]{2024arXiv240813076B}
{Briel}, M.~M., {Metha}, B., {Eldridge}, J.~J., {Moriya}, T.~J., \& {Trenti},
  M. 2024, arXiv e-prints, arXiv:2408.13076

\bibitem[{{Briel} {et~al.}(2023){Briel}, {Stevance}, \&
  {Eldridge}}]{2023MNRAS.520.5724B}
{Briel}, M.~M., {Stevance}, H.~F., \& {Eldridge}, J.~J. 2023, \mnras, 520, 5724

\bibitem[{{Broekgaarden} {et~al.}(2022){Broekgaarden}, {Berger}, {Stevenson},
  {Justham}, {Mandel}, {Chru{\'s}li{\'n}ska}, {van Son}, {Wagg},
  {Vigna-G{\'o}mez}, {de Mink}, {Chattopadhyay}, \&
  {Neijssel}}]{2022MNRAS.516.5737B}
{Broekgaarden}, F.~S., {Berger}, E., {Stevenson}, S., {et~al.} 2022, \mnras,
  516, 5737

\bibitem[{{Bucciantini} {et~al.}(2012){Bucciantini}, {Metzger}, {Thompson}, \&
  {Quataert}}]{2012MNRAS.419.1537B}
{Bucciantini}, N., {Metzger}, B.~D., {Thompson}, T.~A., \& {Quataert}, E. 2012,
  \mnras, 419, 1537

\bibitem[{{Caito} {et~al.}(2009){Caito}, {Bernardini}, {Bianco}, {Dainotti},
  {Guida}, \& {Ruffini}}]{2009A&A...498..501C}
{Caito}, L., {Bernardini}, M.~G., {Bianco}, C.~L., {et~al.} 2009, \aap, 498,
  501

\bibitem[{{Chen} {et~al.}(2023){Chen}, {Tauris}, {Chen}, \&
  {Han}}]{2023ApJ...951...91C}
{Chen}, H.-L., {Tauris}, T.~M., {Chen}, X., \& {Han}, Z. 2023, \apj, 951, 91

\bibitem[{{Chen} {et~al.}(2024){Chen}, {Shen}, {Tan}, {Wang}, {Xiong}, {Chen},
  \& {Zhang}}]{2024ApJ...973L..33C}
{Chen}, J., {Shen}, R.-F., {Tan}, W.-J., {et~al.} 2024, \apjl, 973, L33

\bibitem[{{Cheong} {et~al.}(2025){Cheong}, {Pitik}, {Longo Micchi}, \&
  {Radice}}]{2025ApJ...978L..38C}
{Cheong}, P. C.-K., {Pitik}, T., {Longo Micchi}, L.~F., \& {Radice}, D. 2025,
  \apjl, 978, L38

\bibitem[{{Chrimes} {et~al.}(2020){Chrimes}, {Stanway}, \&
  {Eldridge}}]{2020MNRAS.491.3479C}
{Chrimes}, A.~A., {Stanway}, E.~R., \& {Eldridge}, J.~J. 2020, \mnras, 491,
  3479

\bibitem[{{Chruslinska} \& {Nelemans}(2019)}]{2019MNRAS.488.5300C}
{Chruslinska}, M. \& {Nelemans}, G. 2019, \mnras, 488, 5300

\bibitem[{{Chruslinska} {et~al.}(2019){Chruslinska}, {Nelemans}, \&
  {Belczynski}}]{2019MNRAS.482.5012C}
{Chruslinska}, M., {Nelemans}, G., \& {Belczynski}, K. 2019, \mnras, 482, 5012

\bibitem[{{Church} {et~al.}(2011){Church}, {Levan}, {Davies}, \&
  {Tanvir}}]{2011MNRAS.413.2004C}
{Church}, R.~P., {Levan}, A.~J., {Davies}, M.~B., \& {Tanvir}, N. 2011, \mnras,
  413, 2004

\bibitem[{{Church} {et~al.}(2017){Church}, {Strader}, {Davies}, \&
  {Bobrick}}]{2017ApJ...851L...4C}
{Church}, R.~P., {Strader}, J., {Davies}, M.~B., \& {Bobrick}, A. 2017, \apjl,
  851, L4

\bibitem[{{Claeys} {et~al.}(2014){Claeys}, {Pols}, {Izzard}, {Vink}, \&
  {Verbunt}}]{2014A&A...563A..83C}
{Claeys}, J.~S.~W., {Pols}, O.~R., {Izzard}, R.~G., {Vink}, J., \& {Verbunt},
  F.~W.~M. 2014, \aap, 563, A83

\bibitem[{{Colpi} {et~al.}(2024){Colpi}, {Danzmann}, {Hewitson},
  {Holley-Bockelmann}, {Jetzer}, {Nelemans}, {Petiteau}, {Shoemaker},
  {Sopuerta}, {Stebbins}, {Tanvir}, {Ward}, {Weber}, {Thorpe}, {Daurskikh},
  {Deep}, {Fern{\'a}ndez N{\'u}{\~n}ez}, {Garc{\'\i}a Marirrodriga}, {Gehler},
  {Halain}, {Jennrich}, {Lammers}, {Larra{\~n}aga}, {Lieser},
  {L{\"u}tzgendorf}, {Martens}, {Mondin}, {Piris Ni{\~n}o}, {Amaro-Seoane},
  {Arca Sedda}, {Auclair}, {Babak}, {Baghi}, {Baibhav}, {Baker}, {Bayle},
  {Berry}, {Berti}, {Boileau}, {Bonetti}, {Brito}, {Buscicchio}, {Calcagni},
  {Capelo}, {Caprini}, {Caputo}, {Castelli}, {Chen}, {Chen}, {Chua}, {Davies},
  {Derdzinski}, {Domcke}, {Doneva}, {Dvorkin}, {Mar{\'\i}a Ezquiaga}, {Gair},
  {Haiman}, {Harry}, {Hartwig}, {Hees}, {Heffernan}, {Husa},
  {Izquierdo-Villalba}, {Karnesis}, {Klein}, {Korol}, {Korsakova}, {Kupfer},
  {Laghi}, {Lamberts}, {Larson}, {Le Jeune}, {Lewicki}, {Littenberg}, {Madge},
  {Mangiagli}, {Marsat}, {Vilchez}, {Maselli}, {Mathews}, {van de Meent},
  {Muratore}, {Nardini}, {Pani}, {Peloso}, {Pieroni}, {Pound},
  {Quelquejay-Leclere}, {Ricciardone}, {Rossi}, {Sartirana}, {Savalle},
  {Sberna}, {Sesana}, {Shoemaker}, {Slutsky}, {Sotiriou}, {Speri}, {Staab},
  {Steer}, {Tamanini}, {Tasinato}, {Torrado}, {Torres-Orjuela}, {Toubiana},
  {Vallisneri}, {Vecchio}, {Volonteri}, {Yagi}, \&
  {Zwick}}]{2024arXiv240207571C}
{Colpi}, M., {Danzmann}, K., {Hewitson}, M., {et~al.} 2024, arXiv e-prints,
  arXiv:2402.07571

\bibitem[{{Cooray}(2004)}]{2004MNRAS.354...25C}
{Cooray}, A. 2004, \mnras, 354, 25

\bibitem[{{Cowperthwaite} {et~al.}(2017){Cowperthwaite}, {Berger}, {Villar},
  {Metzger}, {Nicholl}, {Chornock}, {Blanchard}, {Fong}, {Margutti},
  {Soares-Santos}, {Alexander}, {Allam}, {Annis}, {Brout}, {Brown}, {Butler},
  {Chen}, {Diehl}, {Doctor}, {Drout}, {Eftekhari}, {Farr}, {Finley}, {Foley},
  {Frieman}, {Fryer}, {Garc{\'\i}a-Bellido}, {Gill}, {Guillochon}, {Herner},
  {Holz}, {Kasen}, {Kessler}, {Marriner}, {Matheson}, {Neilsen}, {Quataert},
  {Palmese}, {Rest}, {Sako}, {Scolnic}, {Smith}, {Tucker}, {Williams},
  {Balbinot}, {Carlin}, {Cook}, {Durret}, {Li}, {Lopes}, {Louren{\c{c}}o},
  {Marshall}, {Medina}, {Muir}, {Mu{\~n}oz}, {Sauseda}, {Schlegel}, {Secco},
  {Vivas}, {Wester}, {Zenteno}, {Zhang}, {Abbott}, {Banerji}, {Bechtol},
  {Benoit-L{\'e}vy}, {Bertin}, {Buckley-Geer}, {Burke}, {Capozzi}, {Carnero
  Rosell}, {Carrasco Kind}, {Castander}, {Crocce}, {Cunha}, {D'Andrea}, {da
  Costa}, {Davis}, {DePoy}, {Desai}, {Dietrich}, {Drlica-Wagner}, {Eifler},
  {Evrard}, {Fernandez}, {Flaugher}, {Fosalba}, {Gaztanaga}, {Gerdes},
  {Giannantonio}, {Goldstein}, {Gruen}, {Gruendl}, {Gutierrez}, {Honscheid},
  {Jain}, {James}, {Jeltema}, {Johnson}, {Johnson}, {Kent}, {Krause}, {Kron},
  {Kuehn}, {Nuropatkin}, {Lahav}, {Lima}, {Lin}, {Maia}, {March}, {Martini},
  {McMahon}, {Menanteau}, {Miller}, {Miquel}, {Mohr}, {Neilsen}, {Nichol},
  {Ogando}, {Plazas}, {Roe}, {Romer}, {Roodman}, {Rykoff}, {Sanchez},
  {Scarpine}, {Schindler}, {Schubnell}, {Sevilla-Noarbe}, {Smith}, {Smith},
  {Sobreira}, {Suchyta}, {Swanson}, {Tarle}, {Thomas}, {Thomas}, {Troxel},
  {Vikram}, {Walker}, {Wechsler}, {Weller}, {Yanny}, \&
  {Zuntz}}]{2017ApJ...848L..17C}
{Cowperthwaite}, P.~S., {Berger}, E., {Villar}, V.~A., {et~al.} 2017, \apjl,
  848, L17

\bibitem[{{Davies} {et~al.}(2002){Davies}, {Ritter}, \&
  {King}}]{2002MNRAS.335..369D}
{Davies}, M.~B., {Ritter}, H., \& {King}, A. 2002, \mnras, 335, 369

\bibitem[{{de S{\'a}} {et~al.}(2024){de S{\'a}}, {Rocha}, {Bernardo},
  {Bachega}, \& {Horvath}}]{2024arXiv241001451D}
{de S{\'a}}, L.~M., {Rocha}, L.~S., {Bernardo}, A., {Bachega}, R.~R.~A., \&
  {Horvath}, J.~E. 2024, arXiv e-prints, arXiv:2410.01451

\bibitem[{{Desai} {et~al.}(2019){Desai}, {Metzger}, \&
  {Foucart}}]{2019MNRAS.485.4404D}
{Desai}, D., {Metzger}, B.~D., \& {Foucart}, F. 2019, \mnras, 485, 4404

\bibitem[{{Dessart} {et~al.}(2007){Dessart}, {Burrows}, {Livne}, \&
  {Ott}}]{2007ApJ...669..585D}
{Dessart}, L., {Burrows}, A., {Livne}, E., \& {Ott}, C.~D. 2007, \apj, 669, 585

\bibitem[{{Dessart} {et~al.}(2006){Dessart}, {Burrows}, {Ott}, {Livne}, {Yoon},
  \& {Langer}}]{2006ApJ...644.1063D}
{Dessart}, L., {Burrows}, A., {Ott}, C.~D., {et~al.} 2006, \apj, 644, 1063

\bibitem[{{Disberg} {et~al.}(2024){Disberg}, {Gaspari}, \&
  {Levan}}]{2024A&A...689A.348D}
{Disberg}, P., {Gaspari}, N., \& {Levan}, A.~J. 2024, \aap, 689, A348

\bibitem[{{Disberg} {et~al.}(2025){Disberg}, {Gaspari}, \&
  {Levan}}]{2025arXiv250301429D}
{Disberg}, P., {Gaspari}, N., \& {Levan}, A.~J. 2025, arXiv e-prints,
  arXiv:2503.01429

\bibitem[{{Dong} {et~al.}(2018){Dong}, {Gu}, {Liu}, \&
  {Wang}}]{2018MNRAS.475L.101D}
{Dong}, Y.-Z., {Gu}, W.-M., {Liu}, T., \& {Wang}, J. 2018, \mnras, 475, L101

\bibitem[{{Du} {et~al.}(2024){Du}, {L{\"u}}, {Yuan}, {Yang}, \&
  {Liang}}]{2024ApJ...962L..27D}
{Du}, Z., {L{\"u}}, H., {Yuan}, Y., {Yang}, X., \& {Liang}, E. 2024, \apjl,
  962, L27

\bibitem[{{Eggleton}(1971)}]{1971MNRAS.151..351E}
{Eggleton}, P.~P. 1971, \mnras, 151, 351

\bibitem[{{Eldridge} {et~al.}(2006){Eldridge}, {Genet}, {Daigne}, \&
  {Mochkovitch}}]{2006MNRAS.367..186E}
{Eldridge}, J.~J., {Genet}, F., {Daigne}, F., \& {Mochkovitch}, R. 2006,
  \mnras, 367, 186

\bibitem[{{Eldridge} {et~al.}(2019{\natexlab{a}}){Eldridge}, {Guo},
  {Rodrigues}, {Stanway}, \& {Xiao}}]{2019PASA...36...41E}
{Eldridge}, J.~J., {Guo}, N.~Y., {Rodrigues}, N., {Stanway}, E.~R., \& {Xiao},
  L. 2019{\natexlab{a}}, \pasa, 36, e041

\bibitem[{{Eldridge} {et~al.}(2008){Eldridge}, {Izzard}, \&
  {Tout}}]{2008MNRAS.384.1109E}
{Eldridge}, J.~J., {Izzard}, R.~G., \& {Tout}, C.~A. 2008, \mnras, 384, 1109

\bibitem[{{Eldridge} {et~al.}(2019{\natexlab{b}}){Eldridge}, {Stanway}, \&
  {Tang}}]{2019MNRAS.482..870E}
{Eldridge}, J.~J., {Stanway}, E.~R., \& {Tang}, P.~N. 2019{\natexlab{b}},
  \mnras, 482, 870

\bibitem[{{Eldridge} {et~al.}(2017){Eldridge}, {Stanway}, {Xiao}, {McClelland},
  {Taylor}, {Ng}, {Greis}, \& {Bray}}]{2017PASA...34...58E}
{Eldridge}, J.~J., {Stanway}, E.~R., {Xiao}, L., {et~al.} 2017, \pasa, 34, e058

\bibitem[{{Eldridge} \& {Tout}(2004)}]{2004MNRAS.353...87E}
{Eldridge}, J.~J. \& {Tout}, C.~A. 2004, \mnras, 353, 87

\bibitem[{{Eldridge} {et~al.}(2018){Eldridge}, {Xiao}, {Stanway}, {Rodrigues},
  \& {Guo}}]{2018PASA...35...49E}
{Eldridge}, J.~J., {Xiao}, L., {Stanway}, E.~R., {Rodrigues}, N., \& {Guo},
  N.~Y. 2018, \pasa, 35, e049

\bibitem[{{Eyles-Ferris} {et~al.}(2024){Eyles-Ferris}, {Nixon}, {Coughlin}, \&
  {O'Brien}}]{2024ApJ...965L..20E}
{Eyles-Ferris}, R.~A.~J., {Nixon}, C.~J., {Coughlin}, E.~R., \& {O'Brien},
  P.~T. 2024, \apjl, 965, L20

\bibitem[{{Fong} \& {Berger}(2013)}]{2013ApJ...776...18F}
{Fong}, W. \& {Berger}, E. 2013, \apj, 776, 18

\bibitem[{{Fong} {et~al.}(2022){Fong}, {Nugent}, {Dong}, {Berger}, {Paterson},
  {Chornock}, {Levan}, {Blanchard}, {Alexander}, {Andrews}, {Cobb},
  {Cucchiara}, {Fox}, {Fryer}, {Gordon}, {Kilpatrick}, {Lunnan}, {Margutti},
  {Miller}, {Milne}, {Nicholl}, {Perley}, {Rastinejad}, {Escorial},
  {Schroeder}, {Smith}, {Tanvir}, \& {Terreran}}]{2022ApJ...940...56F}
{Fong}, W.-f., {Nugent}, A.~E., {Dong}, Y., {et~al.} 2022, \apj, 940, 56

\bibitem[{{Fragos} {et~al.}(2023){Fragos}, {Andrews}, {Bavera}, {Berry},
  {Coughlin}, {Dotter}, {Giri}, {Kalogera}, {Katsaggelos}, {Kovlakas},
  {Lalvani}, {Misra}, {Srivastava}, {Qin}, {Rocha}, {Rom{\'a}n-Garza}, {Serra},
  {Stahle}, {Sun}, {Teng}, {Trajcevski}, {Tran}, {Xing}, {Zapartas}, \&
  {Zevin}}]{2023ApJS..264...45F}
{Fragos}, T., {Andrews}, J.~J., {Bavera}, S.~S., {et~al.} 2023, \apjs, 264, 45

\bibitem[{{Fruchter} {et~al.}(2006){Fruchter}, {Levan}, {Strolger},
  {Vreeswijk}, {Thorsett}, {Bersier}, {Burud}, {Castro Cer{\'o}n},
  {Castro-Tirado}, {Conselice}, {Dahlen}, {Ferguson}, {Fynbo}, {Garnavich},
  {Gibbons}, {Gorosabel}, {Gull}, {Hjorth}, {Holland}, {Kouveliotou}, {Levay},
  {Livio}, {Metzger}, {Nugent}, {Petro}, {Pian}, {Rhoads}, {Riess}, {Sahu},
  {Smette}, {Tanvir}, {Wijers}, \& {Woosley}}]{2006Natur.441..463F}
{Fruchter}, A.~S., {Levan}, A.~J., {Strolger}, L., {et~al.} 2006, \nat, 441,
  463

\bibitem[{{Fryer} {et~al.}(1999{\natexlab{a}}){Fryer}, {Benz}, {Herant}, \&
  {Colgate}}]{1999ApJ...516..892F}
{Fryer}, C., {Benz}, W., {Herant}, M., \& {Colgate}, S.~A. 1999{\natexlab{a}},
  \apj, 516, 892

\bibitem[{{Fryer} {et~al.}(2024){Fryer}, {Burns}, {Ho}, {Lien}, {Perley},
  {Vail}, \& {Villar}}]{2024arXiv241010378F}
{Fryer}, C.~L., {Burns}, E., {Ho}, A. Y.~Q., {et~al.} 2024, arXiv e-prints,
  arXiv:2410.10378

\bibitem[{{Fryer} {et~al.}(1999{\natexlab{b}}){Fryer}, {Woosley}, \&
  {Hartmann}}]{1999ApJ...526..152F}
{Fryer}, C.~L., {Woosley}, S.~E., \& {Hartmann}, D.~H. 1999{\natexlab{b}},
  \apj, 526, 152

\bibitem[{{Fryer} {et~al.}(1999{\natexlab{c}}){Fryer}, {Woosley}, {Herant}, \&
  {Davies}}]{1999ApJ...520..650F}
{Fryer}, C.~L., {Woosley}, S.~E., {Herant}, M., \& {Davies}, M.~B.
  1999{\natexlab{c}}, \apj, 520, 650

\bibitem[{{Fynbo} {et~al.}(2006){Fynbo}, {Watson}, {Th{\"o}ne}, {Sollerman},
  {Bloom}, {Davis}, {Hjorth}, {Jakobsson}, {J{\o}rgensen}, {Graham},
  {Fruchter}, {Bersier}, {Kewley}, {Cassan}, {Castro Cer{\'o}n}, {Foley},
  {Gorosabel}, {Hinse}, {Horne}, {Jensen}, {Klose}, {Kocevski}, {Marquette},
  {Perley}, {Ramirez-Ruiz}, {Stritzinger}, {Vreeswijk}, {Wijers}, {Woller},
  {Xu}, \& {Zub}}]{2006Natur.444.1047F}
{Fynbo}, J. P.~U., {Watson}, D., {Th{\"o}ne}, C.~C., {et~al.} 2006, \nat, 444,
  1047

\bibitem[{{Galama} {et~al.}(1998){Galama}, {Vreeswijk}, {van Paradijs},
  {Kouveliotou}, {Augusteijn}, {B{\"o}hnhardt}, {Brewer}, {Doublier},
  {Gonzalez}, {Leibundgut}, {Lidman}, {Hainaut}, {Patat}, {Heise}, {in't Zand},
  {Hurley}, {Groot}, {Strom}, {Mazzali}, {Iwamoto}, {Nomoto}, {Umeda},
  {Nakamura}, {Young}, {Suzuki}, {Shigeyama}, {Koshut}, {Kippen}, {Robinson},
  {de Wildt}, {Wijers}, {Tanvir}, {Greiner}, {Pian}, {Palazzi}, {Frontera},
  {Masetti}, {Nicastro}, {Feroci}, {Costa}, {Piro}, {Peterson}, {Tinney},
  {Boyle}, {Cannon}, {Stathakis}, {Sadler}, {Begam}, \&
  {Ianna}}]{1998Natur.395..670G}
{Galama}, T.~J., {Vreeswijk}, P.~M., {van Paradijs}, J., {et~al.} 1998, \nat,
  395, 670

\bibitem[{{Gaspari} {et~al.}(2024{\natexlab{a}}){Gaspari}, {Levan}, {Chrimes},
  \& {Nelemans}}]{2024MNRAS.527.1101G}
{Gaspari}, N., {Levan}, A.~J., {Chrimes}, A.~A., \& {Nelemans}, G.
  2024{\natexlab{a}}, \mnras, 527, 1101

\bibitem[{{Gaspari} {et~al.}(2025){Gaspari}, {Levan}, {Chrimes}, \&
  {Nugent}}]{2025arXiv250404825G}
{Gaspari}, N., {Levan}, A.~J., {Chrimes}, A.~A., \& {Nugent}, A.~E. 2025, arXiv
  e-prints, arXiv:2504.04825

\bibitem[{{Gaspari} {et~al.}(2024{\natexlab{b}}){Gaspari}, {Stevance}, {Levan},
  {Chrimes}, \& {Lyman}}]{2024A&A...692A..21G}
{Gaspari}, N., {Stevance}, H.~F., {Levan}, A.~J., {Chrimes}, A.~A., \& {Lyman},
  J.~D. 2024{\natexlab{b}}, \aap, 692, A21

\bibitem[{{Gehrels} {et~al.}(2006){Gehrels}, {Norris}, {Barthelmy}, {Granot},
  {Kaneko}, {Kouveliotou}, {Markwardt}, {M{\'e}sz{\'a}ros}, {Nakar}, {Nousek},
  {O'Brien}, {Page}, {Palmer}, {Parsons}, {Roming}, {Sakamoto}, {Sarazin},
  {Schady}, {Stamatikos}, \& {Woosley}}]{2006Natur.444.1044G}
{Gehrels}, N., {Norris}, J.~P., {Barthelmy}, S.~D., {et~al.} 2006, \nat, 444,
  1044

\bibitem[{{Ghirlanda} \& {Salvaterra}(2022)}]{2022ApJ...932...10G}
{Ghirlanda}, G. \& {Salvaterra}, R. 2022, \apj, 932, 10

\bibitem[{{Ghodla} \& {Eldridge}(2023)}]{2023MNRAS.523.1711G}
{Ghodla}, S. \& {Eldridge}, J.~J. 2023, \mnras, 523, 1711

\bibitem[{{Ghodla} {et~al.}(2022){Ghodla}, {van Zeist}, {Eldridge}, {Stevance},
  \& {Stanway}}]{2022MNRAS.511.1201G}
{Ghodla}, S., {van Zeist}, W. G.~J., {Eldridge}, J.~J., {Stevance}, H.~F., \&
  {Stanway}, E.~R. 2022, \mnras, 511, 1201

\bibitem[{{Gillanders} {et~al.}(2020){Gillanders}, {Sim}, \&
  {Smartt}}]{2020MNRAS.497..246G}
{Gillanders}, J.~H., {Sim}, S.~A., \& {Smartt}, S.~J. 2020, \mnras, 497, 246

\bibitem[{{Gillanders} {et~al.}(2023){Gillanders}, {Troja}, {Fryer}, {Ristic},
  {O'Connor}, {Fontes}, {Yang}, {Domoto}, {Rahmouni}, {Tanaka}, {Fox}, \&
  {Dichiara}}]{2023arXiv230800633G}
{Gillanders}, J.~H., {Troja}, E., {Fryer}, C.~L., {et~al.} 2023, arXiv
  e-prints, arXiv:2308.00633

\bibitem[{{Gompertz} {et~al.}(2020){Gompertz}, {Levan}, \&
  {Tanvir}}]{2020ApJ...895...58G}
{Gompertz}, B.~P., {Levan}, A.~J., \& {Tanvir}, N.~R. 2020, \apj, 895, 58

\bibitem[{{Gompertz} {et~al.}(2018){Gompertz}, {Levan}, {Tanvir}, {Hjorth},
  {Covino}, {Evans}, {Fruchter}, {Gonz{\'a}lez-Fern{\'a}ndez}, {Jin}, {Lyman},
  {Oates}, {O'Brien}, \& {Wiersema}}]{2018ApJ...860...62G}
{Gompertz}, B.~P., {Levan}, A.~J., {Tanvir}, N.~R., {et~al.} 2018, \apj, 860,
  62

\bibitem[{{Gompertz} {et~al.}(2023){Gompertz}, {Nicholl}, {Smith},
  {Harisankar}, {Pratten}, {Schmidt}, \& {Smith}}]{2023MNRAS.526.4585G}
{Gompertz}, B.~P., {Nicholl}, M., {Smith}, J.~C., {et~al.} 2023, \mnras, 526,
  4585

\bibitem[{{Gompertz} {et~al.}(2014){Gompertz}, {O'Brien}, \&
  {Wynn}}]{2014MNRAS.438..240G}
{Gompertz}, B.~P., {O'Brien}, P.~T., \& {Wynn}, G.~A. 2014, \mnras, 438, 240

\bibitem[{{Gompertz} {et~al.}(2013){Gompertz}, {O'Brien}, {Wynn}, \&
  {Rowlinson}}]{2013MNRAS.431.1745G}
{Gompertz}, B.~P., {O'Brien}, P.~T., {Wynn}, G.~A., \& {Rowlinson}, A. 2013,
  \mnras, 431, 1745

\bibitem[{{Gottlieb} {et~al.}(2025){Gottlieb}, {Metzger}, {Foucart}, \&
  {Ramirez-Ruiz}}]{2025ApJ...984...77G}
{Gottlieb}, O., {Metzger}, B.~D., {Foucart}, F., \& {Ramirez-Ruiz}, E. 2025,
  \apj, 984, 77

\bibitem[{{Gottlieb} {et~al.}(2023){Gottlieb}, {Metzger}, {Quataert}, {Issa},
  {Martineau}, {Foucart}, {Duez}, {Kidder}, {Pfeiffer}, \&
  {Scheel}}]{2023ApJ...958L..33G}
{Gottlieb}, O., {Metzger}, B.~D., {Quataert}, E., {et~al.} 2023, \apjl, 958,
  L33

\bibitem[{{Harris} {et~al.}(2020){Harris}, {Millman}, {van der Walt},
  {Gommers}, {Virtanen}, {Cournapeau}, {Wieser}, {Taylor}, {Berg}, {Smith},
  {Kern}, {Picus}, {Hoyer}, {van Kerkwijk}, {Brett}, {Haldane}, {del R{\'\i}o},
  {Wiebe}, {Peterson}, {G{\'e}rard-Marchant}, {Sheppard}, {Reddy}, {Weckesser},
  {Abbasi}, {Gohlke}, \& {Oliphant}}]{2020Natur.585..357H}
{Harris}, C.~R., {Millman}, K.~J., {van der Walt}, S.~J., {et~al.} 2020, \nat,
  585, 357

\bibitem[{{Hjorth} {et~al.}(2003){Hjorth}, {Sollerman}, {M{\o}ller}, {Fynbo},
  {Woosley}, {Kouveliotou}, {Tanvir}, {Greiner}, {Andersen}, {Castro-Tirado},
  {Castro Cer{\'o}n}, {Fruchter}, {Gorosabel}, {Jakobsson}, {Kaper}, {Klose},
  {Masetti}, {Pedersen}, {Pedersen}, {Pian}, {Palazzi}, {Rhoads}, {Rol}, {van
  den Heuvel}, {Vreeswijk}, {Watson}, \& {Wijers}}]{2003Natur.423..847H}
{Hjorth}, J., {Sollerman}, J., {M{\o}ller}, P., {et~al.} 2003, \nat, 423, 847

\bibitem[{{Hobbs} {et~al.}(2005){Hobbs}, {Lorimer}, {Lyne}, \&
  {Kramer}}]{2005MNRAS.360..974H}
{Hobbs}, G., {Lorimer}, D.~R., {Lyne}, A.~G., \& {Kramer}, M. 2005, \mnras,
  360, 974

\bibitem[{{Hunter}(2007)}]{2007CSE.....9...90H}
{Hunter}, J.~D. 2007, Computing in Science and Engineering, 9, 90

\bibitem[{{Hurley} {et~al.}(2002){Hurley}, {Tout}, \&
  {Pols}}]{2002MNRAS.329..897H}
{Hurley}, J.~R., {Tout}, C.~A., \& {Pols}, O.~R. 2002, \mnras, 329, 897

\bibitem[{{Igoshev}(2020)}]{2020MNRAS.494.3663I}
{Igoshev}, A.~P. 2020, \mnras, 494, 3663

\bibitem[{{Iorio} {et~al.}(2023){Iorio}, {Mapelli}, {Costa}, {Spera},
  {Escobar}, {Sgalletta}, {Trani}, {Korb}, {Santoliquido}, {Dall'Amico},
  {Gaspari}, \& {Bressan}}]{2023MNRAS.524..426I}
{Iorio}, G., {Mapelli}, M., {Costa}, G., {et~al.} 2023, \mnras, 524, 426

\bibitem[{{Ivanova} {et~al.}(2013){Ivanova}, {Justham}, {Chen}, {De Marco},
  {Fryer}, {Gaburov}, {Ge}, {Glebbeek}, {Han}, {Li}, {Lu}, {Marsh},
  {Podsiadlowski}, {Potter}, {Soker}, {Taam}, {Tauris}, {van den Heuvel}, \&
  {Webbink}}]{2013A&ARv..21...59I}
{Ivanova}, N., {Justham}, S., {Chen}, X., {et~al.} 2013, \aapr, 21, 59

\bibitem[{{Jia} \& {Li}(2014)}]{2014ApJ...791..127J}
{Jia}, K. \& {Li}, X.~D. 2014, \apj, 791, 127

\bibitem[{{Johnson} {et~al.}(2021){Johnson}, {Leja}, {Conroy}, \&
  {Speagle}}]{2021ApJS..254...22J}
{Johnson}, B.~D., {Leja}, J., {Conroy}, C., \& {Speagle}, J.~S. 2021, \apjs,
  254, 22

\bibitem[{{Kaltenborn} {et~al.}(2023){Kaltenborn}, {Fryer}, {Wollaeger},
  {Belczynski}, {Even}, \& {Kouveliotou}}]{2023ApJ...956...71K}
{Kaltenborn}, M. A.~R., {Fryer}, C.~L., {Wollaeger}, R.~T., {et~al.} 2023,
  \apj, 956, 71

\bibitem[{{Kapil} {et~al.}(2023){Kapil}, {Mandel}, {Berti}, \&
  {M{\"u}ller}}]{2023MNRAS.519.5893K}
{Kapil}, V., {Mandel}, I., {Berti}, E., \& {M{\"u}ller}, B. 2023, \mnras, 519,
  5893

\bibitem[{{Kelley} {et~al.}(2010){Kelley}, {Ramirez-Ruiz}, {Zemp}, {Diemand},
  \& {Mandel}}]{2010ApJ...725L..91K}
{Kelley}, L.~Z., {Ramirez-Ruiz}, E., {Zemp}, M., {Diemand}, J., \& {Mandel}, I.
  2010, \apjl, 725, L91

\bibitem[{{Kim} {et~al.}(2004){Kim}, {Kalogera}, {Lorimer}, \&
  {White}}]{2004ApJ...616.1109K}
{Kim}, C., {Kalogera}, V., {Lorimer}, D.~R., \& {White}, T. 2004, \apj, 616,
  1109

\bibitem[{{King} {et~al.}(2007){King}, {Olsson}, \&
  {Davies}}]{2007MNRAS.374L..34K}
{King}, A., {Olsson}, E., \& {Davies}, M.~B. 2007, \mnras, 374, L34

\bibitem[{{Korol} {et~al.}(2024){Korol}, {Igoshev}, {Toonen}, {Karnesis},
  {Moore}, {Finch}, \& {Klein}}]{2024MNRAS.530..844K}
{Korol}, V., {Igoshev}, A.~P., {Toonen}, S., {et~al.} 2024, \mnras, 530, 844

\bibitem[{{Kouveliotou} {et~al.}(1993){Kouveliotou}, {Meegan}, {Fishman},
  {Bhat}, {Briggs}, {Koshut}, {Paciesas}, \& {Pendleton}}]{1993ApJ...413L.101K}
{Kouveliotou}, C., {Meegan}, C.~A., {Fishman}, G.~J., {et~al.} 1993, \apjl,
  413, L101

\bibitem[{{Kremer} {et~al.}(2018){Kremer}, {Chatterjee}, {Breivik},
  {Rodriguez}, {Larson}, \& {Rasio}}]{2018PhRvL.120s1103K}
{Kremer}, K., {Chatterjee}, S., {Breivik}, K., {et~al.} 2018, \prl, 120, 191103

\bibitem[{{Kroupa}(2001)}]{2001MNRAS.322..231K}
{Kroupa}, P. 2001, \mnras, 322, 231

\bibitem[{{Langer} \& {Norman}(2006)}]{2006ApJ...638L..63L}
{Langer}, N. \& {Norman}, C.~A. 2006, \apjl, 638, L63

\bibitem[{{Leja} {et~al.}(2019){Leja}, {Carnall}, {Johnson}, {Conroy}, \&
  {Speagle}}]{2019ApJ...876....3L}
{Leja}, J., {Carnall}, A.~C., {Johnson}, B.~D., {Conroy}, C., \& {Speagle},
  J.~S. 2019, \apj, 876, 3

\bibitem[{{Levan} {et~al.}(2024){Levan}, {Gompertz}, {Salafia}, {Bulla},
  {Burns}, {Hotokezaka}, {Izzo}, {Lamb}, {Malesani}, {Oates}, {Ravasio}, {Rouco
  Escorial}, {Schneider}, {Sarin}, {Schulze}, {Tanvir}, {Ackley}, {Anderson},
  {Brammer}, {Christensen}, {Dhillon}, {Evans}, {Fausnaugh}, {Fong},
  {Fruchter}, {Fryer}, {Fynbo}, {Gaspari}, {Heintz}, {Hjorth}, {Kennea},
  {Kennedy}, {Laskar}, {Leloudas}, {Mandel}, {Martin-Carrillo}, {Metzger},
  {Nicholl}, {Nugent}, {Palmerio}, {Pugliese}, {Rastinejad}, {Rhodes}, {Rossi},
  {Saccardi}, {Smartt}, {Stevance}, {Tohuvavohu}, {van der Horst}, {Vergani},
  {Watson}, {Barclay}, {Bhirombhakdi}, {Breedt}, {Breeveld}, {Brown},
  {Campana}, {Chrimes}, {D'Avanzo}, {D'Elia}, {De Pasquale}, {Dyer},
  {Galloway}, {Garbutt}, {Green}, {Hartmann}, {Jakobsson}, {Kerry},
  {Kouveliotou}, {Langeroodi}, {Le Floc'h}, {Leung}, {Littlefair}, {Munday},
  {O'Brien}, {Parsons}, {Pelisoli}, {Sahman}, {Salvaterra}, {Sbarufatti},
  {Steeghs}, {Tagliaferri}, {Th{\"o}ne}, {de Ugarte Postigo}, \&
  {Kann}}]{2024Natur.626..737L}
{Levan}, A.~J., {Gompertz}, B.~P., {Salafia}, O.~S., {et~al.} 2024, \nat, 626,
  737

\bibitem[{{Levan} {et~al.}(2007){Levan}, {Jakobsson}, {Hurkett}, {Tanvir},
  {Gorosabel}, {Vreeswijk}, {Rol}, {Chapman}, {Gehrels}, {O'Brien}, {Osborne},
  {Priddey}, {Kouveliotou}, {Starling}, {vanden Berk}, \&
  {Wiersema}}]{2007MNRAS.378.1439L}
{Levan}, A.~J., {Jakobsson}, P., {Hurkett}, C., {et~al.} 2007, \mnras, 378,
  1439

\bibitem[{{Levan} {et~al.}(2023){Levan}, {Malesani}, {Gompertz}, {Nugent},
  {Nicholl}, {Oates}, {Perley}, {Rastinejad}, {Metzger}, {Schulze}, {Stanway},
  {Inkenhaag}, {Zafar}, {Ag{\"u}{\'\i} Fern{\'a}ndez}, {Chrimes},
  {Bhirombhakdi}, {de Ugarte Postigo}, {Fong}, {Fruchter}, {Fragione}, {Fynbo},
  {Gaspari}, {Heintz}, {Hjorth}, {Jakobsson}, {Jonker}, {Lamb}, {Mandel},
  {Mandhai}, {Ravasio}, {Sollerman}, \& {Tanvir}}]{2023NatAs...7..976L}
{Levan}, A.~J., {Malesani}, D.~B., {Gompertz}, B.~P., {et~al.} 2023, Nature
  Astronomy, 7, 976

\bibitem[{{Levan} {et~al.}(2006){Levan}, {Wynn}, {Chapman}, {Davies}, {King},
  {Priddey}, \& {Tanvir}}]{2006MNRAS.368L...1L}
{Levan}, A.~J., {Wynn}, G.~A., {Chapman}, R., {et~al.} 2006, \mnras, 368, L1

\bibitem[{{Lloyd-Ronning} {et~al.}(2024){Lloyd-Ronning}, {Johnson}, {Upton
  Sanderbeck}, {Silva}, \& {Cheng}}]{2024MNRAS.535.2800L}
{Lloyd-Ronning}, N.~M., {Johnson}, J., {Upton Sanderbeck}, P., {Silva}, M., \&
  {Cheng}, R.~M. 2024, \mnras, 535, 2800

\bibitem[{{Lomel{\'\i}-N{\'u}{\~n}ez}
  {et~al.}(2022){Lomel{\'\i}-N{\'u}{\~n}ez}, {Mayya}, {Rodr{\'\i}guez-Merino},
  {Ovando}, \& {Rosa-Gonz{\'a}lez}}]{2022MNRAS.509..180L}
{Lomel{\'\i}-N{\'u}{\~n}ez}, L., {Mayya}, Y.~D., {Rodr{\'\i}guez-Merino},
  L.~H., {Ovando}, P.~A., \& {Rosa-Gonz{\'a}lez}, D. 2022, \mnras, 509, 180

\bibitem[{{Lyman} {et~al.}(2018){Lyman}, {Lamb}, {Levan}, {Mandel}, {Tanvir},
  {Kobayashi}, {Gompertz}, {Hjorth}, {Fruchter}, {Kangas}, {Steeghs}, {Steele},
  {Cano}, {Copperwheat}, {Evans}, {Fynbo}, {Gall}, {Im}, {Izzo}, {Jakobsson},
  {Milvang-Jensen}, {O'Brien}, {Osborne}, {Palazzi}, {Perley}, {Pian},
  {Rosswog}, {Rowlinson}, {Schulze}, {Stanway}, {Sutton}, {Th{\"o}ne}, {de
  Ugarte Postigo}, {Watson}, {Wiersema}, \& {Wijers}}]{2018NatAs...2..751L}
{Lyman}, J.~D., {Lamb}, G.~P., {Levan}, A.~J., {et~al.} 2018, Nature Astronomy,
  2, 751

\bibitem[{{Lyman} {et~al.}(2014){Lyman}, {Levan}, {Church}, {Davies}, \&
  {Tanvir}}]{2014MNRAS.444.2157L}
{Lyman}, J.~D., {Levan}, A.~J., {Church}, R.~P., {Davies}, M.~B., \& {Tanvir},
  N.~R. 2014, \mnras, 444, 2157

\bibitem[{{Madau} \& {Dickinson}(2014)}]{2014ARA&A..52..415M}
{Madau}, P. \& {Dickinson}, M. 2014, \araa, 52, 415

\bibitem[{{Mandel}(2016)}]{2016MNRAS.456..578M}
{Mandel}, I. 2016, \mnras, 456, 578

\bibitem[{{Mandel}(2021)}]{2021RNAAS...5..223M}
{Mandel}, I. 2021, Research Notes of the American Astronomical Society, 5, 223

\bibitem[{{Mandel} \& {Broekgaarden}(2022)}]{2022LRR....25....1M}
{Mandel}, I. \& {Broekgaarden}, F.~S. 2022, Living Reviews in Relativity, 25, 1

\bibitem[{{Mandhai} {et~al.}(2022){Mandhai}, {Lamb}, {Tanvir}, {Bray}, {Nixon},
  {Eyles-Ferris}, {Levan}, \& {Gompertz}}]{2022MNRAS.514.2716M}
{Mandhai}, S., {Lamb}, G.~P., {Tanvir}, N.~R., {et~al.} 2022, \mnras, 514, 2716

\bibitem[{{Margalit} \& {Metzger}(2016)}]{2016MNRAS.461.1154M}
{Margalit}, B. \& {Metzger}, B.~D. 2016, \mnras, 461, 1154

\bibitem[{{McBrien} {et~al.}(2019){McBrien}, {Smartt}, {Chen}, {Inserra},
  {Gillanders}, {Sim}, {Jerkstrand}, {Rest}, {Valenti}, {Roy}, {Gromadzki},
  {Taubenberger}, {Fl{\"o}rs}, {Huber}, {Chambers}, {Gal-Yam}, {Young},
  {Nicholl}, {Kankare}, {Smith}, {Maguire}, {Mandel}, {Prentice},
  {Rodr{\'\i}guez}, {Pineda Garcia}, {Guti{\'e}rrez}, {Galbany}, {Barbarino},
  {Clark}, {Sollerman}, {Kulkarni}, {De}, {Buckley}, \&
  {Rau}}]{2019ApJ...885L..23M}
{McBrien}, O.~R., {Smartt}, S.~J., {Chen}, T.-W., {et~al.} 2019, \apjl, 885,
  L23

\bibitem[{{McCleery} {et~al.}(2020){McCleery}, {Tremblay}, {Gentile Fusillo},
  {Hollands}, {G{\"a}nsicke}, {Izquierdo}, {Toonen}, {Cunningham}, \&
  {Rebassa-Mansergas}}]{2020MNRAS.499.1890M}
{McCleery}, J., {Tremblay}, P.-E., {Gentile Fusillo}, N.~P., {et~al.} 2020,
  \mnras, 499, 1890

\bibitem[{{Metzger}(2012)}]{2012MNRAS.419..827M}
{Metzger}, B.~D. 2012, \mnras, 419, 827

\bibitem[{{Metzger} {et~al.}(2008){Metzger}, {Quataert}, \&
  {Thompson}}]{2008MNRAS.385.1455M}
{Metzger}, B.~D., {Quataert}, E., \& {Thompson}, T.~A. 2008, \mnras, 385, 1455

\bibitem[{{Micha{\l}owski} {et~al.}(2018){Micha{\l}owski}, {Xu}, {Stevens},
  {Levan}, {Yang}, {Paragi}, {Kamble}, {Tsai}, {Dannerbauer}, {van der Horst},
  {Shao}, {Crosby}, {Gentile}, {Stanway}, {Wiersema}, {Fynbo}, {Tanvir},
  {Kamphuis}, {Garrett}, \& {Bartczak}}]{2018A&A...616A.169M}
{Micha{\l}owski}, M.~J., {Xu}, D., {Stevens}, J., {et~al.} 2018, \aap, 616,
  A169

\bibitem[{{Moe} \& {Di Stefano}(2017)}]{2017ApJS..230...15M}
{Moe}, M. \& {Di Stefano}, R. 2017, \apjs, 230, 15

\bibitem[{{Mor{\'a}n-Fraile} {et~al.}(2024){Mor{\'a}n-Fraile}, {R{\"o}pke},
  {Pakmor}, {Aloy}, {Ohlmann}, {Schneider}, {Leidi}, \&
  {Lioutas}}]{2024A&A...681A..41M}
{Mor{\'a}n-Fraile}, J., {R{\"o}pke}, F.~K., {Pakmor}, R., {et~al.} 2024, \aap,
  681, A41

\bibitem[{{Nagarajan} \& {El-Badry}(2025)}]{2025PASP..137c4203N}
{Nagarajan}, P. \& {El-Badry}, K. 2025, \pasp, 137, 034203

\bibitem[{{Narayana Bhat} {et~al.}(2016){Narayana Bhat}, {Meegan}, {von
  Kienlin}, {Paciesas}, {Briggs}, {Burgess}, {Burns}, {Chaplin}, {Cleveland},
  {Collazzi}, {Connaughton}, {Diekmann}, {Fitzpatrick}, {Gibby}, {Giles},
  {Goldstein}, {Greiner}, {Jenke}, {Kippen}, {Kouveliotou}, {Mailyan},
  {McBreen}, {Pelassa}, {Preece}, {Roberts}, {Sparke}, {Stanbro}, {Veres},
  {Wilson-Hodge}, {Xiong}, {Younes}, {Yu}, \& {Zhang}}]{2016ApJS..223...28N}
{Narayana Bhat}, P., {Meegan}, C.~A., {von Kienlin}, A., {et~al.} 2016, \apjs,
  223, 28

\bibitem[{{Nelemans} {et~al.}(2000){Nelemans}, {Verbunt}, {Yungelson}, \&
  {Portegies Zwart}}]{2000A&A...360.1011N}
{Nelemans}, G., {Verbunt}, F., {Yungelson}, L.~R., \& {Portegies Zwart}, S.~F.
  2000, \aap, 360, 1011

\bibitem[{{Nelemans} {et~al.}(2001){Nelemans}, {Yungelson}, \& {Portegies
  Zwart}}]{2001AandA...375..890N}
{Nelemans}, G., {Yungelson}, L.~R., \& {Portegies Zwart}, S.~F. 2001, \aap,
  375, 890

\bibitem[{{Nugent} {et~al.}(2024){Nugent}, {Fong}, {Castrejon}, {Leja},
  {Zevin}, \& {Ji}}]{2024ApJ...962....5N}
{Nugent}, A.~E., {Fong}, W.-f., {Castrejon}, C., {et~al.} 2024, \apj, 962, 5

\bibitem[{{Nugent} {et~al.}(2022){Nugent}, {Fong}, {Dong}, {Leja}, {Berger},
  {Zevin}, {Chornock}, {Cobb}, {Kelley}, {Kilpatrick}, {Levan}, {Margutti},
  {Paterson}, {Perley}, {Escorial}, {Smith}, \& {Tanvir}}]{2022ApJ...940...57N}
{Nugent}, A.~E., {Fong}, W.-F., {Dong}, Y., {et~al.} 2022, \apj, 940, 57

\bibitem[{{Nugent} {et~al.}(2025){Nugent}, {Ji}, {Fong}, {Shah}, \& {van de
  Voort}}]{2025ApJ...982..144N}
{Nugent}, A.~E., {Ji}, A.~P., {Fong}, W.-f., {Shah}, H., \& {van de Voort}, F.
  2025, \apj, 982, 144

\bibitem[{{O'Connor} {et~al.}(2022){O'Connor}, {Troja}, {Dichiara},
  {Beniamini}, {Cenko}, {Kouveliotou}, {Gonz{\'a}lez}, {Durbak}, {Gatkine},
  {Kutyrev}, {Sakamoto}, {S{\'a}nchez-Ram{\'\i}rez}, \&
  {Veilleux}}]{2022MNRAS.515.4890O}
{O'Connor}, B., {Troja}, E., {Dichiara}, S., {et~al.} 2022, \mnras, 515, 4890

\bibitem[{{Oke} \& {Gunn}(1983)}]{1983ApJ...266..713O}
{Oke}, J.~B. \& {Gunn}, J.~E. 1983, \apj, 266, 713

\bibitem[{{Paschalidis} {et~al.}(2009){Paschalidis}, {MacLeod}, {Baumgarte}, \&
  {Shapiro}}]{2009PhRvD..80b4006P}
{Paschalidis}, V., {MacLeod}, M., {Baumgarte}, T.~W., \& {Shapiro}, S.~L. 2009,
  \prd, 80, 024006

\bibitem[{{Paxton} {et~al.}(2011){Paxton}, {Bildsten}, {Dotter}, {Herwig},
  {Lesaffre}, \& {Timmes}}]{2011ApJS..192....3P}
{Paxton}, B., {Bildsten}, L., {Dotter}, A., {et~al.} 2011, \apjs, 192, 3

\bibitem[{{Peng} {et~al.}(2024){Peng}, {Liu}, {Zhang}, \&
  {Gao}}]{2024ApJ...967..156P}
{Peng}, Z.-k., {Liu}, Z.-k., {Zhang}, B.-B., \& {Gao}, H. 2024, \apj, 967, 156

\bibitem[{{Perets} \& {Beniamini}(2021)}]{2021MNRAS.503.5997P}
{Perets}, H.~B. \& {Beniamini}, P. 2021, \mnras, 503, 5997

\bibitem[{{Perez} \& {Granger}(2007)}]{2007CSE.....9c..21P}
{Perez}, F. \& {Granger}, B.~E. 2007, Computing in Science and Engineering, 9,
  21

\bibitem[{{Perley} {et~al.}(2016){Perley}, {Kr{\"u}hler}, {Schulze}, {de Ugarte
  Postigo}, {Hjorth}, {Berger}, {Cenko}, {Chary}, {Cucchiara}, {Ellis}, {Fong},
  {Fynbo}, {Gorosabel}, {Greiner}, {Jakobsson}, {Kim}, {Laskar}, {Levan},
  {Micha{\l}owski}, {Milvang-Jensen}, {Tanvir}, {Th{\"o}ne}, \&
  {Wiersema}}]{2016ApJ...817....7P}
{Perley}, D.~A., {Kr{\"u}hler}, T., {Schulze}, S., {et~al.} 2016, \apj, 817, 7

\bibitem[{{Petrosian} \& {Dainotti}(2024)}]{2024ApJ...963L..12P}
{Petrosian}, V. \& {Dainotti}, M.~G. 2024, \apjl, 963, L12

\bibitem[{{Piran}(2004)}]{2004RvMP...76.1143P}
{Piran}, T. 2004, Reviews of Modern Physics, 76, 1143

\bibitem[{{Planck Collaboration} {et~al.}(2020){Planck Collaboration},
  {Aghanim}, {Akrami}, {Ashdown}, {Aumont}, {Baccigalupi}, {Ballardini},
  {Banday}, {Barreiro}, {Bartolo}, {Basak}, {Battye}, {Benabed}, {Bernard},
  {Bersanelli}, {Bielewicz}, {Bock}, {Bond}, {Borrill}, {Bouchet}, {Boulanger},
  {Bucher}, {Burigana}, {Butler}, {Calabrese}, {Cardoso}, {Carron},
  {Challinor}, {Chiang}, {Chluba}, {Colombo}, {Combet}, {Contreras}, {Crill},
  {Cuttaia}, {de Bernardis}, {de Zotti}, {Delabrouille}, {Delouis}, {Di
  Valentino}, {Diego}, {Dor{\'e}}, {Douspis}, {Ducout}, {Dupac}, {Dusini},
  {Efstathiou}, {Elsner}, {En{\ss}lin}, {Eriksen}, {Fantaye}, {Farhang},
  {Fergusson}, {Fernandez-Cobos}, {Finelli}, {Forastieri}, {Frailis},
  {Fraisse}, {Franceschi}, {Frolov}, {Galeotta}, {Galli}, {Ganga},
  {G{\'e}nova-Santos}, {Gerbino}, {Ghosh}, {Gonz{\'a}lez-Nuevo}, {G{\'o}rski},
  {Gratton}, {Gruppuso}, {Gudmundsson}, {Hamann}, {Handley}, {Hansen},
  {Herranz}, {Hildebrandt}, {Hivon}, {Huang}, {Jaffe}, {Jones}, {Karakci},
  {Keih{\"a}nen}, {Keskitalo}, {Kiiveri}, {Kim}, {Kisner}, {Knox},
  {Krachmalnicoff}, {Kunz}, {Kurki-Suonio}, {Lagache}, {Lamarre}, {Lasenby},
  {Lattanzi}, {Lawrence}, {Le Jeune}, {Lemos}, {Lesgourgues}, {Levrier},
  {Lewis}, {Liguori}, {Lilje}, {Lilley}, {Lindholm}, {L{\'o}pez-Caniego},
  {Lubin}, {Ma}, {Mac{\'\i}as-P{\'e}rez}, {Maggio}, {Maino}, {Mandolesi},
  {Mangilli}, {Marcos-Caballero}, {Maris}, {Martin}, {Martinelli},
  {Mart{\'\i}nez-Gonz{\'a}lez}, {Matarrese}, {Mauri}, {McEwen}, {Meinhold},
  {Melchiorri}, {Mennella}, {Migliaccio}, {Millea}, {Mitra},
  {Miville-Desch{\^e}nes}, {Molinari}, {Montier}, {Morgante}, {Moss}, {Natoli},
  {N{\o}rgaard-Nielsen}, {Pagano}, {Paoletti}, {Partridge}, {Patanchon},
  {Peiris}, {Perrotta}, {Pettorino}, {Piacentini}, {Polastri}, {Polenta},
  {Puget}, {Rachen}, {Reinecke}, {Remazeilles}, {Renzi}, {Rocha}, {Rosset},
  {Roudier}, {Rubi{\~n}o-Mart{\'\i}n}, {Ruiz-Granados}, {Salvati}, {Sandri},
  {Savelainen}, {Scott}, {Shellard}, {Sirignano}, {Sirri}, {Spencer},
  {Sunyaev}, {Suur-Uski}, {Tauber}, {Tavagnacco}, {Tenti}, {Toffolatti},
  {Tomasi}, {Trombetti}, {Valenziano}, {Valiviita}, {Van Tent}, {Vibert},
  {Vielva}, {Villa}, {Vittorio}, {Wandelt}, {Wehus}, {White}, {White},
  {Zacchei}, \& {Zonca}}]{2020A&A...641A...6P}
{Planck Collaboration}, {Aghanim}, N., {Akrami}, Y., {et~al.} 2020, \aap, 641,
  A6

\bibitem[{{Price-Whelan} {et~al.}(2018){Price-Whelan}, {Sip{\H{o}}cz},
  {G{\"u}nther}, {Lim}, {Crawford}, {Conseil}, {Shupe}, {Craig}, {Dencheva},
  {Ginsburg}, {VanderPlas}, {Bradley}, {P{\'e}rez-Su{\'a}rez}, {de Val-Borro},
  {Paper Contributors}, {Aldcroft}, {Cruz}, {Robitaille}, {Tollerud},
  {Coordination Committee}, {Ardelean}, {Babej}, {Bach}, {Bachetti}, {Bakanov},
  {Bamford}, {Barentsen}, {Barmby}, {Baumbach}, {Berry}, {Biscani}, {Boquien},
  {Bostroem}, {Bouma}, {Brammer}, {Bray}, {Breytenbach}, {Buddelmeijer},
  {Burke}, {Calderone}, {Cano Rodr{\'\i}guez}, {Cara}, {Cardoso}, {Cheedella},
  {Copin}, {Corrales}, {Crichton}, {D{\textquoteright}Avella}, {Deil},
  {Depagne}, {Dietrich}, {Donath}, {Droettboom}, {Earl}, {Erben}, {Fabbro},
  {Ferreira}, {Finethy}, {Fox}, {Garrison}, {Gibbons}, {Goldstein}, {Gommers},
  {Greco}, {Greenfield}, {Groener}, {Grollier}, {Hagen}, {Hirst}, {Homeier},
  {Horton}, {Hosseinzadeh}, {Hu}, {Hunkeler}, {Ivezi{\'c}}, {Jain}, {Jenness},
  {Kanarek}, {Kendrew}, {Kern}, {Kerzendorf}, {Khvalko}, {King}, {Kirkby},
  {Kulkarni}, {Kumar}, {Lee}, {Lenz}, {Littlefair}, {Ma}, {Macleod},
  {Mastropietro}, {McCully}, {Montagnac}, {Morris}, {Mueller}, {Mumford},
  {Muna}, {Murphy}, {Nelson}, {Nguyen}, {Ninan}, {N{\"o}the}, {Ogaz}, {Oh},
  {Parejko}, {Parley}, {Pascual}, {Patil}, {Patil}, {Plunkett}, {Prochaska},
  {Rastogi}, {Reddy Janga}, {Sabater}, {Sakurikar}, {Seifert}, {Sherbert},
  {Sherwood-Taylor}, {Shih}, {Sick}, {Silbiger}, {Singanamalla}, {Singer},
  {Sladen}, {Sooley}, {Sornarajah}, {Streicher}, {Teuben}, {Thomas},
  {Tremblay}, {Turner}, {Terr{\'o}n}, {van Kerkwijk}, {de la Vega}, {Watkins},
  {Weaver}, {Whitmore}, {Woillez}, {Zabalza}, \& {Contributors}}]{astropy:2018}
{Price-Whelan}, A.~M., {Sip{\H{o}}cz}, B.~M., {G{\"u}nther}, H.~M., {et~al.}
  2018, \aj, 156, 123

\bibitem[{{Rastinejad} {et~al.}(2025){Rastinejad}, {Fong}, {Kilpatrick},
  {Nicholl}, \& {Metzger}}]{2025ApJ...979..190R}
{Rastinejad}, J.~C., {Fong}, W., {Kilpatrick}, C.~D., {Nicholl}, M., \&
  {Metzger}, B.~D. 2025, \apj, 979, 190

\bibitem[{{Rastinejad} {et~al.}(2021){Rastinejad}, {Fong}, {Kilpatrick},
  {Paterson}, {Tanvir}, {Levan}, {Metzger}, {Berger}, {Chornock}, {Cobb},
  {Laskar}, {Milne}, {Nugent}, \& {Smith}}]{2021ApJ...916...89R}
{Rastinejad}, J.~C., {Fong}, W., {Kilpatrick}, C.~D., {et~al.} 2021, \apj, 916,
  89

\bibitem[{{Rastinejad} {et~al.}(2022){Rastinejad}, {Gompertz}, {Levan}, {Fong},
  {Nicholl}, {Lamb}, {Malesani}, {Nugent}, {Oates}, {Tanvir}, {de Ugarte
  Postigo}, {Kilpatrick}, {Moore}, {Metzger}, {Ravasio}, {Rossi}, {Schroeder},
  {Jencson}, {Sand}, {Smith}, {Ag{\"u}{\'\i} Fern{\'a}ndez}, {Berger},
  {Blanchard}, {Chornock}, {Cobb}, {De Pasquale}, {Fynbo}, {Izzo}, {Kann},
  {Laskar}, {Marini}, {Paterson}, {Escorial}, {Sears}, \&
  {Th{\"o}ne}}]{2022Natur.612..223R}
{Rastinejad}, J.~C., {Gompertz}, B.~P., {Levan}, A.~J., {et~al.} 2022, \nat,
  612, 223

\bibitem[{{Repetto} {et~al.}(2017){Repetto}, {Igoshev}, \&
  {Nelemans}}]{2017MNRAS.467..298R}
{Repetto}, S., {Igoshev}, A.~P., \& {Nelemans}, G. 2017, \mnras, 467, 298

\bibitem[{{Richards} {et~al.}(2023){Richards}, {Eldridge}, {Briel}, {Stevance},
  \& {Willcox}}]{2023MNRAS.522.3972R}
{Richards}, S.~M., {Eldridge}, J.~J., {Briel}, M.~M., {Stevance}, H.~F., \&
  {Willcox}, R. 2023, \mnras, 522, 3972

\bibitem[{{Riess} {et~al.}(2018){Riess}, {Casertano}, {Yuan}, {Macri},
  {Bucciarelli}, {Lattanzi}, {MacKenty}, {Bowers}, {Zheng}, {Filippenko},
  {Huang}, \& {Anderson}}]{2018ApJ...861..126R}
{Riess}, A.~G., {Casertano}, S., {Yuan}, W., {et~al.} 2018, \apj, 861, 126

\bibitem[{{Rodrigo} \& {Solano}(2020)}]{2020sea..confE.182R}
{Rodrigo}, C. \& {Solano}, E. 2020, in XIV.0 Scientific Meeting (virtual) of
  the Spanish Astronomical Society, 182

\bibitem[{{Rodrigo} {et~al.}(2012){Rodrigo}, {Solano}, \&
  {Bayo}}]{2012ivoa.rept.1015R}
{Rodrigo}, C., {Solano}, E., \& {Bayo}, A. 2012, {SVO Filter Profile Service
  Version 1.0}, IVOA Working Draft 15 October 2012

\bibitem[{{Rossi} {et~al.}(2020){Rossi}, {Stratta}, {Maiorano}, {Spighi},
  {Masetti}, {Palazzi}, {Gardini}, {Melandri}, {Nicastro}, {Pian}, {Branchesi},
  {Dadina}, {Testa}, {Brocato}, {Benetti}, {Ciolfi}, {Covino}, {D'Elia},
  {Grado}, {Izzo}, {Perego}, {Piranomonte}, {Salvaterra}, {Selsing},
  {Tomasella}, {Yang}, {Vergani}, {Amati}, \& {Stephen}}]{2020MNRAS.493.3379R}
{Rossi}, A., {Stratta}, G., {Maiorano}, E., {et~al.} 2020, \mnras, 493, 3379

\bibitem[{{Rosswog}(2007)}]{2007MNRAS.376L..48R}
{Rosswog}, S. 2007, \mnras, 376, L48

\bibitem[{{Ruiter} {et~al.}(2019){Ruiter}, {Ferrario}, {Belczynski},
  {Seitenzahl}, {Crocker}, \& {Karakas}}]{2019MNRAS.484..698R}
{Ruiter}, A.~J., {Ferrario}, L., {Belczynski}, K., {et~al.} 2019, \mnras, 484,
  698

\bibitem[{{Sadowski} {et~al.}(2008){Sadowski}, {Belczynski}, {Bulik},
  {Ivanova}, {Rasio}, \& {O'Shaughnessy}}]{2008ApJ...676.1162S}
{Sadowski}, A., {Belczynski}, K., {Bulik}, T., {et~al.} 2008, \apj, 676, 1162

\bibitem[{{Salafia} {et~al.}(2023){Salafia}, {Ravasio}, {Ghirlanda}, \&
  {Mandel}}]{2023A&A...680A..45S}
{Salafia}, O.~S., {Ravasio}, M.~E., {Ghirlanda}, G., \& {Mandel}, I. 2023,
  \aap, 680, A45

\bibitem[{{S{\'e}rsic}(1963)}]{1963BAAA....6...41S}
{S{\'e}rsic}, J.~L. 1963, Boletin de la Asociacion Argentina de Astronomia La
  Plata Argentina, 6, 41

\bibitem[{{Shao} \& {Li}(2021)}]{2021ApJ...920...81S}
{Shao}, Y. \& {Li}, X.-D. 2021, \apj, 920, 81

\bibitem[{{Shara} \& {Hurley}(2006)}]{2006ApJ...646..464S}
{Shara}, M.~M. \& {Hurley}, J.~R. 2006, \apj, 646, 464

\bibitem[{{Skibba} {et~al.}(2009){Skibba}, {Bamford}, {Nichol}, {Lintott},
  {Andreescu}, {Edmondson}, {Murray}, {Raddick}, {Schawinski}, {Slosar},
  {Szalay}, {Thomas}, \& {Vandenberg}}]{2009MNRAS.399..966S}
{Skibba}, R.~A., {Bamford}, S.~P., {Nichol}, R.~C., {et~al.} 2009, \mnras, 399,
  966

\bibitem[{{Spera} {et~al.}(2015){Spera}, {Mapelli}, \&
  {Bressan}}]{2015MNRAS.451.4086S}
{Spera}, M., {Mapelli}, M., \& {Bressan}, A. 2015, \mnras, 451, 4086

\bibitem[{{Spitler} {et~al.}(2008){Spitler}, {Forbes}, {Strader}, {Brodie}, \&
  {Gallagher}}]{2008MNRAS.385..361S}
{Spitler}, L.~R., {Forbes}, D.~A., {Strader}, J., {Brodie}, J.~P., \&
  {Gallagher}, J.~S. 2008, \mnras, 385, 361

\bibitem[{{Stanway} {et~al.}(2020){Stanway}, {Chrimes}, {Eldridge}, \&
  {Stevance}}]{2020MNRAS.495.4605S}
{Stanway}, E.~R., {Chrimes}, A.~A., {Eldridge}, J.~J., \& {Stevance}, H.~F.
  2020, \mnras, 495, 4605

\bibitem[{{Stanway} \& {Eldridge}(2018)}]{2018MNRAS.479...75S}
{Stanway}, E.~R. \& {Eldridge}, J.~J. 2018, \mnras, 479, 75

\bibitem[{{Stanway} \& {Eldridge}(2023)}]{2023MNRAS.522.4430S}
{Stanway}, E.~R. \& {Eldridge}, J.~J. 2023, \mnras, 522, 4430

\bibitem[{{Stevance} {et~al.}(2020){Stevance}, {Eldridge}, \&
  {Stanway}}]{2020JOSS....5.1987S}
{Stevance}, H., {Eldridge}, J., \& {Stanway}, E. 2020, The Journal of Open
  Source Software, 5, 1987

\bibitem[{{Stevance} \& {Eldridge}(2021)}]{2021MNRAS.504L..51S}
{Stevance}, H.~F. \& {Eldridge}, J.~J. 2021, \mnras, 504, L51

\bibitem[{{Stevance} {et~al.}(2023){Stevance}, {Eldridge}, {Stanway}, {Lyman},
  {McLeod}, \& {Levan}}]{2023NatAs...7..444S}
{Stevance}, H.~F., {Eldridge}, J.~J., {Stanway}, E.~R., {et~al.} 2023, Nature
  Astronomy, 7, 444

\bibitem[{{Stratta} {et~al.}(2024){Stratta}, {Nicuesa Guelbenzu}, {Klose},
  {Rossi}, {Singh}, {Palazzi}, {Guidorzi}, {Camisasca}, {Bernuzzi}, {Rau},
  {Bulla}, {Ragosta}, {Maiorano}, \& {Paris}}]{2024arXiv241204059S}
{Stratta}, G., {Nicuesa Guelbenzu}, A.~M., {Klose}, S., {et~al.} 2024, arXiv
  e-prints, arXiv:2412.04059

\bibitem[{{Tang} {et~al.}(2024{\natexlab{a}}){Tang}, {Eldridge}, {Meyer},
  {Lamberts}, {Boileau}, \& {van Zeist}}]{2024MNRAS.534.1707T}
{Tang}, P., {Eldridge}, J.~J., {Meyer}, R., {et~al.} 2024{\natexlab{a}},
  \mnras, 534, 1707

\bibitem[{{Tang} {et~al.}(2024{\natexlab{b}}){Tang}, {Meyer}, \&
  {Eldridge}}]{2024arXiv241102563T}
{Tang}, P., {Meyer}, R., \& {Eldridge}, J. 2024{\natexlab{b}}, arXiv e-prints,
  arXiv:2411.02563

\bibitem[{{Tanvir} {et~al.}(2013){Tanvir}, {Levan}, {Fruchter}, {Hjorth},
  {Hounsell}, {Wiersema}, \& {Tunnicliffe}}]{2013Natur.500..547T}
{Tanvir}, N.~R., {Levan}, A.~J., {Fruchter}, A.~S., {et~al.} 2013, \nat, 500,
  547

\bibitem[{{Tanvir} {et~al.}(2017){Tanvir}, {Levan},
  {Gonz{\'a}lez-Fern{\'a}ndez}, {Korobkin}, {Mandel}, {Rosswog}, {Hjorth},
  {D'Avanzo}, {Fruchter}, {Fryer}, {Kangas}, {Milvang-Jensen}, {Rosetti},
  {Steeghs}, {Wollaeger}, {Cano}, {Copperwheat}, {Covino}, {D'Elia}, {de Ugarte
  Postigo}, {Evans}, {Even}, {Fairhurst}, {Figuera Jaimes}, {Fontes}, {Fujii},
  {Fynbo}, {Gompertz}, {Greiner}, {Hodosan}, {Irwin}, {Jakobsson},
  {J{\o}rgensen}, {Kann}, {Lyman}, {Malesani}, {McMahon}, {Melandri},
  {O'Brien}, {Osborne}, {Palazzi}, {Perley}, {Pian}, {Piranomonte}, {Rabus},
  {Rol}, {Rowlinson}, {Schulze}, {Sutton}, {Th{\"o}ne}, {Ulaczyk}, {Watson},
  {Wiersema}, \& {Wijers}}]{2017ApJ...848L..27T}
{Tanvir}, N.~R., {Levan}, A.~J., {Gonz{\'a}lez-Fern{\'a}ndez}, C., {et~al.}
  2017, \apjl, 848, L27

\bibitem[{{Tauris} {et~al.}(2017){Tauris}, {Kramer}, {Freire}, {Wex}, {Janka},
  {Langer}, {Podsiadlowski}, {Bozzo}, {Chaty}, {Kruckow}, {van den Heuvel},
  {Antoniadis}, {Breton}, \& {Champion}}]{2017ApJ...846..170T}
{Tauris}, T.~M., {Kramer}, M., {Freire}, P.~C.~C., {et~al.} 2017, \apj, 846,
  170

\bibitem[{{Tauris} \& {Takens}(1998)}]{1998A&A...330.1047T}
{Tauris}, T.~M. \& {Takens}, R.~J. 1998, \aap, 330, 1047

\bibitem[{{Thomas} {et~al.}(2009){Thomas}, {Saglia}, {Bender}, {Thomas},
  {Gebhardt}, {Magorrian}, {Corsini}, \& {Wegner}}]{2009ApJ...691..770T}
{Thomas}, J., {Saglia}, R.~P., {Bender}, R., {et~al.} 2009, \apj, 691, 770

\bibitem[{{Thorne} \& {Zytkow}(1977)}]{1977ApJ...212..832T}
{Thorne}, K.~S. \& {Zytkow}, A.~N. 1977, \apj, 212, 832

\bibitem[{{Toonen} {et~al.}(2012){Toonen}, {Nelemans}, \& {Portegies
  Zwart}}]{2012A&A...546A..70T}
{Toonen}, S., {Nelemans}, G., \& {Portegies Zwart}, S. 2012, \aap, 546, A70

\bibitem[{{Toonen} {et~al.}(2018){Toonen}, {Perets}, {Igoshev}, {Michaely}, \&
  {Zenati}}]{2018AandA...619A..53T}
{Toonen}, S., {Perets}, H.~B., {Igoshev}, A.~P., {Michaely}, E., \& {Zenati},
  Y. 2018, \aap, 619, A53

\bibitem[{{Troja} {et~al.}(2022){Troja}, {Fryer}, {O'Connor}, {Ryan},
  {Dichiara}, {Kumar}, {Ito}, {Gupta}, {Wollaeger}, {Norris}, {Kawai},
  {Butler}, {Aryan}, {Misra}, {Hosokawa}, {Murata}, {Niwano}, {Pandey},
  {Kutyrev}, {van Eerten}, {Chase}, {Hu}, {Caballero-Garcia}, \&
  {Castro-Tirado}}]{2022Natur.612..228T}
{Troja}, E., {Fryer}, C.~L., {O'Connor}, B., {et~al.} 2022, \nat, 612, 228

\bibitem[{{Tunnicliffe} {et~al.}(2014){Tunnicliffe}, {Levan}, {Tanvir},
  {Rowlinson}, {Perley}, {Bloom}, {Cenko}, {O'Brien}, {Cobb}, {Wiersema},
  {Malesani}, {de Ugarte Postigo}, {Hjorth}, {Fynbo}, \&
  {Jakobsson}}]{2014MNRAS.437.1495T}
{Tunnicliffe}, R.~L., {Levan}, A.~J., {Tanvir}, N.~R., {et~al.} 2014, \mnras,
  437, 1495

\bibitem[{{van Haaften} {et~al.}(2012){van Haaften}, {Nelemans}, {Voss},
  {Wood}, \& {Kuijpers}}]{2012A&A...537A.104V}
{van Haaften}, L.~M., {Nelemans}, G., {Voss}, R., {Wood}, M.~A., \& {Kuijpers},
  J. 2012, \aap, 537, A104

\bibitem[{{van Marle} {et~al.}(2006){van Marle}, {Langer}, {Achterberg}, \&
  {Garc{\'\i}a-Segura}}]{2006A&A...460..105V}
{van Marle}, A.~J., {Langer}, N., {Achterberg}, A., \& {Garc{\'\i}a-Segura}, G.
  2006, \aap, 460, 105

\bibitem[{{van Son} {et~al.}(2025){van Son}, {Roy}, {Mandel}, {Farr}, {Lam},
  {Merritt}, {Broekgaarden}, {Sander}, \& {Andrews}}]{2025ApJ...979..209V}
{van Son}, L.~A.~C., {Roy}, S.~K., {Mandel}, I., {et~al.} 2025, \apj, 979, 209

\bibitem[{{van Zeist} {et~al.}(2023){van Zeist}, {Eldridge}, \&
  {Tang}}]{2023MNRAS.524.2836V}
{van Zeist}, W. G.~J., {Eldridge}, J.~J., \& {Tang}, P.~N. 2023, \mnras, 524,
  2836

\bibitem[{{van Zeist} {et~al.}(2024){van Zeist}, {Nelemans}, {Portegies Zwart},
  \& {Eldridge}}]{2024A&A...691A.316V}
{van Zeist}, W. G.~J., {Nelemans}, G., {Portegies Zwart}, S.~F., \& {Eldridge},
  J.~J. 2024, \aap, 691, A316

\bibitem[{{van Zeist} {et~al.}(2025){van Zeist}, {van Roestel}, {Nelemans},
  {Eldridge}, {Korol}, \& {Toonen}}]{2025A&A...699A.172V}
{van Zeist}, W. G.~J., {van Roestel}, J., {Nelemans}, G., {et~al.} 2025, \aap,
  699, A172

\bibitem[{{Verbunt} {et~al.}(2017){Verbunt}, {Igoshev}, \&
  {Cator}}]{2017A&A...608A..57V}
{Verbunt}, F., {Igoshev}, A., \& {Cator}, E. 2017, \aap, 608, A57

\bibitem[{{Veres} {et~al.}(2023){Veres}, {Bhat}, {Burns}, {Hamburg}, {Fraija},
  {Kocevski}, {Preece}, {Poolakkil}, {Christensen}, {Bizouard}, {Dal Canton},
  {Bala}, {Bissaldi}, {Briggs}, {Cleveland}, {Goldstein}, {Hristov}, {Hui},
  {Lesage}, {Mailyan}, {Roberts}, \& {Wilson-Hodge}}]{2023ApJ...954L...5V}
{Veres}, P., {Bhat}, P.~N., {Burns}, E., {et~al.} 2023, \apjl, 954, L5

\bibitem[{{Vink}(2008)}]{2008NewAR..52..419V}
{Vink}, J.~S. 2008, \nar, 52, 419

\bibitem[{{Vink} \& {de Koter}(2005)}]{2005A&A...442..587V}
{Vink}, J.~S. \& {de Koter}, A. 2005, \aap, 442, 587

\bibitem[{{Virtanen} {et~al.}(2020){Virtanen}, {Gommers}, {Oliphant},
  {Haberland}, {Reddy}, {Cournapeau}, {Burovski}, {Peterson}, {Weckesser},
  {Bright}, {van der Walt}, {Brett}, {Wilson}, {Millman}, {Mayorov}, {Nelson},
  {Jones}, {Kern}, {Larson}, {Carey}, {Polat}, {Feng}, {Moore}, {VanderPlas},
  {Laxalde}, {Perktold}, {Cimrman}, {Henriksen}, {Quintero}, {Harris},
  {Archibald}, {Ribeiro}, {Pedregosa}, {van Mulbregt}, \& {SciPy 1. 0
  Contributors}}]{2020NatMe..17..261V}
{Virtanen}, P., {Gommers}, R., {Oliphant}, T.~E., {et~al.} 2020, Nature
  Methods, 17, 261

\bibitem[{{von Kienlin} {et~al.}(2020){von Kienlin}, {Meegan}, {Paciesas},
  {Bhat}, {Bissaldi}, {Briggs}, {Burns}, {Cleveland}, {Gibby}, {Giles},
  {Goldstein}, {Hamburg}, {Hui}, {Kocevski}, {Mailyan}, {Malacaria},
  {Poolakkil}, {Preece}, {Roberts}, {Veres}, \&
  {Wilson-Hodge}}]{2020ApJ...893...46V}
{von Kienlin}, A., {Meegan}, C.~A., {Paciesas}, W.~S., {et~al.} 2020, \apj,
  893, 46

\bibitem[{{Wagg} {et~al.}(2024){Wagg}, {Breivik}, {Renzo}, \&
  {Price-Whelan}}]{2024arXiv240904543W}
{Wagg}, T., {Breivik}, K., {Renzo}, M., \& {Price-Whelan}, A.~M. 2024, arXiv
  e-prints, arXiv:2409.04543

\bibitem[{{Wang} {et~al.}(2024{\natexlab{a}}){Wang}, {L{\"u}}, {Yuan}, {Yuan},
  {Rice}, {Chen}, \& {Liang}}]{2024ApJ...963..156W}
{Wang}, S.-N., {L{\"u}}, H.-J., {Yuan}, Y., {et~al.} 2024{\natexlab{a}}, \apj,
  963, 156

\bibitem[{{Wang} {et~al.}(2024{\natexlab{b}}){Wang}, {Yu}, {Ren}, {Yang},
  {Zou}, \& {Zhu}}]{2024ApJ...964L...9W}
{Wang}, X.~I., {Yu}, Y.-W., {Ren}, J., {et~al.} 2024{\natexlab{b}}, \apjl, 964,
  L9

\bibitem[{{Washabaugh} \& {Bregman}(2013)}]{2013ApJ...762....1W}
{Washabaugh}, P.~C. \& {Bregman}, J.~N. 2013, \apj, 762, 1

\bibitem[{Waskom(2021)}]{Waskom2021}
Waskom, M.~L. 2021, Journal of Open Source Software, 6, 3021

\bibitem[{{Weatherford} {et~al.}(2023){Weatherford}, {K{\i}ro{\u{g}}lu},
  {Fragione}, {Chatterjee}, {Kremer}, \& {Rasio}}]{2023ApJ...946..104W}
{Weatherford}, N.~C., {K{\i}ro{\u{g}}lu}, F., {Fragione}, G., {et~al.} 2023,
  \apj, 946, 104

\bibitem[{{Wiggins} {et~al.}(2018){Wiggins}, {Fryer}, {Smidt}, {Hartmann},
  {Lloyd-Ronning}, \& {Belcynski}}]{2018ApJ...865...27W}
{Wiggins}, B.~K., {Fryer}, C.~L., {Smidt}, J.~M., {et~al.} 2018, \apj, 865, 27

\bibitem[{{Woosley}(1993)}]{1993ApJ...405..273W}
{Woosley}, S.~E. 1993, \apj, 405, 273

\bibitem[{{Yang} {et~al.}(2022){Yang}, {Ai}, {Zhang}, {Zhang}, {Liu}, {Wang},
  {Yang}, {Yin}, {Li}, \& {L{\"u}}}]{2022Natur.612..232Y}
{Yang}, J., {Ai}, S., {Zhang}, B.-B., {et~al.} 2022, \nat, 612, 232

\bibitem[{{Yang} {et~al.}(2024){Yang}, {Troja}, {O'Connor}, {Fryer}, {Im},
  {Durbak}, {Paek}, {Ricci}, {Bom}, {Gillanders}, {Castro-Tirado}, {Peng},
  {Dichiara}, {Ryan}, {van Eerten}, {Dai}, {Chang}, {Choi}, {De}, {Hu},
  {Kilpatrick}, {Kutyrev}, {Jeong}, {Lee}, {Makler}, {Navarete}, \&
  {P{\'e}rez-Garc{\'\i}a}}]{2024Natur.626..742Y}
{Yang}, Y.-H., {Troja}, E., {O'Connor}, B., {et~al.} 2024, \nat, 626, 742

\bibitem[{{Ye} {et~al.}(2020){Ye}, {Fong}, {Kremer}, {Rodriguez}, {Chatterjee},
  {Fragione}, \& {Rasio}}]{2020ApJ...888L..10Y}
{Ye}, C.~S., {Fong}, W.-f., {Kremer}, K., {et~al.} 2020, \apjl, 888, L10

\bibitem[{{Zenati} {et~al.}(2020){Zenati}, {Bobrick}, \&
  {Perets}}]{2020MNRAS.493.3956Z}
{Zenati}, Y., {Bobrick}, A., \& {Perets}, H.~B. 2020, \mnras, 493, 3956

\bibitem[{{Zenati} {et~al.}(2019){Zenati}, {Perets}, \&
  {Toonen}}]{2019MNRAS.486.1805Z}
{Zenati}, Y., {Perets}, H.~B., \& {Toonen}, S. 2019, \mnras, 486, 1805

\bibitem[{{Zhang}(2024)}]{2025arXiv250100239Z}
{Zhang}, B. 2024, arXiv e-prints, arXiv:2501.00239

\bibitem[{{Zhong} {et~al.}(2023){Zhong}, {Li}, \& {Dai}}]{2023ApJ...947L..21Z}
{Zhong}, S.-Q., {Li}, L., \& {Dai}, Z.-G. 2023, \apjl, 947, L21

\end{thebibliography}
%

\begin{appendix} 
\section{Additional figures} \label{app:A}
Here we provide equivalents of Figures \ref{fig:delay_sep} and \ref{fig:vsystdelay}, but using \citet{2005MNRAS.360..974H} and \citet{2016MNRAS.461.3747B} natal kicks.

\begin{figure*}[!ht]
\centering
\includegraphics[width=0.49\textwidth]{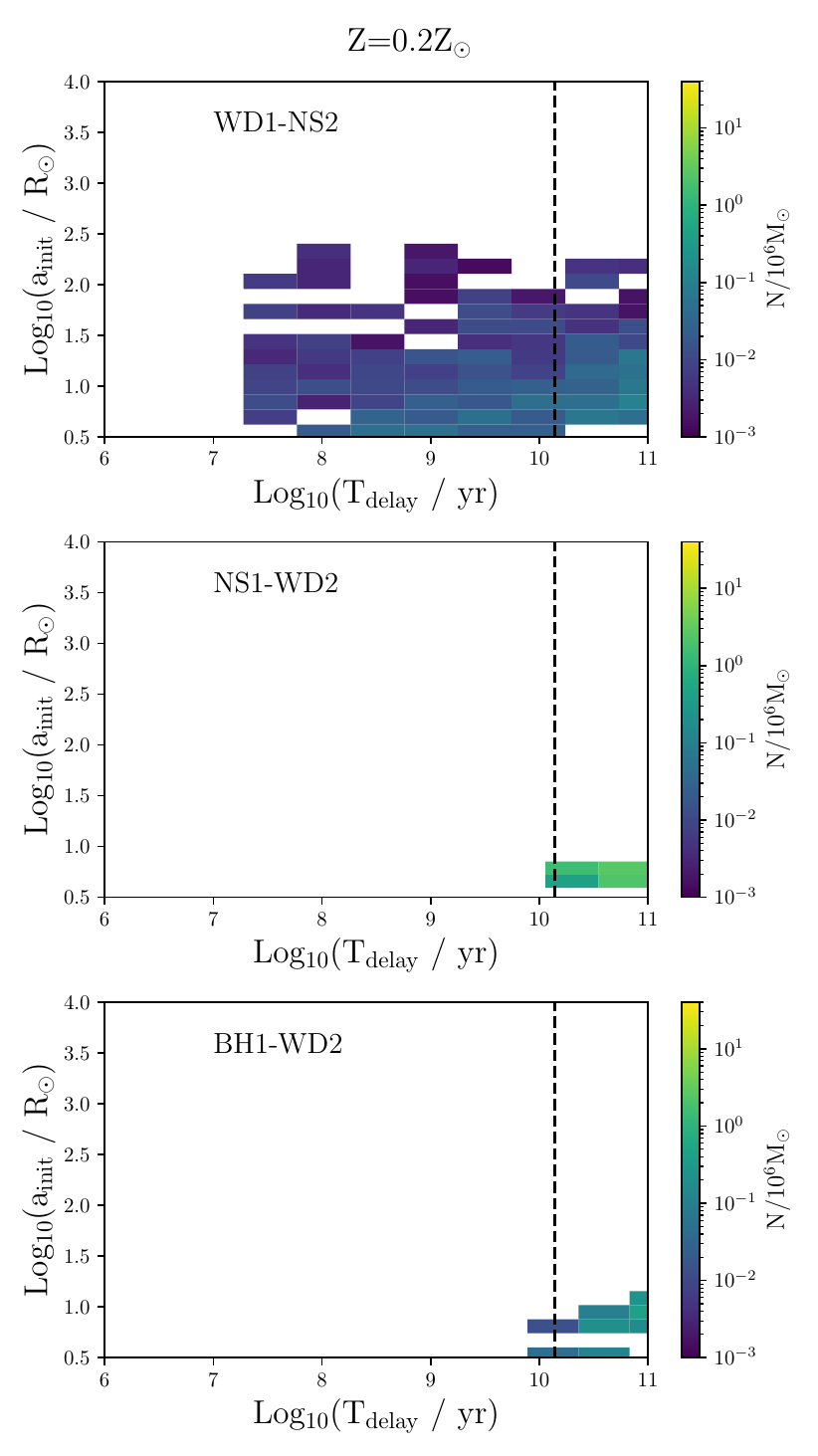}
\includegraphics[width=0.49\textwidth]{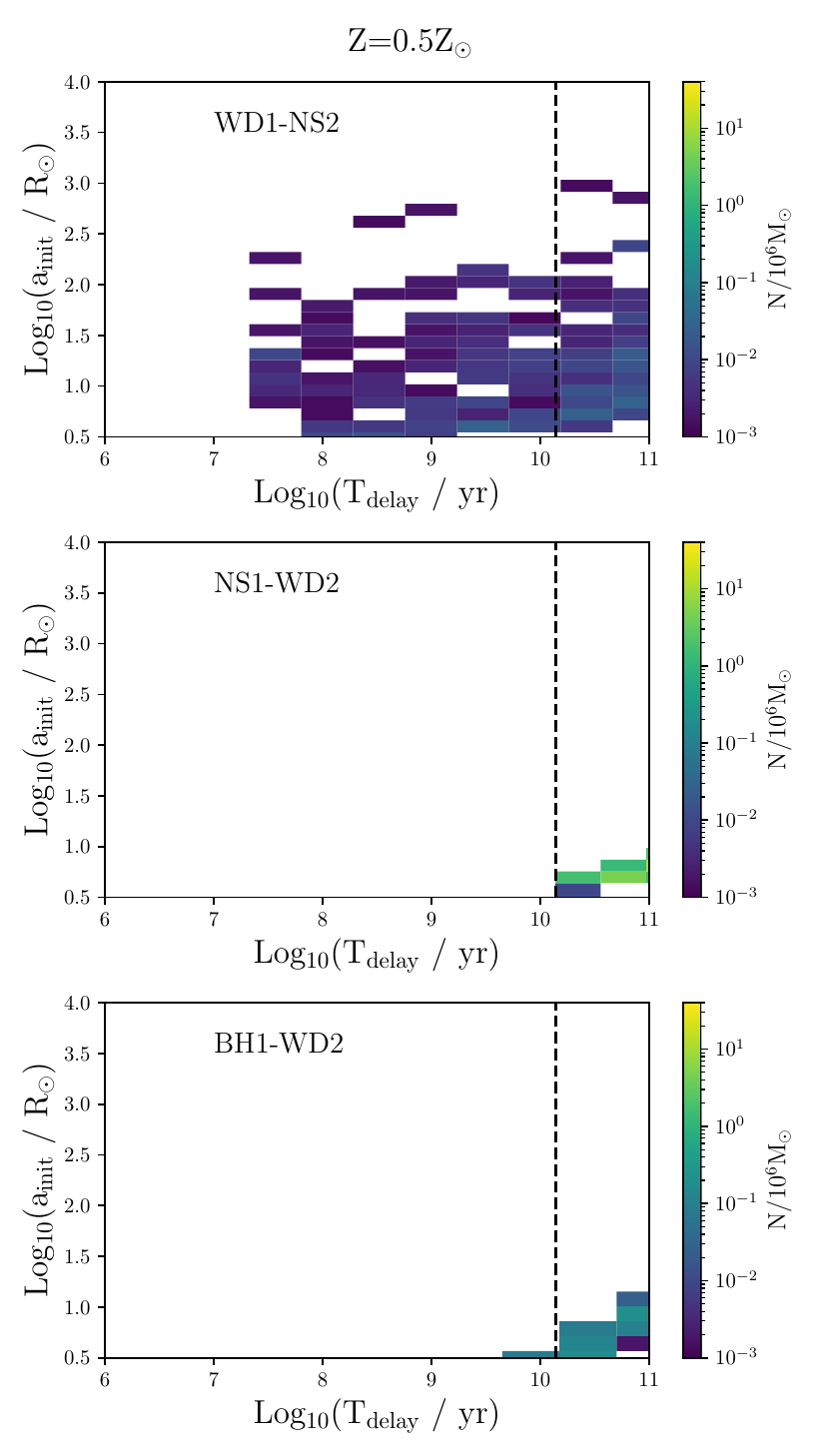}
\caption{As for Figure \ref{fig:delay_sep} but with the natal kicks of \citet{2005MNRAS.360..974H}.}
\label{fig:delay_sep_hobbs}
\end{figure*}

\begin{figure*}[!ht]
\centering
\includegraphics[width=0.49\textwidth]{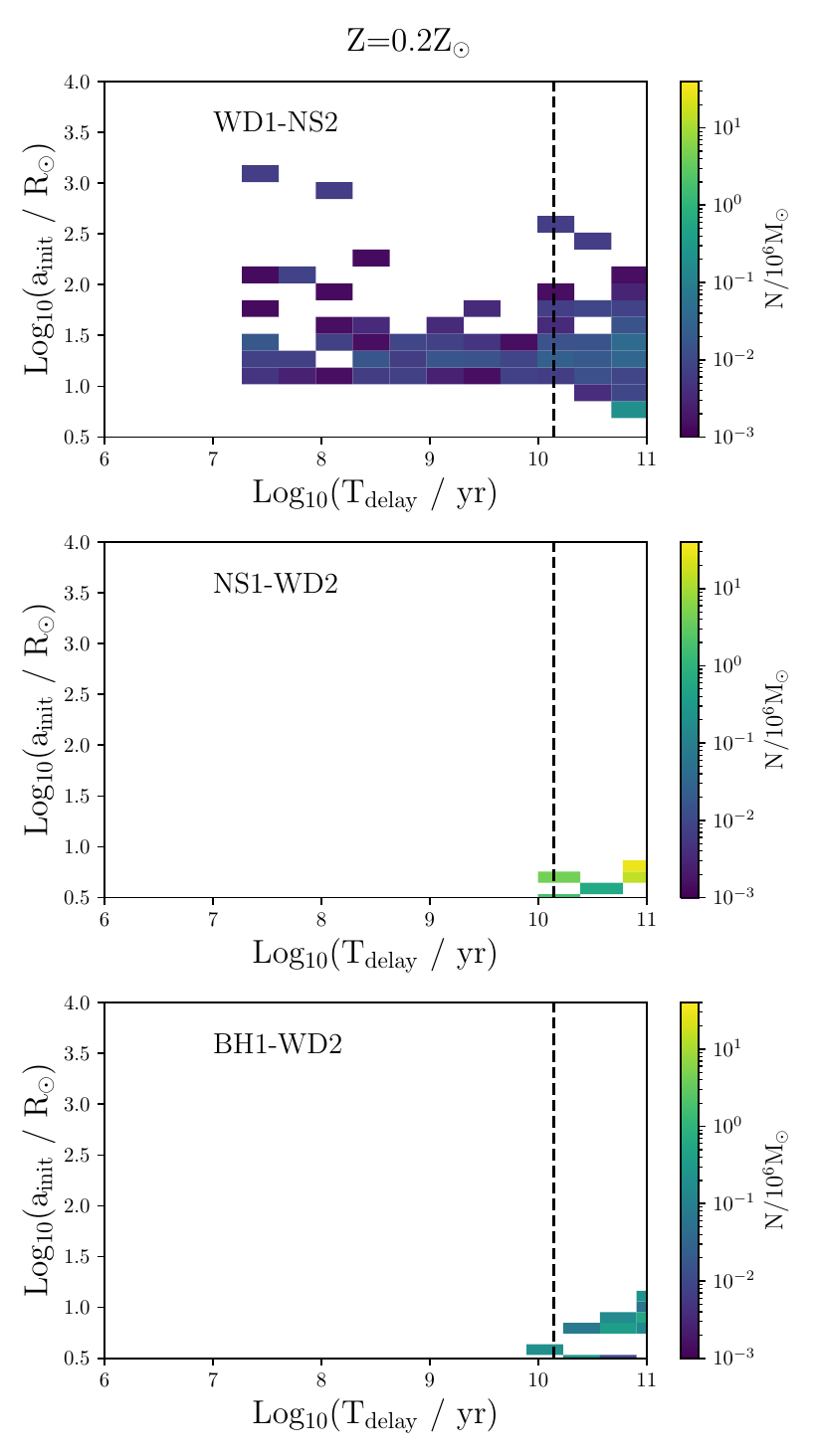}
\includegraphics[width=0.49\textwidth]{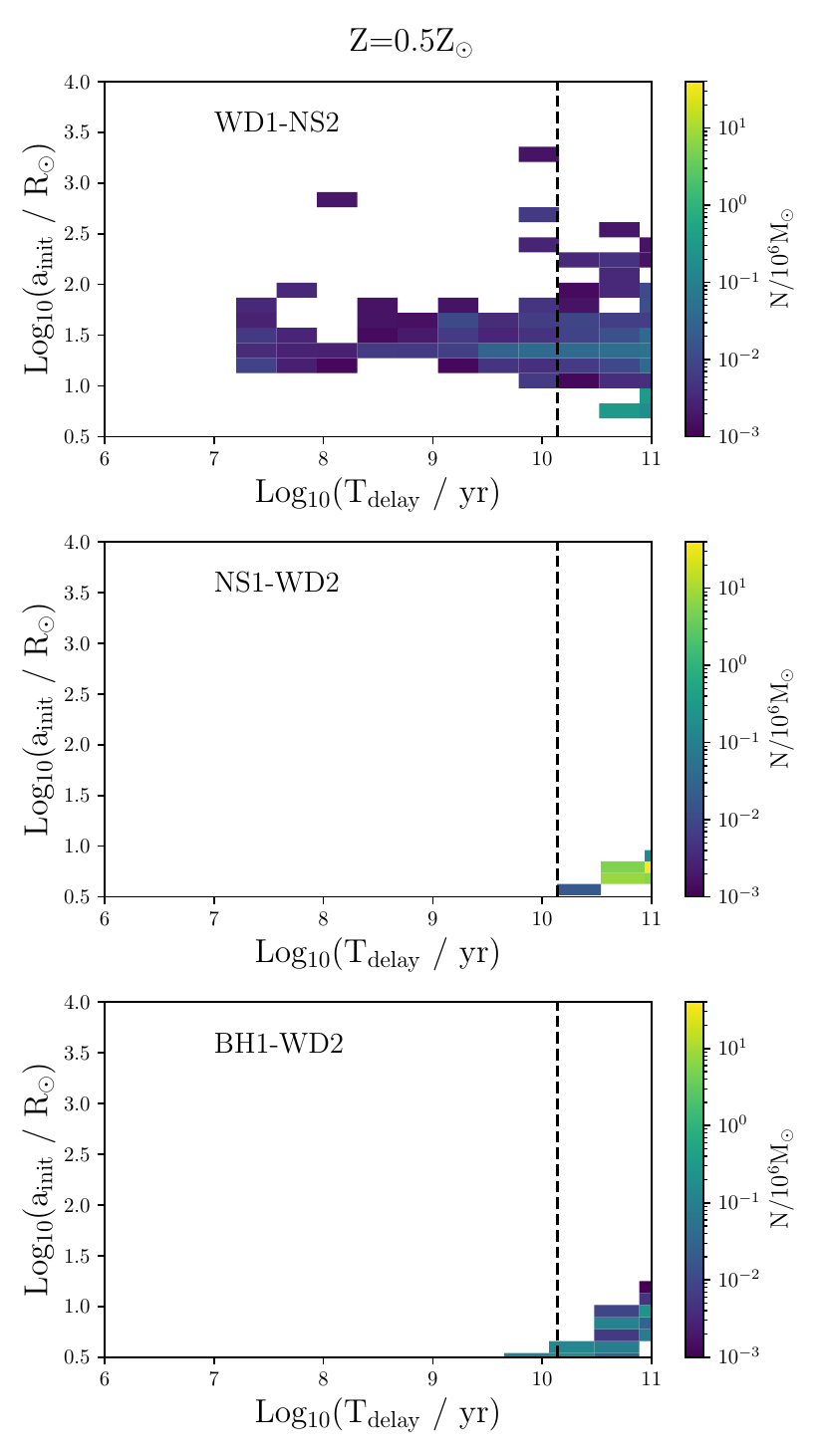}
\caption{As for Figure \ref{fig:delay_sep} but with the natal kick prescription of \citet{2016MNRAS.461.3747B}.}
\label{fig:delay_sep_bray}
\end{figure*}

\begin{figure*}[!ht]
\centering
\includegraphics[width=\textwidth]{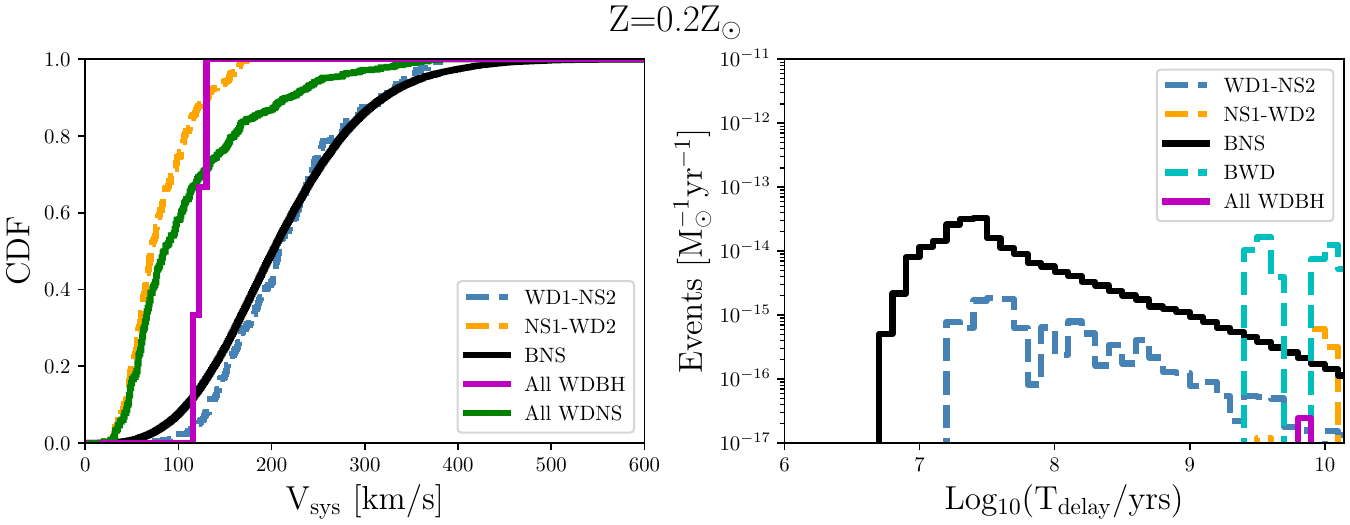}
\includegraphics[width=\textwidth]{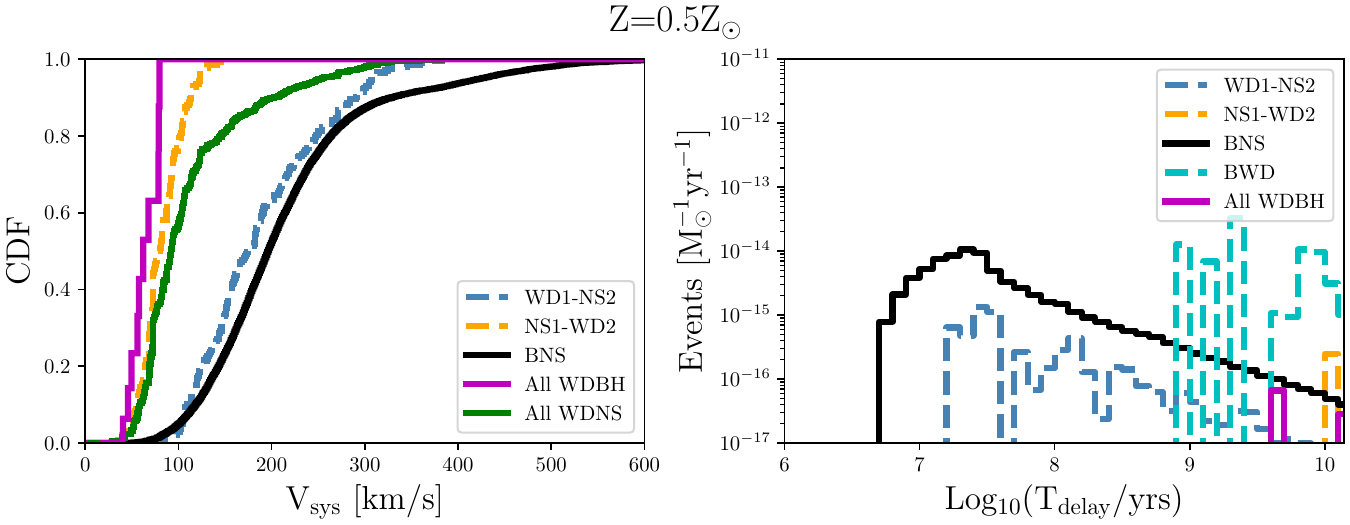}
\caption{As for Figure \ref{fig:vsystdelay} but with the natal kick distribution of \citet{2005MNRAS.360..974H}.}
\label{fig:apx_hobbs}
\end{figure*}

\begin{figure*}[!ht]
\centering
\includegraphics[width=\textwidth]{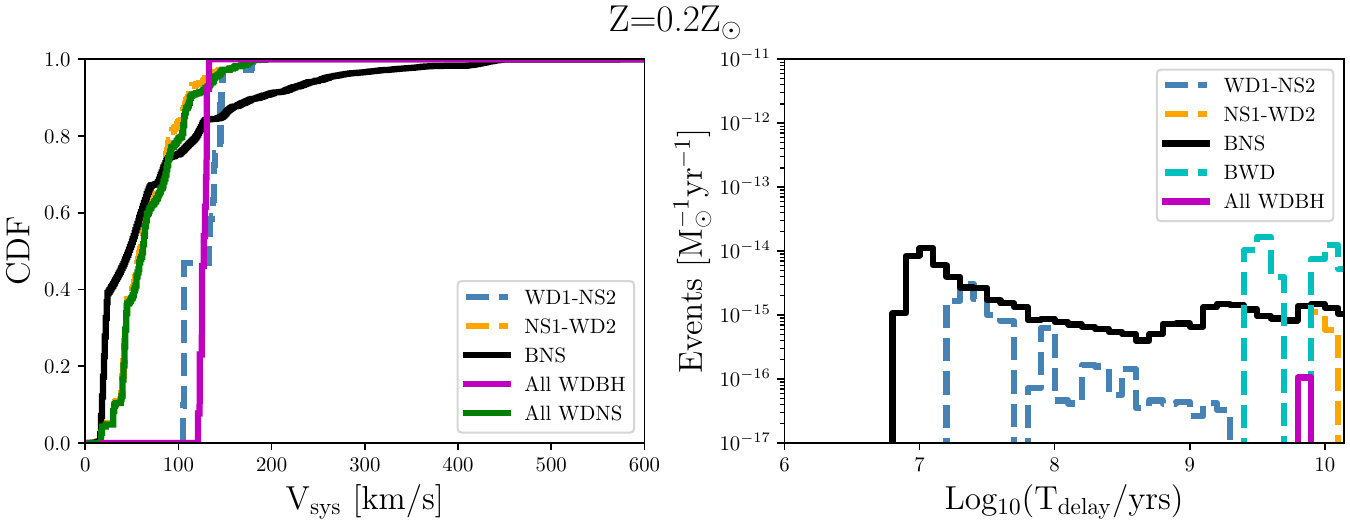}
\includegraphics[width=\textwidth]{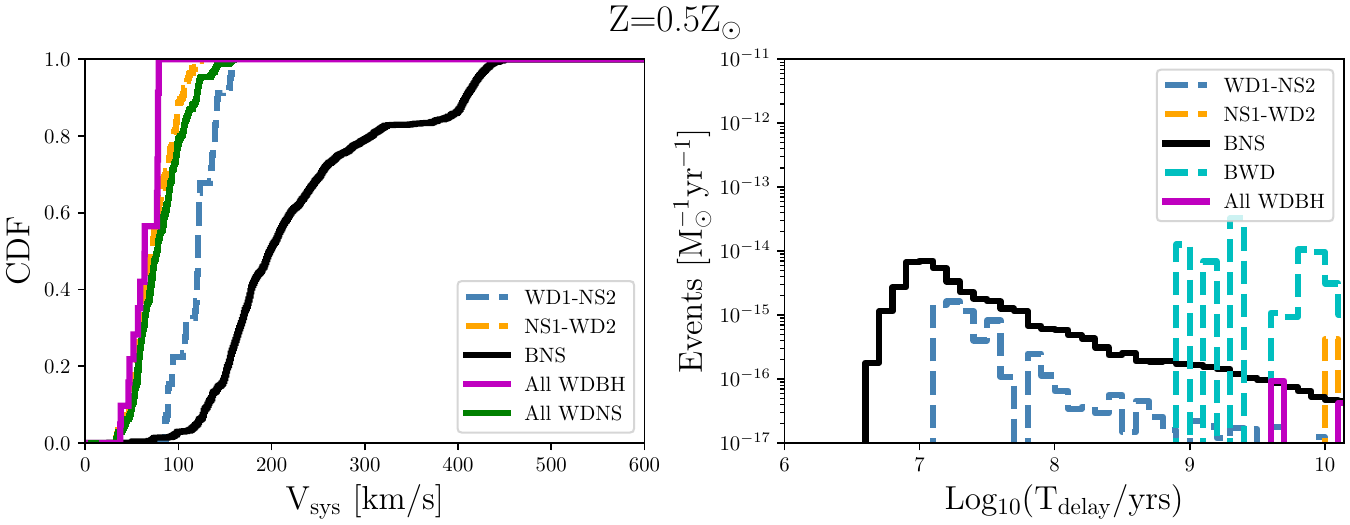}
\caption{As for Figure \ref{fig:vsystdelay} but with the natal kick prescription of \citet{2016MNRAS.461.3747B}.}
\label{fig:apx_bray}
\end{figure*}

\section{Merger rate comparisons} \label{app:B}
In Table \ref{tab:ratecompare} we compare our predicted WDNS and WDBH merger rates, both with other population synthesis models and with rates derived from observations of Galactic binaries. The {\sc bpass} WDNS predictions lie between those of SeBa and rates derived from observations of WDNS binaries (detached and interacting). Comparing with SeBa, the substantially lower {\sc bpass} merger rate can also be seen in the binary WD population, and is due to different treatments of mass transfer stability and the common envelope phase \citep{2025A&A...699A.172V}. This ultimately produces compact binaries with wider initial orbits than SeBa, and hence lower merger rates (due to longer GW in-spiral times), even though {\sc bpass} predicts {\it more} compact binaries containing a WD overall \citep{2024MNRAS.534.1707T,2024arXiv241102563T}. 

\begin{table}
	\centering
	\caption{Literature predictions for WDNS and WDBH merger rates $R$, based on the discussion of \citet{2018AandA...619A..53T}. Population synthesis rates are provided above the horizontal line, and unless otherwise indicated, assume a constant star formation rate of 3\,M$_{\odot}$\,yr$^{-1}$. We provide {\sc bpass} merger rates at metallicity $Z=0.010$ (by mass fraction). Merger rates derived from observed Galactic binary populations are listed below the line.}
	\label{tab:ratecompare}
	\begin{tabular}{llcc} 
		\hline
		\hline
             & & \multicolumn{2}{c}{$R$ [yr$^{-1}$]} \\
		  Code/constraint & Reference & WDNS & WDBH \\
		\hline
            {\sc bpass} & This work & 1.4$\times$10$^{-5}$ & 1.8$\times$10$^{-7}$ \\
            BSE & \citet{2021ApJ...920...81S} & - & $<$3$\times$10$^{-6}$ \\
            SeBa & \citet{2018AandA...619A..53T} & (1-2)$\times$10$^{-4}$ & - \\
            SeBa$\star$ & \citet{2001AandA...375..890N} & 1.4$\times$10$^{-4}$ & 1.9$\times$10$^{-6}$ \\
		\hline
            Pulsar binaries & \citet{2017MNRAS.467.3556B} & 2.6$\times$10$^{-4}$ & - \\
            X-ray binaries & \citet{2004MNRAS.354...25C} & (1-10)$\times$10$^{-6}$ & - \\
            WDNS binaries & \citet{2004ApJ...616.1109K} & 4.1$\times$10$^{-6}$ & - \\
            WDNS binaries & \citet{2002MNRAS.335..369D} & 10$^{-5}$ - 10$^{-4}$ & - \\
		\hline
	\end{tabular}
 \newline
 $\star$ - assumes an exponential SFH with a present day SFR of 3.6\,M$_{\odot}$\,yr$^{-1}$
\end{table}

\end{appendix}

\end{document}